\documentstyle[amssymb,epsfig]{mn2e}

\title[The Globular Cluster Gaussian Initial Mass Function]
{The Origin of the Gaussian Initial Mass Function of Old Globular Cluster Systems}
\author[G.~Parmentier \& G.~Gilmore]{Genevi\`eve Parmentier 
\thanks{E-mail: gparm@ast.cam.ac.uk} \& Gerard Gilmore \\
Institute of Astronomy, University of Cambridge, 
Madingley Road, Cambridge CB3 0HA, United Kingdom}
\begin{document}

\date{Accepted .... Received ... ; in original form ...}

\pagerange{\pageref{firstpage}--\pageref{lastpage}} \pubyear{2004}

\maketitle

\label{firstpage}
 
\begin{abstract}
Evidence favouring a Gaussian initial mass function for systems of old globular 
clusters has accumulated over recent years.  We show that an approximately Gaussian mass 
function is naturally generated from a power-law mass distribution of protoglobular
clouds by expulsion from the protocluster of star forming gas due to supernova activity, 
provided that the power-law mass distribution shows a lower-mass limit.  As a result of gas
loss, the gravitational potential of the protocluster gets weaker and only 
a fraction of the newly formed stars is retained.  The mass fraction of bound stars ranges from zero 
to unity, depending on the local star formation efficiency $\epsilon$.  Assuming that $\epsilon$
is independent of the protoglobular cloud mass, 
we investigate how such variations affect the mapping of a protoglobular 
cloud mass function to the resulting globular cluster initial mass function.  A truncated
power-law cloud mass spectrum generates bell-shaped cluster initial mass functions, 
with a turnover location mostly sensitive to the lower limit of the cloud 
mass range.  Assuming instantaneous gas removal and a slope $\alpha \simeq -1.7$ 
for the cloud mass spectrum, we evolve the derived cluster initial 
mass functions up to an age of 13\,Gyr in a potential like that of the Milky Way.  
We obtain a good match to the Old Halo cluster 
mass function, with a present-day mass mass fraction of clusters in the halo of $2\,\%$, as is observed,
with $m_{low} \simeq 6 \times 10^5\,{\rm M}_{\odot}$, $m_{up} \ge 5 \times 10^6\,{\rm M}_{\odot}$,
$\delta \simeq -2.9$ and $r_c \simeq 0.025$, respectively the lower and upper limits of
the cloud mass range, the slope and the core of the power-law spectrum for the
star formation efficiency.  The steep slope $\delta$ means that most protoglobular 
clouds achieve too low a star formation efficiency
to give rise to bound star clusters following gas removal.  As a result, most newly
formed stars are scattered into the field soon after their formation.  Gas removal during
star formation in massive clouds
is thus likely the prime cause of the predominance of field
stars in the Galactic halo.  The shape of the present-day cluster mass 
function depends weakly on the underlying distribution of the star formation efficiency.    
Finally, we show that a Gaussian mass function for the protoglobular clouds with a mean 
${\rm log}m_G \simeq 6.1-6.2$ and a standard deviation $\sigma \lesssim 0.4$ 
provides results very similar to those resulting from a truncated power-law 
cloud mass spectrum, that is, the distribution function of masses of protoglobular clouds 
influences only weakly the shape of the resulting globular star cluster initial mass function.
The gas removal process and the protoglobular cloud mass-scale dominate the relevant physics.

\end{abstract}


\section{Introduction}
\label{sec:intro}

Globular clusters are dense spherical gravitationally-bound stellar clusters.
By virtue of the age of the oldest clusters ($\simeq 13$\,Gyr), they are invaluable probes into the
earliest evolutionary stages of their host galaxy.  In that context however, the original properties
of (systems of) globular clusters will have been modified by the effects of a Hubble-time of evolution in their galactic environment.  It is essential to disentangle formation from evolution.  
How the present-day globular cluster mass distribution in a large galaxy compares with the initial one 
constitutes one such striking example.  The cluster mass function
\footnote{In what follows, we adopt the nomenclature of McLaughlin \& Pudritz (1996).  We call mass {\sl spectrum} the number of objects per {\sl linear} mass interval ${\rm d}N/{\rm d}m$,
while we refer to the mass {\sl function} to describe the number of objects per {\sl logarithmic} mass interval ${\rm d}N/{\rm dlog}~m$. }, namely, the cluster number per constant 
logarithmic cluster mass interval ${\rm d}N/{\rm dlog}m$, which is proportional 
to the number of objects per magnitude unit, constitutes a primary characteristic of 
any globular cluster system hosted by a massive galaxy.  Intriguingly, it shows only a weak 
dependence on the size, the morphological type or the environment of the host galaxy (Ashman 
\& Zepf 1998, Harris 1999).  This universal globular cluster mass function is 
usually fitted with a Gaussian with a mean of $\overline{{\rm log}m} \simeq 5.2-5.3$ and a 
standard deviation of $\sigma _{{\rm log}m} \simeq 0.5-0.6$.  The underlying mass spectrum 
(i.e., the number of objects per {\it linear} mass interval) is well described by a two-index power-law, 
with exponents $\sim -2$ and  $\sim -0.2$ above and below $\sim 2 \times 10^5$ M$_{\odot}$, 
respectively (McLaughlin 1994).  The peak of the Gaussian function in fact coincides with the
cluster mass at which the slope of the mass spectrum changes.

Individual globular clusters and globular cluster systems having evolved over a 
Hubble-time in their galactic environment, the globular cluster initial mass function
has remained model-dependent, with two competing hypotheses.  It may have 
been a featureless power-law with a slope of $\sim -1$, the Gaussian
function characteristic of old globular cluster populations then 
resulting from evolutionary effects, predominantly the preferential removal 
of the more vulnerable low-mass clusters (Fall \& Zhang 2001).  In that case, the cluster
mass at the turnover of the present-day mass function depends on the age
of the cluster system as well as on the cluster disruption time-scale
in the galaxy of relevance.  The older the cluster system and/or the shorter the disruption time-scale, the higher the cluster mass at the mass function turnover.  
In this situation, Vesperini (2001) shows that it is 
not straightforward to produce (almost) universal globular cluster mass functions 
in very different types of galaxies if starting from an initial featureless power-law.
Alternatively, Vesperini (1998) demonstrates that the present-day cluster
mass function represents an equilibrium state "able to preserve its initial shape 
and parameters for one Hubble-time through a subtle balance between disruption
of clusters and evolution of the masses of the surviving ones", even though
a significant fraction of the initial cluster population has been destroyed.  That is,
the initial cluster mass function may also be a Gaussian similar to that today.  
Should that be the case, the observed Gaussian shape of the cluster mass function
and its universality among galaxies are the preserved imprints of the globular cluster 
formation process. 

Parmentier \& Gilmore (2005) and Vesperini et al.~(2003)
provide evidence for a Gaussian cluster initial mass function in the 
Galactic halo and in the giant elliptical M87, respectively. 
Theoretical support for a general bell-shaped cluster initial mass function has 
been missing so far however.  
   
The cluster initial mass function has often been assumed to mirror
the mass function of the cluster gaseous progenitors (McLaughlin \& Pudritz 1996,
Elmegreen \& Falgarone 1996, Elmegreen \& Efremov 1997).  However,
this is true only if all the cluster forming clouds turn the gas they
are made of into bound star clusters with roughly the same efficiency.  
At its earliest stages, a protostellar cluster is embedded in the left-over star forming gas.
That residual gas is blown away once the newly formed massive 
stars explode as supernovae.  As a result of the corresponding
weakening of its gravitational potential, the protocluster can emerge
unbound from the violent relaxation phase which follows.  By means of analytical computations, 
Hills (1980) predicted that, in the case of instantaneous gas removal 
\footnote{The gas removal takes place on a time-scale $\tau _{gr}$ shorter than a
protocluster crossing-time $\tau _{cross}$, implying that the stars do not have time 
to adjust to the new gravitational potential.  Their velocity dispersions before and immediately after
gas loss are thus considered to be the same.},
initially virialized systems eventually dissolve for star forming
efficiencies (i.e., the mass fraction of gas ending up in stars,
hereafter $\epsilon$) smaller than 50\,\%.  That is, star forming 
clouds must be better than 50\,\% efficient in converting gas into stars 
in order to produce bound stellar clusters.  The limited variations in $\epsilon$
(i.e., less than a factor of 2) may then guarantee that the initial mass function
of the clusters is that of their parent clouds, shifted downwards by a factor 
of $\lesssim 2$.  

Using $N$-body simulations however, Lada, Margulis \& Dearborn (1984) 
revisited that issue and pointed out that a system which becomes globally unbound
due to a less than 50\,\% efficiency is not necessarily completely disrupted and 
may retain a core of bound stars.  Conversely, even a star forming efficiency larger
than 50\,\% does not prevent the protocluster from losing a fraction of its stars.
This is due to the fact that the protocluster stars are characterized
by a velocity distribution.  The stars in the low-velocity tail of the
distribution tend to survive as a gravitationally bound entity even if
$\epsilon \leq 0.5$, while high-velocity stars escape even if $\epsilon > 0.5$. 
As a result, the initial mass of a stellar cluster is not determined
by the star formation efficiency only.  It depends on the mass fraction 
of the cluster parent cloud which is turned into stars {\it which remain 
bound after the dispersal of the gaseous component}.  

Modelling the dynamical evolution of gas-embedded clusters, that is, how they
respond to supernova activity, has since then attracted considerable attention 
(Verschueren 1989, Goodwin 1997, Geyer \& Burkert 2001, Boily \& Kroupa 2003, 
Fellhauer \& Kroupa 2005).  Surprisingly however, whether the early
evolution of gas-embedded protoclusters influences the cluster initial mass function,
that is, the mass function of the emerging gas-free bound groups of stars, 
has remained poorly explored (Kroupa \& Boily 2002).  Since the initial mass of a 
star cluster depends on gas removal, gas mass loss is actually worth being put to the test as a 
possible controlling mechanism of the globular cluster initial mass function.  

The outline of the paper is as follows.  In section \ref{sec:impact}, we present the results of Monte-Carlo simulations highlighting the impact of gas removal upon the shape of the 
globular cluster initial mass function with respect to that of their gaseous progenitors.  We also discuss how variations in the different model parameters affect the cluster initial mass function.  In Sect.~\ref{sec:Gal_Halo}, we apply our model to the specific case of the Galactic stellar halo.  Finally, we present our conclusions in Sec.~\ref{sec:conclu}.  

\section{Impact of Gas Removal on the Cluster Initial Mass Function}
\label{sec:impact}

\begin{figure}
\begin{minipage}[t]{\linewidth}
\centering\epsfig{figure=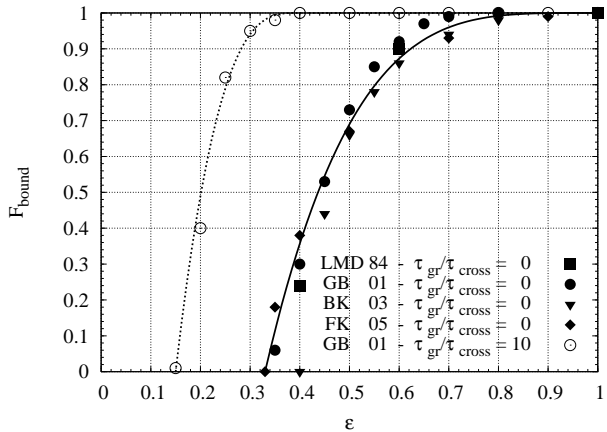, width=\linewidth}
\end{minipage}
\caption{Relations between the fraction $F_{bound}$ of stars remaining bound to the protocluster after gas removal and the star formation efficiency $\epsilon$ achieved by the gaseous progenitor.  The plain/open symbols correspond to the case of rapid/slow gas removal (i.e, $\tau _{gr} << \tau _{cross}$ or $\tau _{gr} >> \tau _{cross}$).  Data are from Lada et al.~(1984), Geyer \& Burkert (2001), Boily \& Kroupa (2003) and Fellhauer \& Kroupa (2005) (respectively quoted as LMD84, GB01, BK03 and FK05).  The solid and dotted lines depict the relations used in our simulations.  }
\label{fig:Fb_SFE} 
\end{figure}

The formation of a stellar cluster is terminated when the newly formed
massive stars go supernovae and blow away the gas left-over by
the star formation process.  Following the dispersal of that residual
gas, the stars suddenly find themselves in a shallower gravitational
potential, entailing either the escape of some of them, or even the complete 
destruction of the protocluster.  If the gas is removed explosively (i.e. $\tau _{gr} << \tau _{cross}$), the star formation efficiency must be larger than a threshold value $\epsilon _{th} = 33$\,\%
(see below and Fig.~\ref{fig:Fb_SFE}) for the protocluster to retain a bound core of stars. 
The mass fraction $F_{bound}$ of stars remaining bound after gas removal ranges from zero (when the efficiency $\epsilon$ is at its threshold value, or lower, i.e. $\epsilon \leq 0.33$) up to unity (when $\epsilon \lesssim 1$, so that gas removal is just a small 
perturbation of the stellar system).  Considering the case of initially virialized gas-embedded
protoclusters, various studies (Lada et al.~1984, 
Geyer \& Burkert 2001, Boily \& Kroupa 2003, Fellhauer \& Kroupa 2005) 
have led to fairly consistent results regarding the $F_{bound}$
vs. $\epsilon$ relation (see the plain symbols in Fig.~1).   
The knowledge of this relation enables us to relate the initial mass $m_{init}$ 
of a gas-free bound star cluster to the mass $m_{cloud}$ of its gaseous progenitor, namely:
\begin{equation}
m_{init} = F_{bound} \times \epsilon \times m_{cloud}\;.
\label{eq:minit}
\end{equation}
As a result of the large variations in the bound star formation efficiency $F_{bound} \times \epsilon$, 
an assumed simple mapping between the mass function of the cluster gaseous precursors 
on the one hand and the initial mass function of the clusters on the other hand 
can no longer be taken for granted.  We now investigate how 
the cluster initial mass function differs with respect to the protoglobular
cloud mass function as a result of gas removal.  

\subsection{From a truncated power-law mass function to a Gaussian mass function}
\label{subsec:PL_G}
In what follows, we assume that the protoglobular cloud mass spectrum 
obeys a power-law 
\begin{equation}
{\rm d}N \propto m^{\alpha} {\rm d}m\,,
\end{equation}
with $\alpha$ varying between $-2.5$ and $-1.5$, as is observed for giant molecular 
clouds and their star forming cores in the Local Group of galaxies (e.g. Rosolowski 2005, see also section \ref{subsec:isoQfGC_mlowmup}).  A power-law mass spectrum may result from the 
coalescence of initially small equal-mass cores into a system of more massive   
objects with a wide range of masses (McLaughlin \& Pudritz 1996).  
Alternatively, Elmegreen \& Falgarone (1996) and Elmegreen \& Efremov (1997) 
suggest that a power-law mass spectrum with $\alpha \lesssim -2$ for the cluster gaseous progenitors 
is an imprint of the fractal structure of the star forming gas.  As for the lower and upper limits of 
the cloud mass range, we adopt $m_{low}=4 \times 10^5 {\rm M}_{\odot}$ and $m_{up}=10^7 {\rm M}_{\odot}$, 
that is, the Jeans mass range (see section \ref{subsec:proto_gx}).  In the next section, we will 
explore how the mass function of the newly-born star clusters depends on these mass limits. 

Star forming regions are characterized by a range in their respective star formation efficiency 
$\epsilon$, so that the protoglobular cloud mass spectrum is convolved with an 
$\epsilon$ probability distribution function, which we describe by a decreasing power-law 
of slope $\delta$ and core $r_c$, that is:
\begin{equation}
\wp(\epsilon) = \frac{{\rm d}N}{{\rm d}\epsilon} = c_1 \left(1 + \frac{\epsilon}{r_c} \right) ^{\delta} + c_4\;.
\label{eq:sfe_prob}
\end{equation}
The two parameters $c_1$ and $c_4$ are determined so as to satisfy the two following constraints: (1) the integration of the probability distribution over the range $\epsilon = $[0,1] is unity, and the probability $\wp(\epsilon)$ is zero when $\epsilon = 1$.  

The formation of a bound star cluster requires its gaseous progenitor to achieve $\epsilon > \epsilon _{th}$ (where "th" stands for "threshold"), i.e., the {\it local} star formation efficiency must be greater than $\simeq 0.3 - 0.4$.  On the scale of a galaxy, star formation proceeds inefficiently, so the {\it global} star formation efficiency may be of order a few per cent only.  The core $r_c$ and the slope $\delta$ of the efficiency distribution $\wp (\epsilon )$ are thus bounded so that the mean star formation efficiency $\overline{{\epsilon}}$, namely, the mass fraction of gas converted into stars for an entire system of protoglobular clouds, is one per cent: 
\begin{equation}
\overline {\epsilon} = \int_{0}^{1} \epsilon ~ {\rm d}N(\epsilon) =0.01\;.
\label{eq_meansfe}
\end{equation}

For a given system of protoglobular clouds, the slope $\delta$ determines the fraction $f_{>th}$ of clouds achieving a star formation efficiency larger than the threshold $\epsilon _{th}$ and, therefore, the initial size of the cluster system.  Combined with the $F_{bound}$ vs. $\epsilon$ relation, it also determines the mean bound star formation efficiency $\overline{{\epsilon _b}}$, 
\begin{equation}
\overline{{\epsilon _b}} = \int_{\epsilon _{th}}^{1} (\epsilon \times F_{bound}) {\rm d}N(\epsilon)\,,
\label{eq_meanbsfe}
\end{equation}
that is, the total mass fraction of gas converted into stars residing in bound systems 
after gas removal.  To illustrate this with specific examples, Table \ref{tab:sfe_dist} 
lists the values of $f_{>th}$ and $\overline{{\epsilon _b}}$ for different triples 
($\delta$, $r_c$, $\epsilon _{th}$) (see section \ref {subsubsec:sfeth} for the case 
of non-explosive gas removal leading to $\epsilon _{th}$ significantly less than 0.33).  \\

\begin{table}
\begin{center}
\caption[]{Fraction $f_{>th}$ of protoglobular clouds giving rise to bound stellar clusters and mean bound 
star formation efficiency $\overline{{\epsilon _b}}$ for various distributions $\wp(\epsilon)$ 
and thresholds $\epsilon _{th}$ of the star formation efficiency, that is, for various triples 
($\delta$, $r_c$, $\epsilon _{th}$).  If $\delta \neq 0$, the core $r_c$ and the slope $\delta$ 
are bounded so that the mean star formation efficiency $\overline {\epsilon}$ across the 
whole galaxy is 0.01.  The threshold $\epsilon _{th}$ depends on the gas removal 
timescale $\tau _{gr}$ (see section \ref {subsubsec:sfeth})       } 
\label{tab:sfe_dist}
\begin{tabular}{  c c c c c  } \hline 
 $\delta$ & $r_c$    & $\epsilon _{th}$ & $f_{>th}$            & $\overline{{\epsilon _b}}$  \\ \hline
  $-4$   & $0.020$   & $0.33$           & $1.7 \times 10^{-4}$ & $3.7 \times 10^{-5}$        \\ 
  $-4$   & $0.020$   & $0.15$           & $1.6 \times 10^{-3}$ & $2.0 \times 10^{-4}$        \\ 
  $-2$   & $0.002$   & $0.33$           & $3.0 \times 10^{-3}$ & $9.0 \times 10^{-4}$        \\ 
  $-2$   & $0.002$   & $0.15$           & $1.0 \times 10^{-2}$ & $2.4 \times 10^{-3}$        \\ 
  $0$    &   -       & $0.33$           & $0.67$               & $0.39$            \\ 
  $0$    &   -       & $0.15$           & $0.85$               & $0.44$       \\ \hline
\end{tabular}
\end{center}
\end{table}   

In our model, each protoglobular cloud is randomly assigned a mass m$_{cloud}$ and a star formation efficiency $\epsilon$, both parameters being drawn independently from their respective distribution, i.e. we assume no correlation
between the mass of a cluster gaseous progenitor and its star formation efficiency.  \\
What the star formation efficiency of a star forming region depends on is still ill-known.  Elmegreen \& Efremov (1997) suggest that, for any given ambient pressure, there is a trend of increasing star formation efficiency for more massive clouds (bottom panel of their Fig.~4).  Considering a large galaxy, such a relation might be useful in any region which is spatially limited enough so that the ambient pressure does not vary significantly, that is, all clouds of that region are bounded by roughly the same pressure.  However, our study is concerned with the formation of an entire star cluster system, spanning several tens of kpc in size and, therefore, spatial variations in the ambient pressure of the protoglobular clouds should be accounted for as well.  Actually, Elmegreen \& Efremov 's (1997) model also predicts that, for any given cloud mass, a higher ambient pressure promotes a larger star formation efficiency.  As a result, two equal-mass protoglobular clouds may undergo markedly different star formation efficiencies due to different spatial locations in the protogalaxy, i.e., different external pressures.  Clearly, our model cannot rely on a simple one-to-one m$_{cloud}$-$\epsilon$ relation. 

What is more, the very existence of a monotonic m$_{cloud}$-$\epsilon$ relation, even within a spatially limited environment, is rather uncertain.  Overplotting the star formation efficiencies inferred for a few star forming regions of the solar neighbourhood (Lada \& Lada 2003, their Table 2) on the m$_{cloud}$-$\epsilon$ relations predicted by  Elmegreen \& Efremov (1997), we note a large scatter and the inability of the m$_{cloud}$-$\epsilon$ relation corresponding to the solar neighbourhood pressure to account for all data.  Moreover, the presumed steady rise of the star formation efficiency with the cloud mass (for a given pressure) stems from an increase of the cloud specific binding energy, hence from an increase of the cloud velocity dispersion (section 4 of Elmegreen \& Efremov 1997).  How the star formation efficiency responds to the latter is not accounted for, however.  Higher gas velocity dispersions are promoted by larger turbulence and/or magnetic field pressures within the star forming cloud (e.g. Harris \& Pudritz 1994), two processes known for {\it hampering} star formation.  Specifically, Schmeja, Klessen \& Froebrich 's (2005, their section 4.2) simulations demonstrate the existence of an inverse correlation of the star formation efficiency with the Mach number, that is, a higher velocity dispersion gas is less efficient at forming stars.  Therefore, two counteracting effects are at work: while the increase of the cloud specific binding energy raises the star formation efficiency, this gets lowered by the corresponding increase of the gas velocity dispersion, equivalently the growth of the magnetic fields and/or turbulence pervading the star forming gas.  Even for a given ambient pressure, the star formation efficiency does not depend on the cloud mass monotonically.  

In the presence of significant variations in the ambient pressure, turbulence and magnetic field pressures from one star forming cloud to another, as may be expected when considering a whole protogalaxy, the m$_{cloud}$-$\epsilon$ relation is likely reduced to a mere scatter, hence justifying our choice of a probability distribution function $\wp (\epsilon)$ independent of any explicit protoglobular cloud mass dependence.  That is, given the lack of a clear
physical prediction, and the observational uncertainties which we review in 
section \ref{subsec:proto_gx} below, in this analysis we adopt a probabilistic approach, to
investigate the extent to which it can be a viable option.  \\

Following the onset of supernova activity, the gas-embedded cluster gets exposed as its 
residual gas is removed.  Not only does the protocluster lose its gaseous component, it 
also loses a fraction of its initial {\it stellar} mass.  We account for this phase by 
matching each efficiency value $\epsilon$ to the corresponding fraction $F_{bound}$ of 
bound stars.  The $F_{bound}$ vs. $\epsilon$ relation we are using is shown as the solid 
line in Fig.~\ref{fig:Fb_SFE}.  It matches the results of Geyer \& Burkert (2001) 
and Fellhauer \& Kroupa (2005) and is characterized by $\epsilon _{th} = 0.33$. 

Finally, the initial mass $m_{init}$ of globular clusters is derived following equation \ref{eq:minit}.  
The corresponding cluster initial mass functions, along with their protoglobular cloud mass functions, 
are shown in the top and middle panels of Fig.~\ref{fig:alpha_delta}, assuming $\delta = -4$ and $\delta = -2$.  
It is worth emphasising that, although the simulations were started with truncated power-law cloud mass 
functions, the newly formed gas-free bound star clusters show bell-shaped mass functions.  That is, little memory 
of the cloud mass function is retained as gas removal generates the low-mass regime of the cluster mass 
distribution.  Therefore, {\it gas removal is supported as a prime candidate mechanism 
responsible for generating bell-shaped globular cluster initial mass functions} and, as such, may be central to our understanding of the origin of the bell-shaped universal globular cluster mass function.  

\subsection{A possible origin for the universal location of the globular cluster mass function turnover}
\label{subsec:TO}

In this section, we investigate how the derived cluster initial mass functions respond to model parameter variations.  The results are illustrated in the panels of Figs.~\ref{fig:alpha_delta} - \ref{fig:avsfe}.  Unless otherwise stated on the panels, we adopt $m_{low} = 4 \times 10^5\,{\rm M}_{\odot}$, $m_{up} = 10^7\,{\rm M}_{\odot}$, $\delta =-2$, $\epsilon _{th} = 0.33$ and $\overline{\epsilon} = 0.01$.  For each simulation, Table \ref{tab:res} lists the values of the input parameters along with the characteristics of the generated cluster initial mass functions.  Specifically, we present the logarithm of the cluster mass at the turnover ${\rm log}m_{\rm TO}$, the mean logaritmic cluster mass $\overline{{\rm log}m}$, the standard deviation $\sigma$, the skewness and the curtosis
\footnote{We remind the reader that the skewness of a distribution characterizes its degree of asymetry while the curtosis measures its relative peakedness or flatness with respect to the null value of a Gaussian (Press et al.~1992).} of the ${\rm log}m$ distribution.
The cluster initial mass functions are not strictly Gaussian, as shown by the top and middle panels of Fig.~\ref{fig:alpha_delta}-\ref{fig:sfeth}, where we have overlaid the high mass regime of each mass function with a Gaussian.  Our model predicts a number of low-mass
(i.e. with a mass smaller than that at the turnover) clusters which is larger than if the cluster mass were drawn from a Gaussian distribution.  As a result, the skewness of the cluster mass distributions is negative.  Regarding the width of the mass function, Table \ref{tab:res} shows that the standard deviation $\sigma$ remains roughly constant regardless of the input parameter values.  It is on the order of 0.6 and, therefore, consistent with the width of observed globular cluster mass functions.  
   
As emphasized in the introduction, if the globular cluster initial mass function is actually a bell-shape similar to that observed today, then the origin of the almost universal cluster mass at the turnover is locked into the cluster formation process.  In the frame of our model, we now explore what the turnover location depends on.  

\subsubsection{The spectral index $\alpha$ of the cloud mass spectrum}
\label{subsubsec:alpha}

We firstly address the effect of varying the spectral index $\alpha$ of the protoglobular cloud mass spectrum.  We consider three values of $\alpha$: $-2.5, -2, -1.5$.  As a result of the greater fraction of high-mass clouds in the case of a shallower cloud mass spectrum, the cluster mass at the turnover may be larger if $\alpha =-1.5$ than if $\alpha =-2.5$.  Such an effect actually shows up in case of a steep star formation efficiency distribution ($\delta =-4$), although the effect remains moderate, with $[{\rm log}m_{TO}]_{\alpha =-1.5}-[{\rm log}m_{TO}]_{\alpha =-2.5} \lesssim 0.2$.  For shallower star formation efficiency distributions ($\delta =-2$ or $\delta =0$), the effect is negligible
(see Fig.~\ref{fig:alpha_delta}).  Memory of a steeper cloud mass spectrum is sometimes 
retained as a more pronounced skewness of the cluster initial mass function.  

\subsubsection{The slope $\delta$ of the star formation efficiency distribution $\wp (\epsilon)$}
\label{subsubsec:delta}
Star forming regions do show variations in their respective star formation efficiency (Lada \& Lada 2003).  In our model, this is accounted for by the functional form $\wp (\epsilon)$, which describes the probability distribution of the star formation efficiency $\epsilon$.  This is likely the most ill-determined ingredient of our model.  Actually, very few $\epsilon$ measurements exist for star forming regions in the Galactic disc since these require estimates of both the gaseous mass and the stellar mass.  We know however that on a galactic scale, equivalent here to the scale of a system of protoglobular clouds, star formation proceeds with a global efficiency of a few per cent only.  Additionally, the vast majority of star forming regions give rise to unbound stellar groups, a process sometimes referred to as "infant mortality".  In the Galactic disc for instance, the birthrate of embedded clusters in molecular cloud cores is observed to be extremely high compared to the birth rate of classical open clusters, thus suggesting that only a small fraction (a few per cent at most) of embedded clusters emerge from their natal cores as bound clusters.  As quoted by Lada \& Lada (2003), this high infant mortality results from the low to modest star formation efficiency and rapid gas dispersal that characterize their birth, that is, 
$\epsilon < \epsilon _{th}$ for most star forming regions. 
Observations of violent star forming regions in interacting and merging galaxies lead to similar conclusions.  For instance, Fall, Chandar \& Whitmore (2005) show that the age distribution of star clusters in the Antennae galaxies (NGC4038/39) declines approximately as ${\rm d}N/{\rm d}\tau \propto \tau ^{-1}$ over the range $10^6 < \tau < 10^9$\,yr.  They interpret this steep decline as evidence for a high rate of infant mortality, that is, most of the young clusters are not tightly gravitationally bound and are disrupted shortly after they form by the energy and momentum input from young massive stars to the residual star forming gas, and massive star mass loss.  

\begin{figure}
\begin{minipage}[t]{\linewidth}
\centering\epsfig{figure=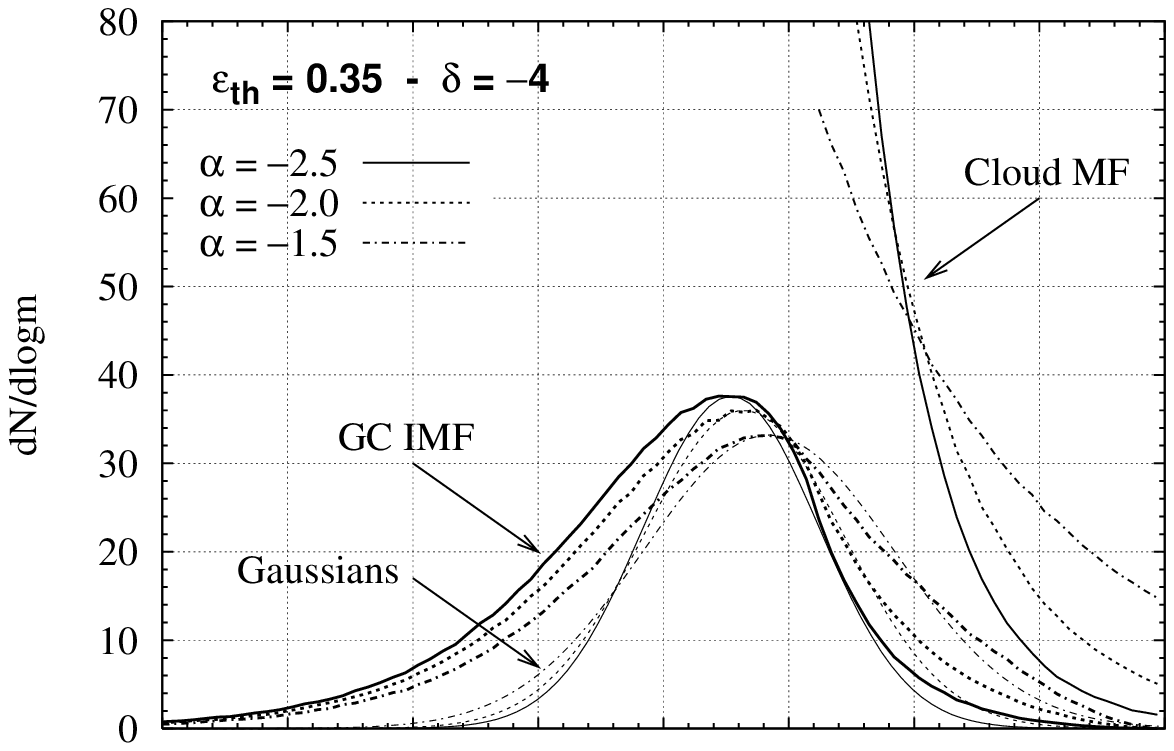, width=\linewidth} 
\end{minipage}
\vfill
\vspace{-10mm}
\begin{minipage}[t]{\linewidth}
\centering\epsfig{figure=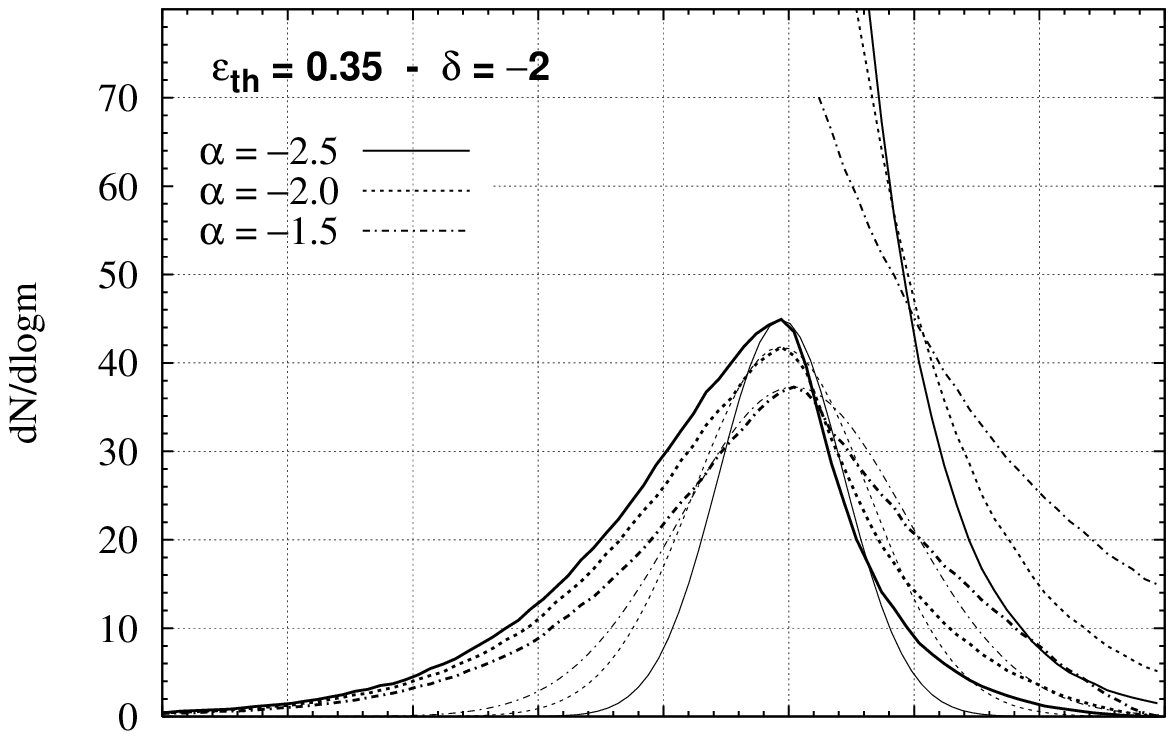, width=\linewidth}
\end{minipage}
\vfill  \vspace{-10mm}
\begin{minipage}[t]{\linewidth}
\centering\epsfig{figure=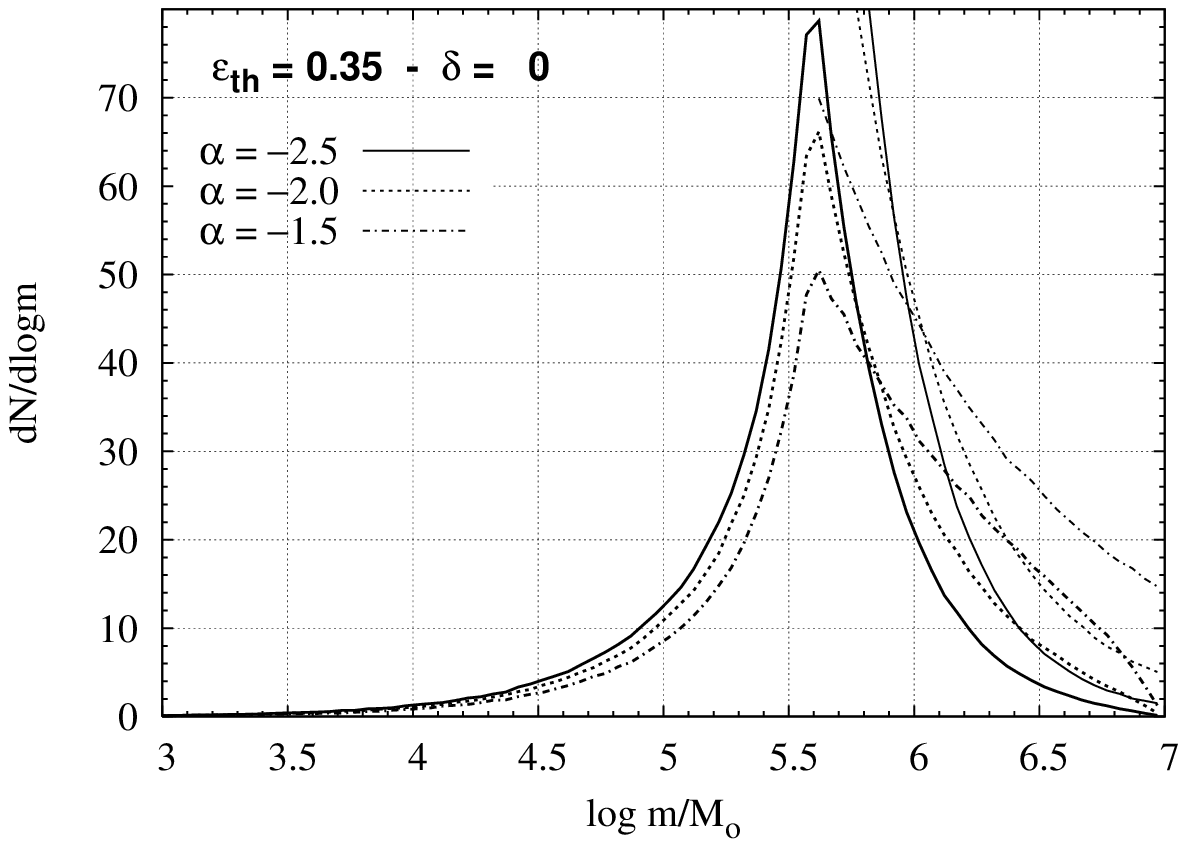, width=\linewidth}
\end{minipage}
\caption{Truncated power-law protoglobular cloud mass functions and resulting bell-shaped cluster initial mass functions in the case of instantaneous gas removal ($\epsilon _{th}=0.33$, solid line in Fig.~\ref{fig:Fb_SFE}).  The cloud mass range is $4 \times 10^5\,{\rm M}_{\odot} < m_{cloud} < 10^7\,{\rm M}_{\odot}$.  The solid, dotted and dashed-dotted lines correspond to $\alpha =-2.5, -2, -1.5$, respectively, where $\alpha$ is the spectral index of the cloud mass spectrum.  The bottom, middle and top panels correspond to three distinct slopes $\delta$ of the probability distribution $\wp (\epsilon)$ for the star formation efficiency, respectively, $\delta =-4, -2$ and $0$.  In the top and middle panels, Gaussian curves overlaid on each cluster initial mass function show that our model predicts more low-mass clusters than does a pure Gaussian mass function.}
\label{fig:alpha_delta} 
\end{figure}

Therefore, even though the overall distribution $\wp (\epsilon)$ remains poorly determined, it actually makes sense to describe it as a decreasing power-law of the efficiency $\epsilon$ with a slope steep enough so that only a small fraction of the protoglobular clouds convert gas into stars with an efficiency $\epsilon$ larger than the threshold $\epsilon _{th}$.  We consider two values for the slope of $\wp (\epsilon)$: $\delta =-4$ and $\delta =-2$.  At the same time, we retain the constraint of a global star formation efficiency $\overline {\epsilon}$ of one per cent and we determine $r_c$ accordingly (see Table \ref{tab:sfe_dist}).  Additionally, we investigate what happens if the star formation efficiency is uniformly distributed (i.e. $\wp (\epsilon)$ is flat).  In that case also,  bell-shaped cluster initial mass functions emerge, although with a much sharper peak.  The corresponding cluster initial mass functions are shown in the three panels of Fig.~\ref{fig:alpha_delta} with, from top to bottom, $\delta =-4$, $\delta =-2$ and $\delta =0$.  

As expected, the steeper the distribution $\wp (\epsilon)$, the smaller the cluster mass at the turnover.  
The dependence of the turnover location on $\delta$ remains limited however, with the steep ($\delta =-4$) and flat ($\delta =0$) $\wp (\epsilon)$ distributions leading to a difference of at most 0.4 in  ${\rm log}m$.  The cluster mass at the turnover is even more robust in case of a shallow cloud mass spectrum (i.e. if $-2 \leq \alpha \leq -1.5$, $[{\rm log}m_{TO}]_{\delta =0}-[{\rm log}m_{TO}]_{\delta =-4} \lesssim 0.3$).  If the mass spectrum of the protoglobular clouds were similar to 
that of the cluster forming cores of molecular clouds in the present-day Galactic disc, for which  $\alpha \simeq -1.7$ (Lada \& Lada 2003), then the uncertainties in the predicted turnover location arising from our misknowledge of $\wp (\epsilon)$ are not larger than the uncertainties in the globular cluster mass-to-light ratio, which is on the order of a factor of 2.  Varying the slope $\delta$ of $\wp (\epsilon)$ mostly affects the curtosis of the cluster initial mass function in the sense that a shallower efficiency distribution increases its peakedness.  

\begin{figure}
\begin{minipage}[t]{\linewidth}
\centering\epsfig{figure=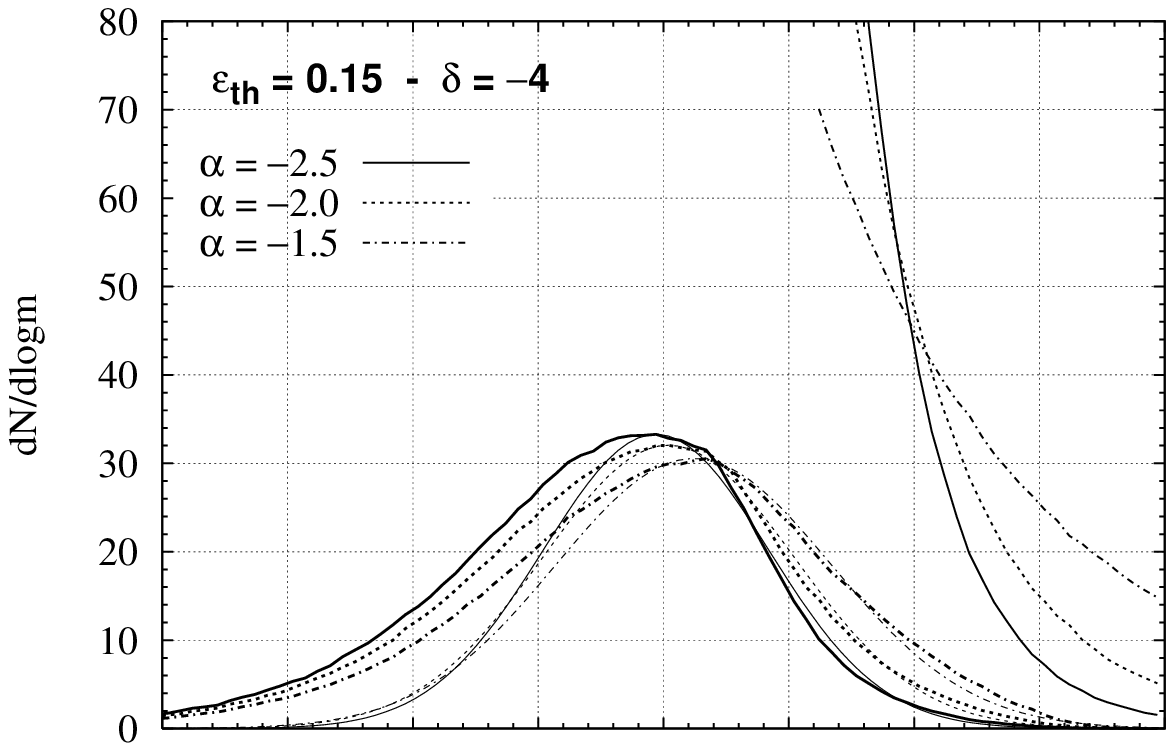, width=\linewidth} 
\end{minipage}
\vfill
\vspace{-10mm}
\begin{minipage}[t]{\linewidth}
\centering\epsfig{figure=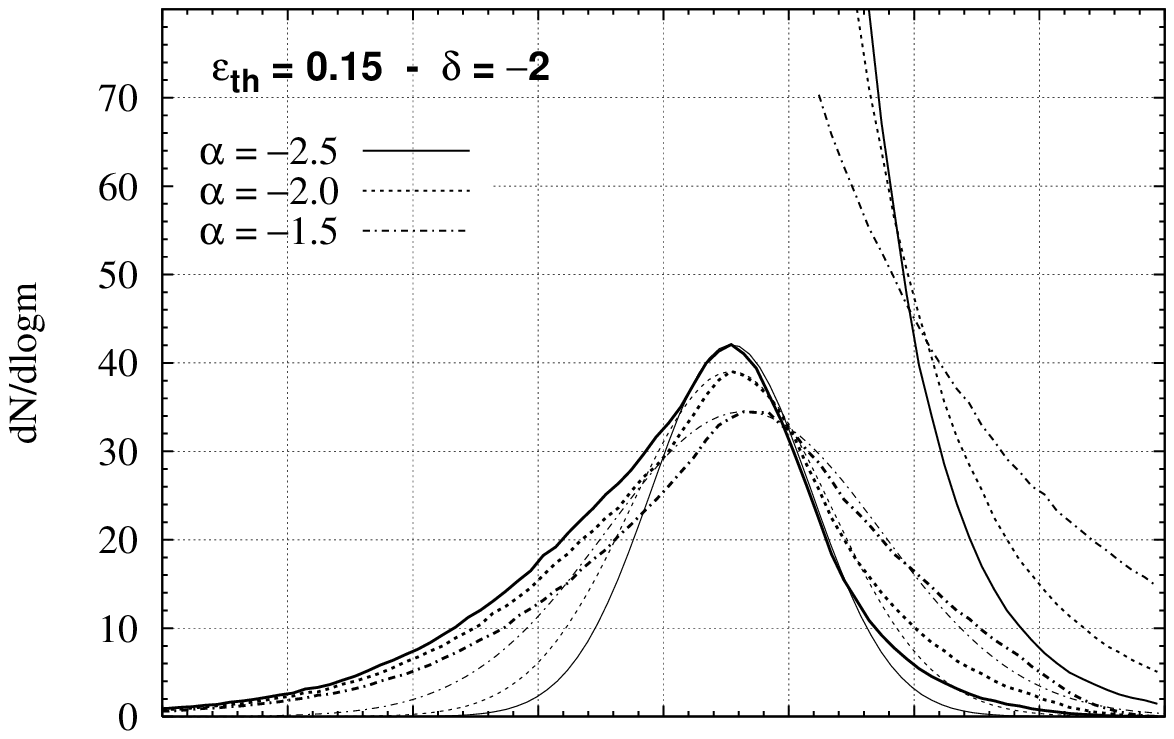, width=\linewidth}
\end{minipage}
\vfill  \vspace{-10mm}
\begin{minipage}[t]{\linewidth}
\centering\epsfig{figure=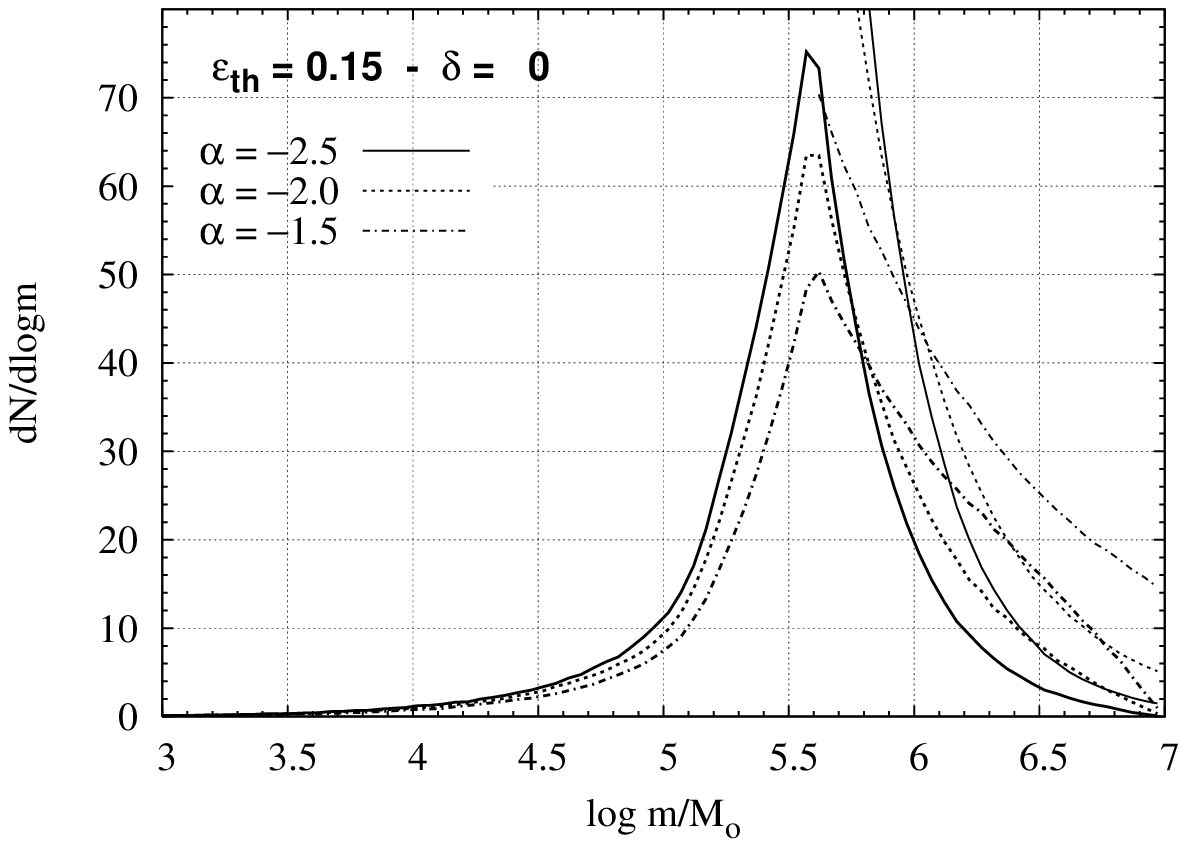, width=\linewidth}
\end{minipage}
\caption{Same as Fig.~\ref{fig:alpha_delta}, but in case of slow gas removal ($\epsilon _{th}=0.15$, dotted line in Fig.~\ref{fig:Fb_SFE}). } 
\label{fig:sfeth} 
\end{figure}

\subsubsection{The star formation efficiency threshold $\epsilon _{th}$}
\label{subsubsec:sfeth}

The ultimate fate of an embedded cluster (i.e. whether it will survive as a bound stellar cluster or not) depends on whether the star formation efficiency $\epsilon$ achieved by the gaseous precursor is larger than the threshold $\epsilon _{th}$ or not.  The efficiency threshold itself depends on the timescale $\tau _{gr}$ for removing the gas out of the protocluster.  If the gas is removed instantaneously (i.e., $\tau _{gr} << \tau _{cross}$), we get $\epsilon _{th} \simeq 0.33$ (see the plain symbols in Fig.~\ref{fig:Fb_SFE}).  For slower gas removal ($\tau _{gr} > \tau _{cross}$) however, the efficiency threshold gets smaller (see the open symbols in Fig.~\ref{fig:Fb_SFE}, data taken from Geyer \& Burkert 2001) because the stars can now adjust adiabatically to 
the new gravitational potential they sit in and expand to a new state of virial equilibrium, even for $\epsilon \lesssim 0.33 $.  If $\tau _{gr} \simeq 10 \times \tau _{cross}$, $\epsilon _{th} \simeq 0.15$.

Although protoclusters containing O stars are expected to remove any residual star forming gas on a timescale shorter than $\tau _{cross}$ (Geyer \& Burkert 2001, Lada \& Lada 2003, Kroupa 2005), it is nevertheless interesting to investigate whether a slow gas dispersal affects our results significantly. 
This situation mimics a weakly bound cluster where mass loss from the most massive stars on a stellar 
evolution timescale ($\sim 10^6$\,yr) can be significant.
Using the dashed curve in Fig.~\ref{fig:Fb_SFE} as the $F_{bound}$ vs. $\epsilon$ relation (i.e. considering $\tau _{gr} = 10\,\tau _{cross}$ and $\epsilon _{th} = 0.15$ instead of $\tau _{gr} << \tau _{cross}$ and $\epsilon _{th} = 0.33$), we obtain the cluster initial mass functions of Fig.~\ref{fig:sfeth}, each panel differing from its counterpart in Fig.~\ref{fig:alpha_delta} by the star formation efficiency threshold only.  With respect to the case of instantaneous gas expulsion, any value of $F_{bound}$ is now coupled with a lower value of $\epsilon$ and we therefore expect a downward shift of the turnover location.  This remains moderate however, with $[{\rm log}m_{TO}]_{{\epsilon}_{th} =0.33}-[{\rm log}m_{TO}]_{{\epsilon}_{th} =0.15} \lesssim 0.25$.  

\begin{figure}
\begin{minipage}[t]{\linewidth}
\centering\epsfig{figure=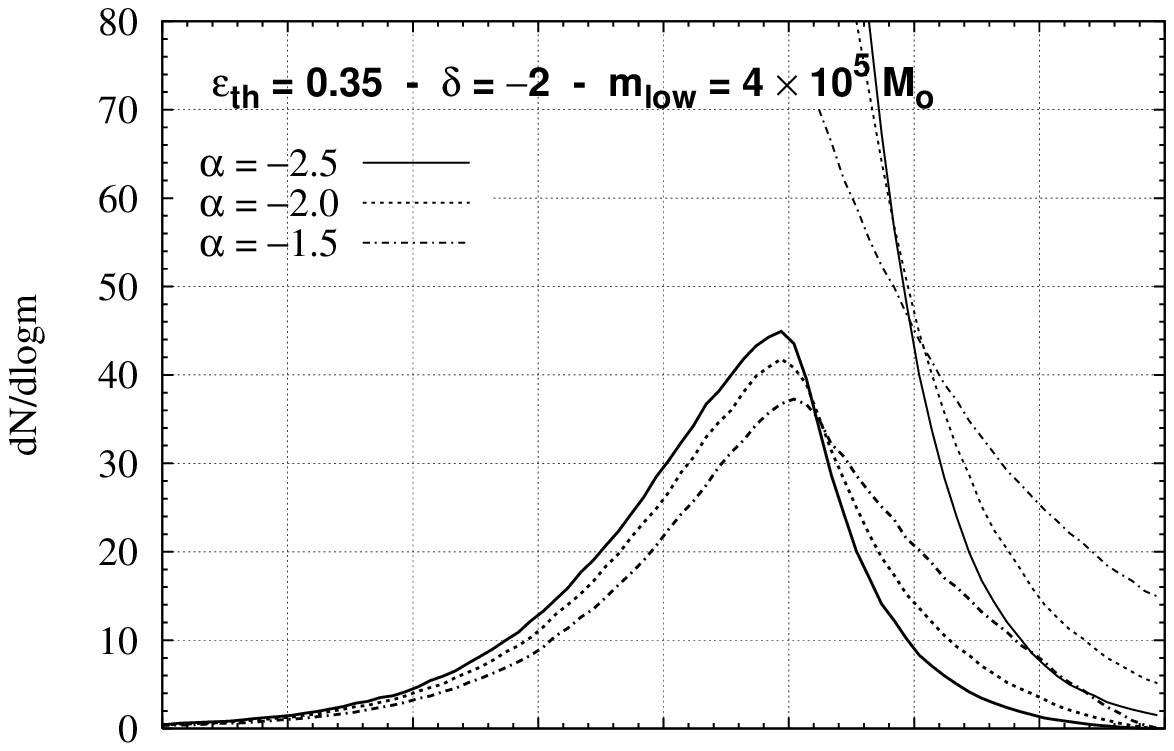, width=\linewidth} 
\end{minipage}
\vfill
\vspace{-10mm}
\begin{minipage}[t]{\linewidth}
\centering\epsfig{figure=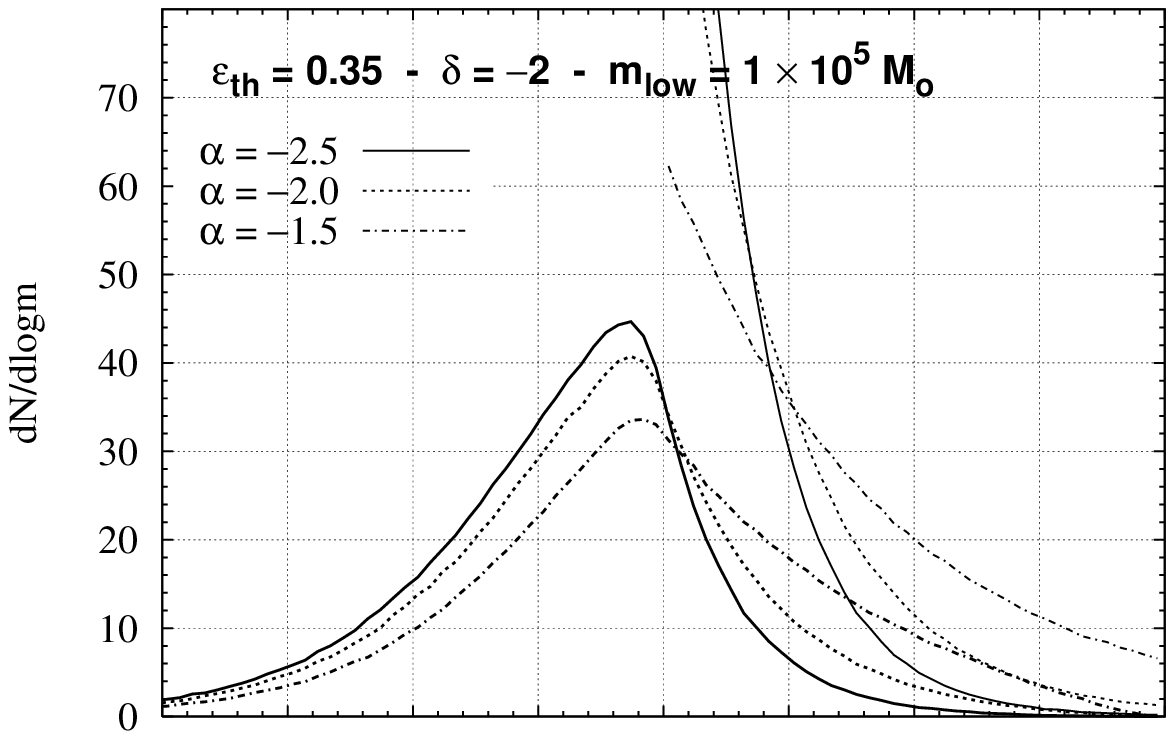, width=\linewidth}
\end{minipage}
\vfill  \vspace{-10mm}
\begin{minipage}[t]{\linewidth}
\centering\epsfig{figure=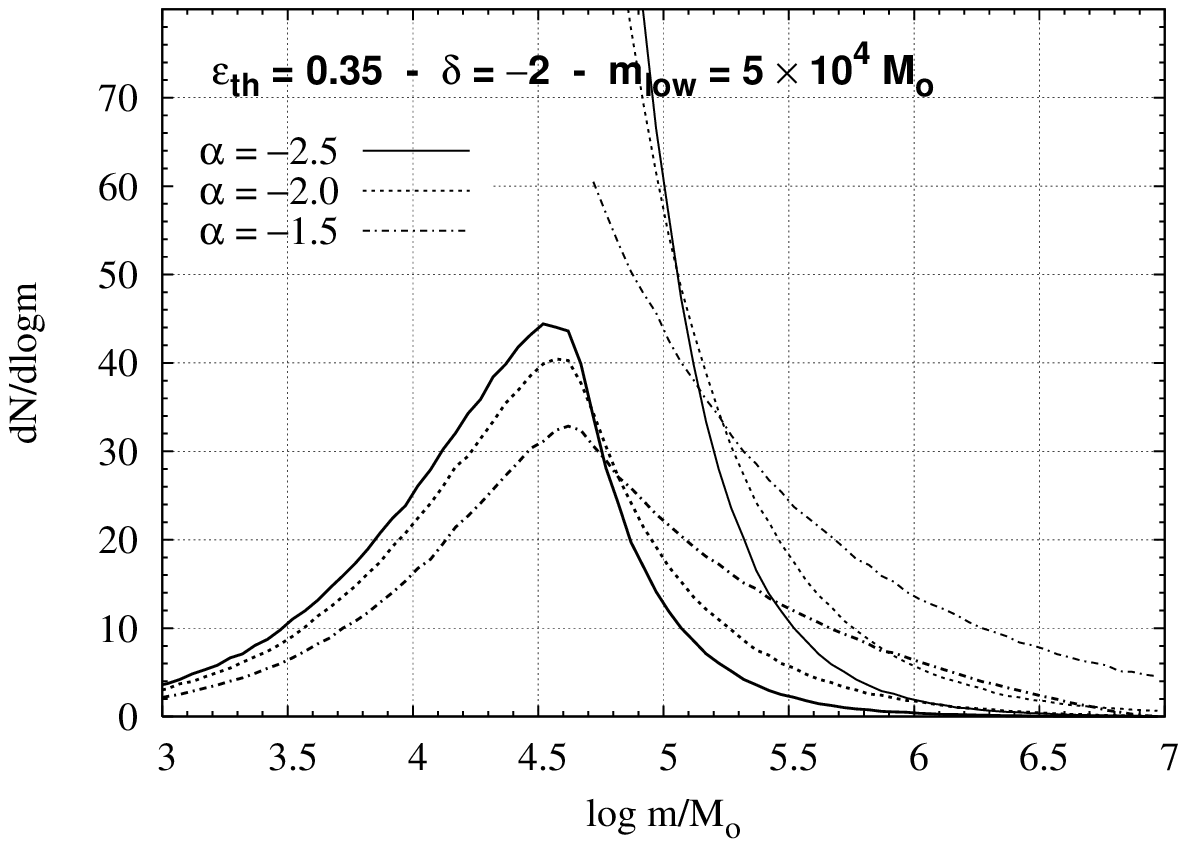, width=\linewidth}
\end{minipage}
\caption{Same as the middle panel of Fig.~\ref{fig:alpha_delta}, but for three different lower limits $m_{low}$ of the cloud mass range.  The mass at the turnover of the cluster initial mass function sensitively depends on $m_{low}$.  }
\label{fig:mlow} 
\end{figure}

\subsubsection{The lower limit $m_{low}$ of the cloud mass range}
\label{subsubsec:mlow}

Unlike $\alpha$ and $\delta$, the lower limit of the protoglobular cloud mass range is a prime controlling 
parameter of the turnover location, as highlighted in Fig.~\ref{fig:mlow}.  For instance, 
a four times smaller lower mass limit (i.e. $10^5\,{\rm M}_{\odot}$ instead of 
$4 \times 10^5\,{\rm M}_{\odot}$) results in a turnover
shifted by $-0.6$ in ${\rm log}m$ (compare top and middle panels of Fig.~\ref{fig:mlow}).  The turnover of the cluster initial mass function thus tracks the lower mass limit of the progenitor clouds, with the cluster mass $m_{TO}$ at the turnover being of order the lower limit $m_{low}$ of the protoglobular cloud mass spectrum.  Specifically, the difference $m_{low}-m_{TO}$ depends on the slope $\alpha$ of the cloud mass spectrum, on the slope $\delta$ of the star formation efficiency distribution $\wp (\epsilon)$ and on the efficiency threshold $\epsilon _{th}$.  A steeper cloud mass spectrum, a steeper star formation efficiency distribution and/or a longer gas removal time-scale increase the offset between the cluster mass at the turnover and the lower cloud mass limit.    

\subsubsection{The upper limit $m_{up}$ of the cloud mass range}
\label{subsubsec:mup}
\begin{figure}
\begin{minipage}[t]{\linewidth}
\centering\epsfig{figure=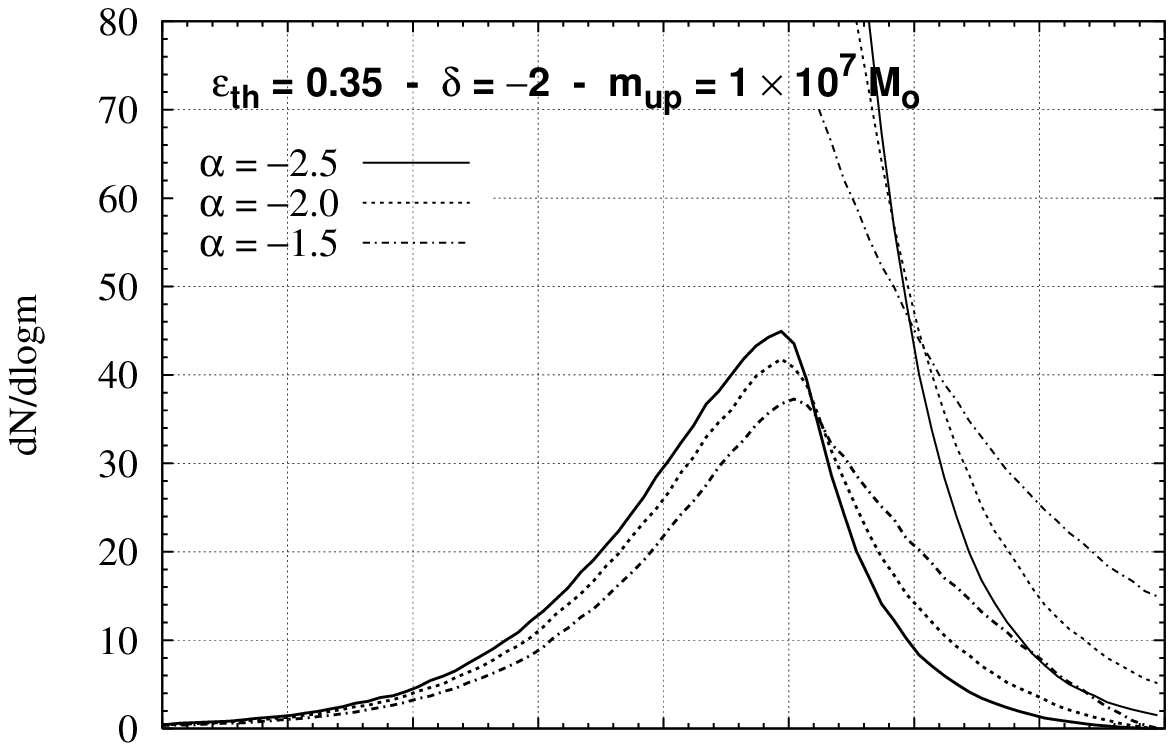, width=\linewidth} 
\end{minipage}
\vfill
\vspace{-10mm}
\begin{minipage}[t]{\linewidth}
\centering\epsfig{figure=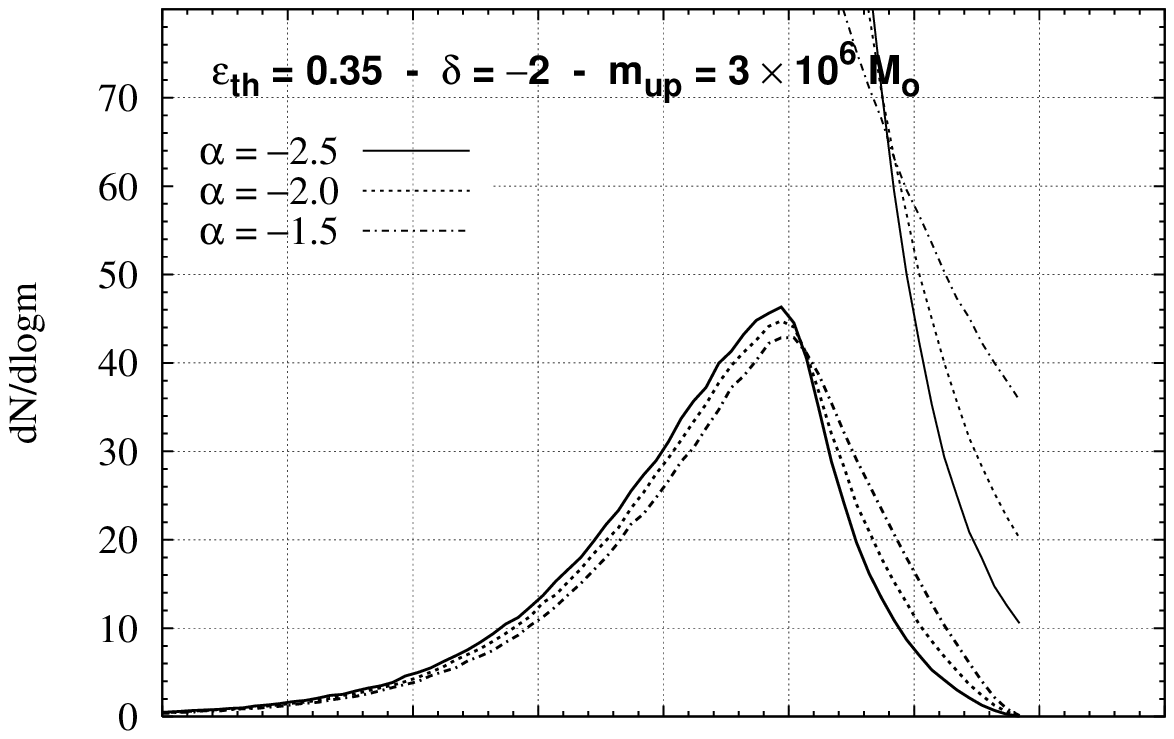, width=\linewidth}
\end{minipage}
\vfill  \vspace{-10mm}
\begin{minipage}[t]{\linewidth}
\centering\epsfig{figure=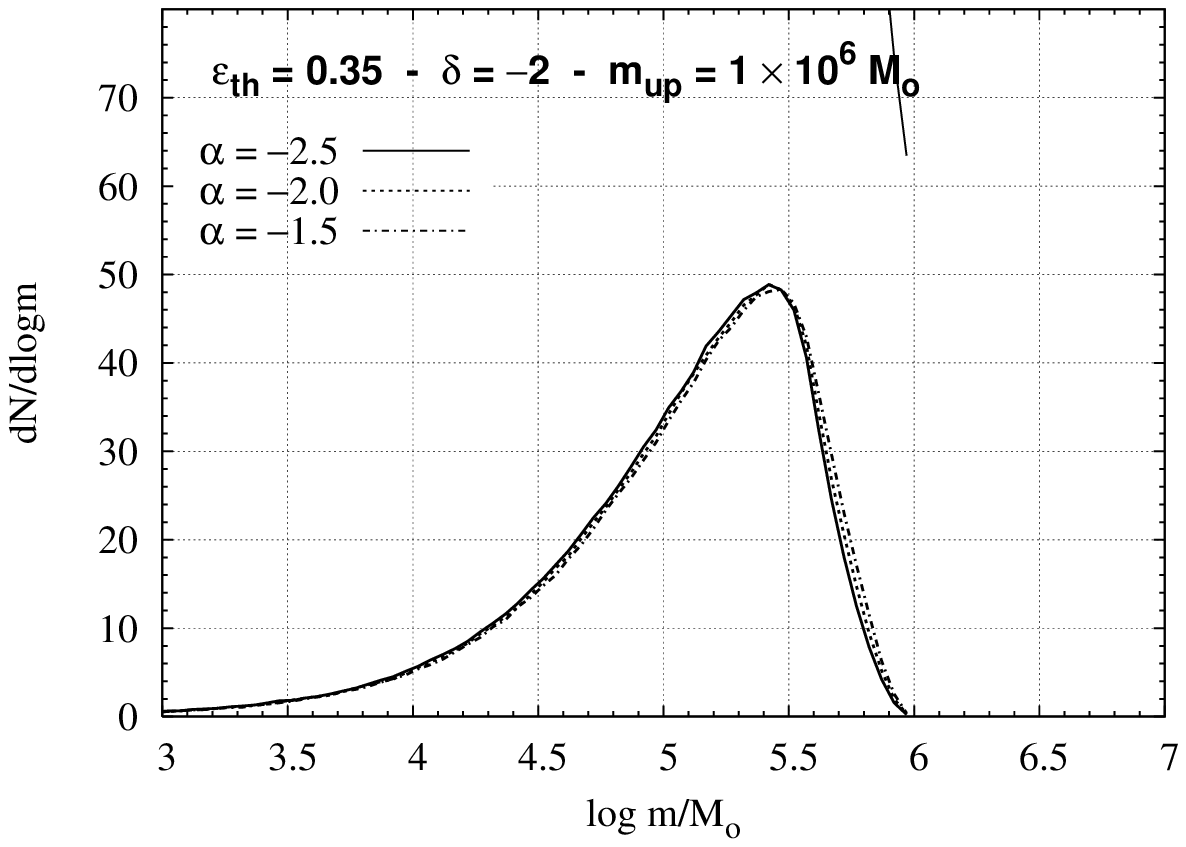, width=\linewidth}
\end{minipage}
\caption{Same as the middle panel of Fig.~\ref{fig:alpha_delta}, but for three different upper limits $m_{up}$ of the cloud mass range.  The turnover location is unaffected by $m_{up}$ variations.} 
\label{fig:mup} 
\end{figure}

We now explore the effect of varying the uppper limit $m_{up}$ of the protoglobular cloud mass range.
In Fig.~\ref{fig:mup}, three values of $m_{up}$ are considered, from top to bottom:  
$10 ^7\,{\rm M}_{\odot}$, $3 \times 10 ^6\,{\rm M}_{\odot}$ and 
$10 ^6\,{\rm M}_{\odot}$.  The turnover location is practically unaffected, the shift in ${\rm log}m$ being $\lesssim 0.05$ when $m_{up}$ is decreased by a factor of ten.  The upper cloud mass limit however influences markedly the high-mass regime of the cluster initial mass function, an effect which we will investigate in detail in section \ref{subsec:mup}.

\subsubsection{The averaged star formation efficiency $\overline{\epsilon}$}
\label{subsubsec:avsfe}
\begin{figure}
\begin{minipage}[t]{\linewidth}
\centering\epsfig{figure=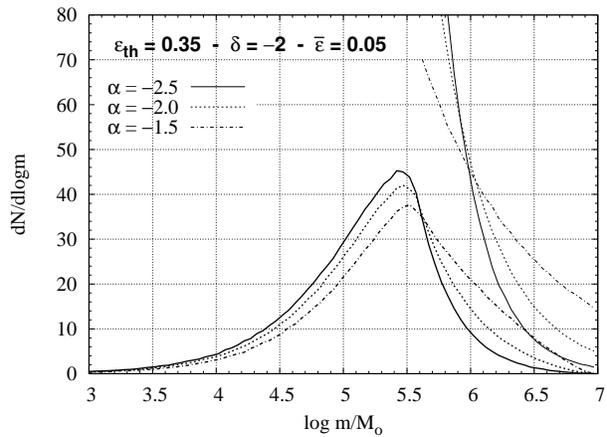, width=\linewidth} 
\end{minipage}
\caption{Same as the middle panel of Fig.~\ref{fig:alpha_delta}, but with the core $r_c$ of the star formation efficiency distribution $\wp (\epsilon)$ adjusted so that the global mean efficiency $\overline {\epsilon}$ is 0.05 instead of 0.01. } 
\label{fig:avsfe} 
\end{figure}

Giant molecular clouds in the present-day Galactic disc turn their gas into stars with a global efficiency in the range between 1 to 5 per cent (Lada \& Lada 2003).  High star formation efficiencies are achieved only locally, in some dense cores.  All the simulations performed up to now assume a global efficiency $\overline{\epsilon}=0.01$.
We now check whether a five times larger value $\overline{\epsilon}=0.05$ affects the shape of the
cluster initial mass function significantly.
Figure \ref{fig:avsfe} is the counterpart of the middle panel of Fig.~\ref{fig:alpha_delta}.  Comparison of both panels demonstrates the robustness of the turnover location.  In fact, 
the increase in $\overline{\epsilon}$ is accounted for by a core radius of the efficiency distribution $\wp (\epsilon)$ ten times greater (i.e, $r_c=0.02$) than that derived if $\overline{\epsilon}=0.01$.  The efficiency distribution $\wp (\epsilon)$ thus remains almost unaffected above the threshold $\epsilon _{th}$ and so does the cluster initial mass function.  \\

It follows from the above simulations that a bell-shaped cluster initial mass function arises from gas removal, provided that the power-law protoglobular cloud mass spectrum shows a lower-mass limit.  Assuming that the star formation efficiency is independent of the protoglobular cloud mass (see section \ref{subsec:PL_G}), the turnover location depends weakly only on all but one of the model parameters, namely, the lower mass-limit of the protoglobular cloud mass spectrum.  To first order, the cloud lower mass limit dictates the turnover location, that is, the lower $m_{low}$, the smaller the cluster mass $m_{TO}$ at the turnover.  As an extreme case, a lower cloud mass limit of, say, $m_{low} = 100\,{\rm M}_{\odot}$ leads to a turnover located at $m_{TO} \lesssim 100\,{\rm M}_{\odot}$.  If the cluster detection limit is higher than $m_{TO}$, then the observed cluster mass spectrum is a power-law whose slope is similar to that of the gaseous progenitors,
by virtue of the mass-independent $\wp (\epsilon)$ distribution.  In section \ref{subsec:proto_gx}, we will discuss further the importance of an appropriate protoglobular cloud mass-scale with respect to explaining the universal Gaussian globular cluster mass function. 

\begin{table*}
\begin{center}
\caption[]{Characteristics of the initial mass functions of gas-free bound star clusters as derived utilising the model of equation \ref{eq:minit} for various sets of model parameters.  The first column gives the star formation efficiency threshold $\epsilon _{th}$ required for a protocluster to retain a bound core of stars (equivalently, the gas removal time-scale, see section \ref{subsubsec:sfeth}).  The slope $\delta$ and the core $r_c$ of the star formation efficiency distribution $\wp (\epsilon)$ are listed in the second and third columns.  Both parameters are adjusted so that the mean star formation efficiency $\overline {\epsilon}$ is 0.01 (except in the case of a flat distribution $\wp (\epsilon)$ for which $\delta =0$).  The next three columns give the lower limit $m_{low}$, the upper limit $m_{up}$ and the slope $\alpha$ of the protoglobular cloud mass spectrum.  The last five columns list the logarithm of the cluster mass at the turnover ${\rm log}m_{\rm TO}$ of the cluster initial mass function, its average value $\overline{{\rm log}m}$, standard deviation $\sigma$, skewness and curtosis.  } 
\label{tab:res}
\begin{tabular}{ c c c c c c c c c c c c } \hline 
$\epsilon _{th}$ & $\delta$ & $r_c$ & ${\overline {\epsilon}}$ & $m_{low}$   &     $m_{up}$    & $\alpha$ & ${\rm log}m_{\rm TO}$ & $\overline{{\rm log}m}$ & $\sigma$ & Skewness &  Curtosis \\ \hline

$0.33$ & $-4$   & $0.020$ & 0.01 & $4 \times 10^5$ &          $10^7$ & $-2.5$   &  $5.22$               &        $5.02$           & $0.60$   & $-0.64$  &  $0.85$ \\ 
$0.33$ & $-4$   & $0.020$ & 0.01 & $4 \times 10^5$ &          $10^7$ & $-2.0$   &  $5.37$               &        $5.12$           & $0.63$   & $-0.53$  &  $0.71$ \\ 
$0.33$ & $-4$   & $0.020$ & 0.01 & $4 \times 10^5$ &          $10^7$ & $-1.5$   &  $5.42$               &        $5.26$           & $0.66$   & $-0.48$  &  $0.50$ \\ \hline

$0.33$ & $-2$   & $0.002$ & 0.01 & $4 \times 10^5$ &          $10^7$ & $-2.5$   &  $5.47$               &        $5.17$           & $0.58$   & $-0.77$  &  $1.28$ \\ 
$0.33$ & $-2$   & $0.002$ & 0.01 & $4 \times 10^5$ &          $10^7$ & $-2.0$   &  $5.47$               &        $5.27$           & $0.61$   & $-0.62$  &  $1.03$ \\ 
$0.33$ & $-2$   & $0.002$ & 0.01 & $4 \times 10^5$ &          $10^7$ & $-1.5$   &  $5.52$               &        $5.41$           & $0.64$   & $-0.56$  &  $0.74$ \\ \hline

$0.33$ & $~~~0$ & - & $0.50$ & $4 \times 10^5$ &          $10^7$ & $-2.5$   &  $5.62$               &        $5.54$           & $0.47$   & $-1.05$  &  $3.40$ \\ 
$0.33$ & $~~~0$ & - & $0.50$ & $4 \times 10^5$ &          $10^7$ & $-2.0$   &  $5.62$               &        $5.64$           & $0.51$   & $-0.77$  &  $2.48$ \\ 
$0.33$ & $~~~0$ & - & $0.50$ & $4 \times 10^5$ &          $10^7$ & $-1.5$   &  $5.62$               &        $5.78$           & $0.55$   & $-0.66$  &  $1.66$ \\ \hline

$0.33$ & $-4$   & $0.020$ & 0.01 & $4 \times 10^5$ & $3 \times 10^6$ & $-2.5$   &  $5.22$               &        $4.99$           & $0.58$   & $-0.81$  &  $0.91$ \\ 
$0.33$ & $-4$   & $0.020$ & 0.01 & $4 \times 10^5$ & $3 \times 10^6$ & $-2.0$   &  $5.32$               &        $5.04$           & $0.59$   & $-0.79$  &  $0.88$ \\ 
$0.33$ & $-4$   & $0.020$ & 0.01 & $4 \times 10^5$ & $3 \times 10^6$ & $-1.5$   &  $5.37$               &        $5.10$           & $0.59$   & $-0.78$  &  $0.85$ \\ \hline

$0.33$ & $-2$   & $0.002$ & 0.01 & $4 \times 10^5$ & $3 \times 10^6$ & $-2.5$   &  $5.47$               &        $5.14$           & $0.55$   & $-0.98$  &  $1.41$ \\ 
$0.33$ & $-2$   & $0.002$ & 0.01 & $4 \times 10^5$ & $3 \times 10^6$ & $-2.0$   &  $5.47$               &        $5.19$           & $0.56$   & $-0.94$  &  $1.35$ \\ 
$0.33$ & $-2$   & $0.002$ & 0.01 & $4 \times 10^5$ & $3 \times 10^6$ & $-1.5$   &  $5.52$               &        $5.25$           & $0.57$   & $-0.93$  &  $1.29$ \\ \hline

$0.33$ & $~~~0$ & - & $0.50$ &  $4 \times 10^5$ & $3 \times 10^6$ & $-2.5$   &  $5.62$               &        $5.50$           & $0.44$   & $-1.48$  &  $4.16$ \\ 
$0.33$ & $~~~0$ & - & $0.50$ & $4 \times 10^5$ & $3 \times 10^6$ & $-2.0$   &  $5.62$               &        $5.56$           & $0.45$   & $-1.40$  &  $3.80$ \\ 
$0.33$ & $~~~0$ & - & $0.50$ & $4 \times 10^5$ & $3 \times 10^6$ & $-1.5$   &  $5.62$               &        $5.62$           & $0.46$   & $-1.36$  &  $3.56$ \\ \hline

$0.33$ & $-4$   &  $0.020$ & 0.01 & $10^5$         &          $10^7$ & $-2.5$   &  $4.67$               &        $4.44$           & $0.59$   & $-0.37$  &  $0.42$ \\ 
$0.33$ & $-4$   & $0.020$ & 0.01 & $ 10^5$         &          $10^7$ & $-2.0$   &  $4.72$               &        $4.57$           & $0.65$   & $-0.15$  &  $0.40$ \\ 
$0.33$ & $-4$   & $0.020$ & 0.01 & $ 10^5$         &          $10^7$ & $-1.5$   &  $4.77$               &        $4.80$           & $0.73$   & $-0.06$  &  $0.04$ \\ \hline

$0.33$ & $-2$   & $0.002$ & 0.01 & $ 10^5$         &          $10^7$ & $-2.5$   &  $4.87$               &        $4.59$           & $0.57$   & $-0.50$  &  $0.83$ \\ 
$0.33$ & $-2$   & $0.002$ & 0.01 & $ 10^5$         &          $10^7$ & $-2.0$   &  $4.87$               &        $4.72$           & $0.63$   & $-0.22$  &  $0.67$ \\ 
$0.33$ & $-2$   & $0.002$ & 0.01 & $ 10^5$         &          $10^7$ & $-1.5$   &  $4.92$               &        $4.95$           & $0.72$   & $-0.10$  &  $0.18$ \\ \hline

$0.33$ & $~~~0$ & - & $0.50$ & $ 10^5$         &          $10^7$ & $-2.5$   &  $5.02$               &        $4.95$           & $0.47$   & $-0.71$  &  $2.75$ \\ 
$0.33$ & $~~~0$ & - & $0.50$ & $ 10^5$         &          $10^7$ & $-2.0$   &  $5.02$               &        $5.08$           & $0.54$   & $-0.24$  &  $1.83$ \\
$0.33$ & $~~~0$ & - & $0.50$ & $ 10^5$         &          $10^7$ & $-1.5$   &  $5.02$               &        $5.31$           & $0.64$   & $-0.07$  &  $0.60$ \\ \hline
 
$0.15$ & $-4$   & $0.020$ & 0.01 & $4 \times 10^5$ &          $10^7$ & $-2.5$   &  $4.97$               &        $4.72$           & $0.63$   & $-0.47$  &  $0.32$ \\ 
$0.15$ & $-4$   & $0.020$ & 0.01 & $4 \times 10^5$ &          $10^7$ & $-2.0$   &  $5.02$               &        $4.82$           & $0.66$   & $-0.39$  &  $0.26$ \\ 
$0.15$ & $-4$   & $0.020$ & 0.01 & $4 \times 10^5$ &          $10^7$ & $-1.5$   &  $5.17$               &        $4.96$           & $0.69$   & $-0.37$  &  $0.16$ \\ \hline

$0.15$ & $-2$   & $0.002$ & 0.01 & $4 \times 10^5$ &          $10^7$ & $-2.5$   &  $5.27$               &        $5.01$           & $0.61$   & $-0.70$  &  $0.86$ \\ 
$0.15$ & $-2$   & $0.002$ & 0.01 & $4 \times 10^5$ &          $10^7$ & $-2.0$   &  $5.27$               &        $5.10$           & $0.64$   & $-0.58$  &  $0.72$ \\ 
$0.15$ & $-2$   & $0.002$ & 0.01 & $4 \times 10^5$ &          $10^7$ & $-1.5$   &  $5.32$               &        $5.25$           & $0.67$   & $-0.54$  &  $0.52$ \\ \hline 

$0.15$ & $~~~0$ & - & $0.50$ & $4 \times 10^5$ &          $10^7$ & $-2.5$   &  $5.57$               &        $5.53$           & $0.45$   & $-1.02$  &  $3.90$ \\ 
$0.15$ & $~~~0$ & - & $0.50$ & $4 \times 10^5$ &          $10^7$ & $-2.0$   &  $5.57$               &        $5.63$           & $0.49$   & $-0.71$  &  $2.70$ \\ 
$0.15$ & $~~~0$ & - & $0.50$ & $4 \times 10^5$ &          $10^7$ & $-1.5$   &  $5.62$               &        $5.78$           & $0.53$   & $-0.61$  &  $1.76$ \\ \hline 

$0.33$ & $-2$   & $0.002$ & $0.05$ & $4 \times 10^5$ &          $10^7$ & $-2.5$   &  $5.42$               &        $5.18$           & $0.57$   & $-0.78$  &  $1.33$ \\ 
$0.33$ & $-2$   & $0.002$ & $0.05$ & $4 \times 10^5$ &          $10^7$ & $-2.0$   &  $5.47$               &        $5.28$           & $0.61$   & $-0.63$  &  $1.07$ \\ 
$0.33$ & $-2$   & $0.002$ & $0.05$ & $4 \times 10^5$ &          $10^7$ & $-1.5$   &  $5.52$               &        $5.42$           & $0.64$   & $-0.57$  &  $0.77$ \\ \hline
   
\end{tabular}
\end{center}
\end{table*}    

\section{An application to the Galactic Stellar Halo}
\label{sec:Gal_Halo}

\subsection{Fitting the halo cluster mass function}
\label{subsec:isoQfGC_mlowmup} 
In this section, we explore whether our model can successfully reproduce 
the observed mass function of the Galactic halo globular cluster system.
In order to evolve up to an age of 13\,Gyr the cluster initial mass functions 
derived in the previous section,  we make use of Baumgardt \& Makino 's (2003) equation 12, 
which they derived by fitting the results of a large set of $N$-body
simulations taking into account the effects of stellar evolution, of
two-body relaxation and of cluster tidal truncation.  This analytical formula 
supplies at any time $t$ the mass of a star cluster with initial mass $m_i$ which 
is moving along a cirular orbit at a galactocentric distance $D$:
\begin{equation}
\frac{m(t)}{m_i} = 0.70 \left(1-\frac{t}{t_{dis}}\right)\;.
\label{eq:m_GC(t)}
\end{equation}
This equation assumes that mass loss due to massive-star stellar evolution 
takes place immediately after cluster formation and 
amounts to 30 per cent of the cluster initial mass.  Later on, globular cluster
mass decreases linearly with time due to the combination of internal 
(two-body relaxation) and external (galactic tidal field) effects.  
The disruption time $t_{dis}$ of a cluster (equation 10 of Baumgardt \& 
Makino 2003) depends on the circular velocity $V_c$ of the host galaxy, on the 
cluster initial mass $m_i$ and on the cluster galactocentric distance $D$.  In what follows, we adopt
$V_c = 220\,{\rm km.s}^{-1}$ and we distribute the globular clusters in space following 
an initial number density profile scaling as $D^{-3.5}$ (Parmentier \& Gilmore 2005).  
Additionally, we account for the non-circularity of the globular cluster orbits by
halving the cluster disruption time-scale, which is equivalent to assuming a mean orbital 
eccentricity $e=0.5$ (Baumgardt \& Makino 2003, their equation 10).  Since the globular
cluster mass function is an equilibrium shape (see section \ref{subsec:eqMF}), that is, 
its shape is practically independent of the age $t$ of the cluster system as well as of the 
exact cluster disruption time-scale $t_{dis}$, the choice of that correction factor for $t_{dis}$ does not significantly affect the goodness of fit of the present-day globular cluster mass function. 
Only the fraction $F_N$ of surviving clusters and the ratio $F_M$ of the final to the 
initial mass in globular clusters are affected (see Table \ref{tab:OHMF_var_PL}).  In order to take 
into account dynamical friction, clusters whose time-scale of orbital decay is smaller than $t$ are removed from the cluster system at that time (Binney \& Tremaine 1987).

In this study, we focus on the mass distribution of the
first generation globular clusters which formed within the 
gravitational potential well of the Galaxy.  Hence, we do not consider 
the more metal-rich, presumably second-generation, bulge/disc 
globular clusters (Zinn 1985).
The halo cluster system itself hosts two distinct populations of clusters,
the so-called Old Halo and Younger Halo (Van den Bergh 1993,
Zinn 1993).  The Old Halo globular clusters might have been formed 'in
situ'.  In contrast, Younger Halo globular clusters
are suspected of having been accreted.  Regardless of their
formation history, the Old Halo globular clusters form a coherent
and well-defined group, well-suited to an analysis of their
properties (see Table 1 in Parmentier \& Grebel 2005 for the definition of our sample). 
Our source for the absolute visual magnitudes $M_v$ of 
the clusters is the McMaster database compiled and maintained by Harris
(1996, updated February 2003).  These are converted into luminous mass estimates 
on the assumption of a constant cluster integrated  
mass-to-light ratio $M/L_v$ of $2.35$ (i.e., the average of the mass-to-light 
ratios of the halo clusters for which Pryor \& Meylan (1993) obtained 
dynamical mass estimates).  

Having detailed the cluster evolutionary model and the cluster sample against
which we will compare the outcomes of our simulations, we now turn to the issue
of deriving a cluster initial mass function which, when evolved to an age
of 13\,Gyr, reproduces the observed Old Halo cluster mass function.  
The shape of the cluster initial mass function depends on the
following: 
\begin{itemize} 
\item [{\it (i)}] the truncated protoglobular cloud mass spectrum, defined by its slope $\alpha$ and its lower 
and upper mass limits $m_{low}$ and $m_{up}$,
\item [{\it (ii)}] the star formation efficiency probability distribution
$\wp (\epsilon)$, defined by its slope $\delta$ and its core $r_c$ 
(equation \ref{eq:sfe_prob}), 
\item [{\it (iii)}] the fraction of stars remaining bound to the protocluster 
after gas removal, as defined by the $F_{bound}$ vs. $\epsilon$ diagram 
(Fig.~\ref{fig:Fb_SFE}).  
\end{itemize}

In what follows, we assume that any left-over star forming gas in each protocluster
is instantaneously removed, and we therefore consider the solid line
in Fig.~\ref{fig:Fb_SFE} as the $F_{bound}$ vs. $\epsilon$ relation.
For the slope of the cloud mass spectrum, we adopt $\alpha = -1.7$, which is
the spectral index reported by surveys of giant molecular clouds 
in a sample of galaxies in the Local Group, namely, the outer disc of the Milky Way, the 
Large Magellanic Cloud, M31, and IC10 (Blitz et al.~2006, their Table 3).
Although, shallower ($\alpha = -1.5$) and steeper ($\alpha = -2.5$) spectral indices 
are reported for the inner disc of the Milky Way and the spiral galaxy M33, respectively,
these are likely ascribed to observational biaises.  As noted by Rosolowski (2005),
in the inner Milky Way, line-of-sight blending will make several less
massive clouds appear as a single more massive one, shifting the index toward
shallower values.  On the other hand, the steep slope of the cloud mass distribution
in M33 can be attributed to fitting a truncated power-law distribution above the 
mass cutoff.  In the present-day Galactic disc, the same spectral index $\alpha = -1.7$
also describes the mass spectrum of the dense cores of giant molecular clouds, to which
embedded clusters are often physically associated (Lada \& Lada 2003 and references therin). 
The lower ($m_{low}$) and upper ($m_{up}$) limits of the protoglobular
cloud mass spectrum are left as free parameters.

As already quoted in section \ref{subsubsec:delta}, the star formation efficiency 
distribution $\wp (\epsilon )$ is poorly determined.  Moreover, even if it were
reasonably known for the present-day Galactic disc, there is no guarantee that this would equally-well 
describe star formation in the Galactic halo some 13\,Gyr ago.  Nevertheless, we attempt to constrain $\wp (\epsilon )$ in the following way.  We assume that the global star formation efficiency is on the 
order of a few per cent, thereby reflecting the ineffectiveness of star formation of galactic scales.  We note in passing that the ratio between the baryonic masses of the stellar halo and of the Galaxy is on that order of magnitude.  This may suggest that the stellar halo formed out of an amount of gas equivalent to the total mass in gas and stars in the present-day Milky Way.  Thus, we adopt the following mean star formation efficiency: 
\begin{equation}
\overline {\epsilon} = \frac{M_{Halo}^{init}}{M_{Gas}^{init}}
                     = \frac{0.7^{-1} \times M_{Halo}^{13\,Gyr}}{M_{Gas+stars~in~MW}^{13\,Gyr}}\,.
\label{eq:mean_SFE}
\end{equation}  
In this equation, $M_{Halo}^{init}$ and $M_{Halo}^{13\,Gyr}$ are the initial
and present-day stellar masses of the Galactic halo, respectively, the factor
$0.7^{-1}$ accounting for the corresponding stellar evolutionary mass loss.
The stellar halo is processed out of an amount of gas $M_{Gas}^{init}$ which, under
the above mentioned assumption, is equivalent to the total mass $M_{Gas+stars~in~MW}^{13\,Gyr}$
in stars and gas in the present-day Milky Way.
We adopt a present-day halo mass $M_{Halo}^{13\,Gyr}$
of $10^9\,M_{\odot}$ (Freeman \& Bland-Hawthorn 2002) and a present-day total mass of stars and gas in the
Milky Way $M_{Gas+stars~in~MW}^{13\,Gyr}$ of $6 \times 10^{10}\,M_{\odot}$, leading 
to a mean star formation efficiency $\overline {\epsilon} = 2.5\,\%$.
We acknowledge that this constitutes a very crude approximation since
equation \ref{eq:mean_SFE} neglects the effects of later mergers and accretions onto the Milky Way.
As we discuss below however (see Table \ref{tab:OHMF_var_PL} and the third panel of Fig.~\ref{fig:OHMF}), the exact value of $\overline {\epsilon}$ does not significantly affect the goodness of fit to the Old Halo cluster mass function. 

\begin{figure}
\begin{minipage}[b]{\linewidth}
\centering\epsfig{figure=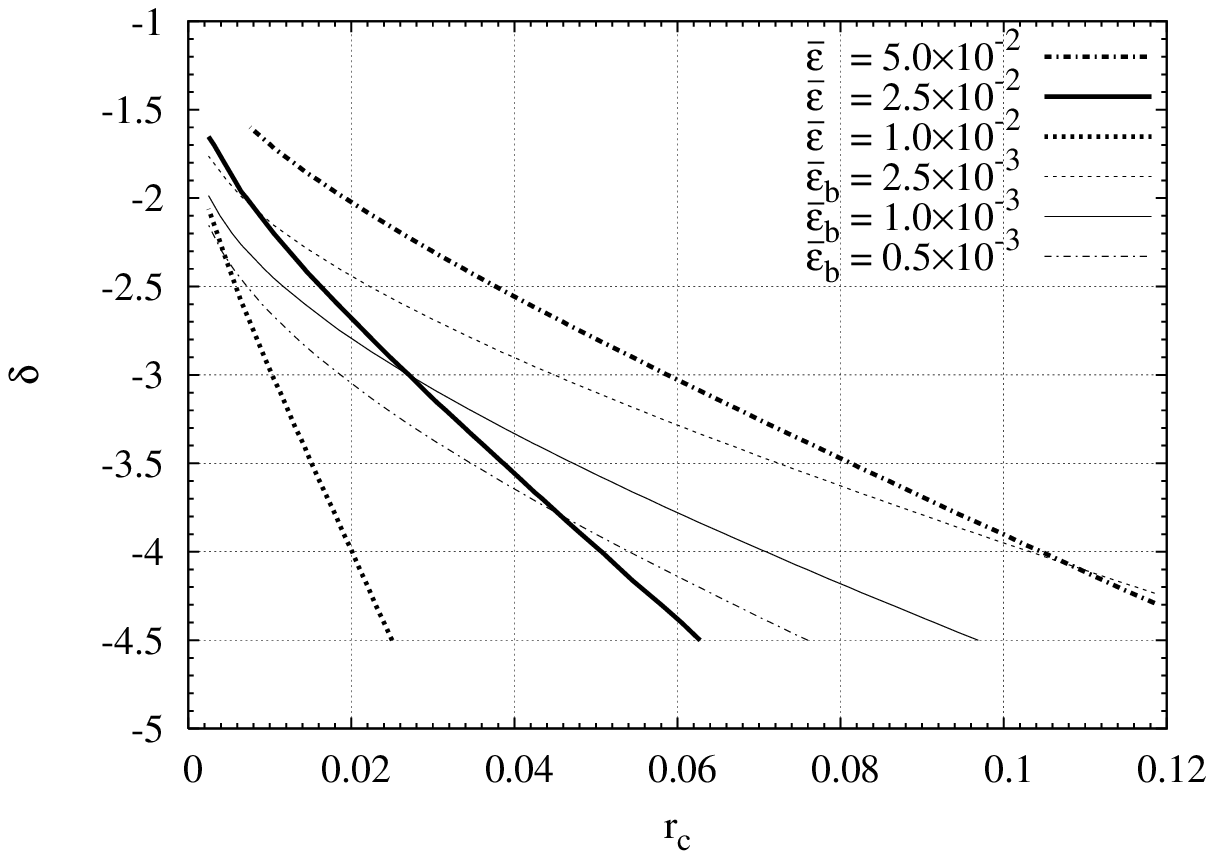, width=\linewidth} 
\end{minipage}
\vfill \vspace{-1mm}
\begin{minipage}[b]{\linewidth}
\centering\epsfig{figure=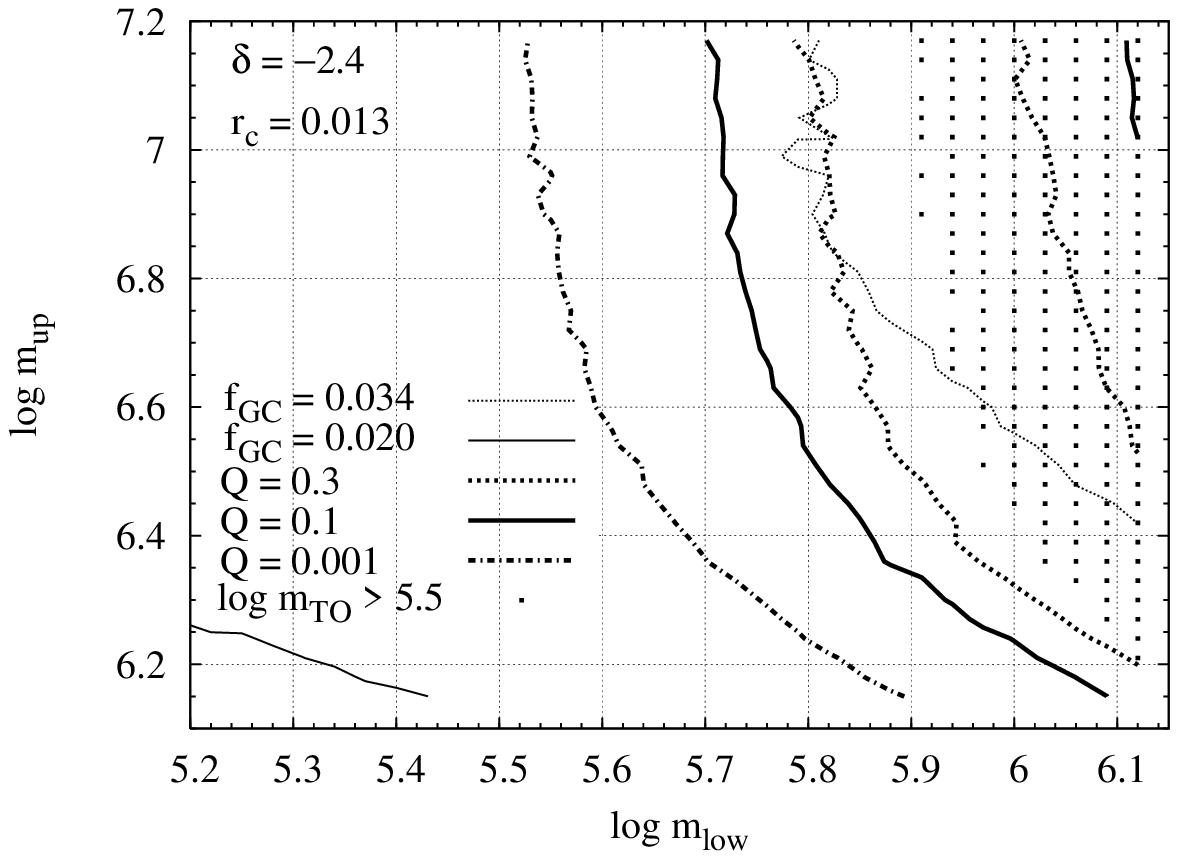, width=\linewidth}
\end{minipage}
\vfill \vspace{-1mm}
\begin{minipage}[b]{\linewidth}
\centering\epsfig{figure=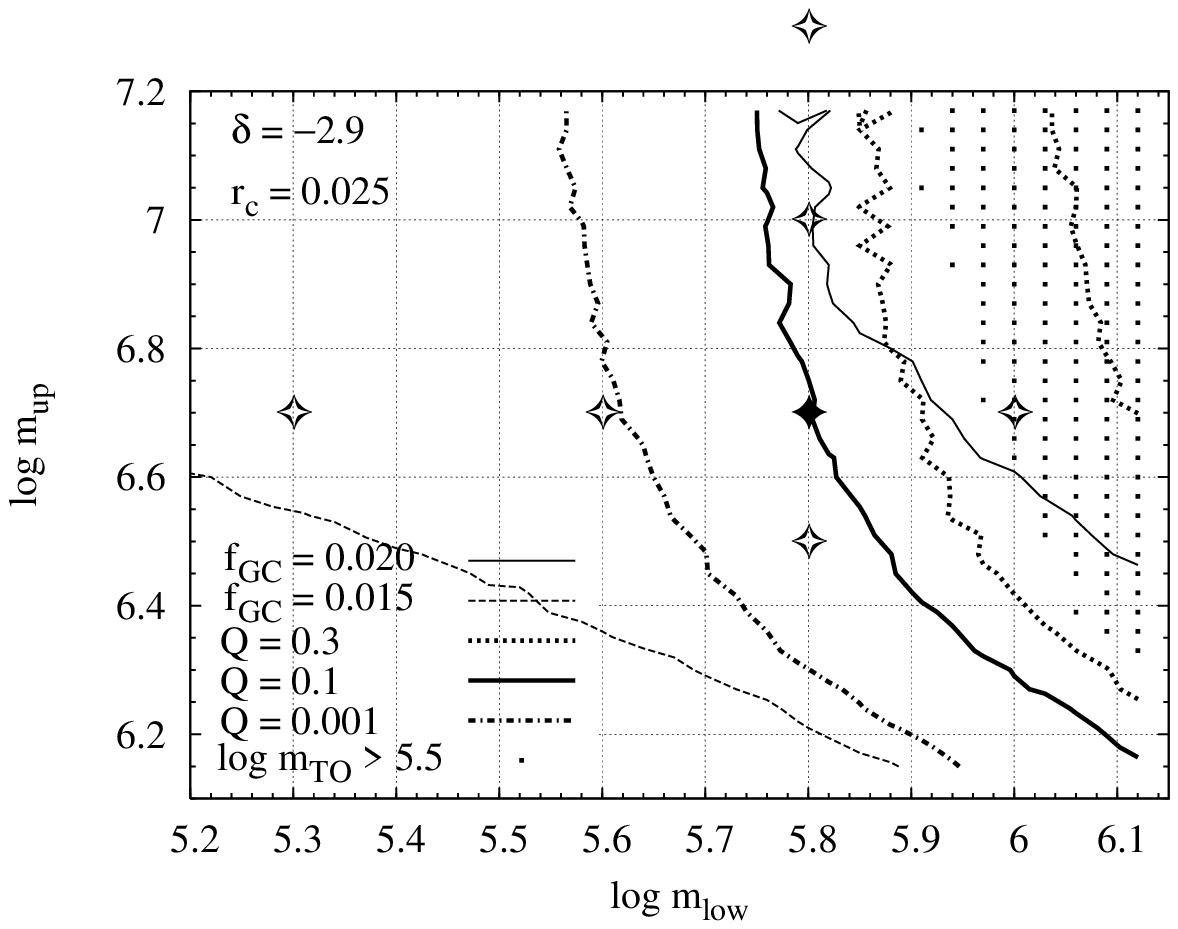, width=\linewidth}
\end{minipage}
\caption{{\it Top panel:} Locii of points with identical mean star formation efficiencies $\overline {\epsilon}$ (thick curves) and identical mean bound star formation efficiencies $\overline {\epsilon _b}$ (thin curves) in the ($r_c, \delta$) plane, where $\delta$ and $r_c$ are the slope and the core of the efficiency distribution $\wp (\epsilon)$.  {\it Middle panel:} Incomplete gamma functions $Q$ of the fit of the evolved cluster mass function to the observed one (thick curves) and cluster mass fractions $f_{GC}$ at an age of 13\,Gyr as predicted by the model (thin curves) versus the lower and upper limits ($m_{low}$, $m_{up}$) of the cloud mass spectrum.  The dotted area corresponds to simulations for which the mass at the turnover of the evolved cluster mass function is larger than $3 \times 10^5\,{\rm M}_{\odot}$ (see text for details).  {\it Bottom panel:} Same as middle panel, but for a steeper $\wp (\epsilon)$ function.  This results in a greater rate of cluster infant mortality, thus lowering the present-day cluster mass fraction for any cloud mass range.  The four-branch stars represent different cases presented in Fig.~\ref{fig:OHMF} and Table \ref{tab:OHMF_var_PL}.  The solid star is our adopted fiducial case (first line in Table \ref{tab:OHMF_var_PL}).  }
\label{fig:isoQfGC_mlowmup} 
\end{figure}

As a second constraint on the star formation efficiency distribution $\wp (\epsilon )$,
we use the present-day mass fraction of Old Halo clusters in the Galactic 
halo, namely: 
\begin{equation}
f_{GC}^{13\,Gyr} = \frac{M_{tot}^{GC}}{M_{Halo}^{13\,Gyr}} = 0.02\,,  
\label{eq:fGC1}
\end{equation}
where $M_{tot}^{GC}$ is the present-day total mass in Old Halo globular clusters,
that is, $2 \times 10^7\,M_{\odot}$ (Parmentier \& Grebel 2005).  The mass of the Galactic halo
is actually dominated by field stars.  In this model, these arise from
{\it (i)} the disruption of protoclusters in the $\epsilon < \epsilon _{th}$ regime 
(infant mortality), {\it (ii)} the escape of stars out of protoclusters  
following gas removal when $\epsilon > \epsilon _{th}$ (infant weight-loss), and
{\it (iii)} the evaporation and disruption of globular clusters 
over a Hubble-time of dynamical evolution in the tidal field of the 
Milky Way.  The present-day cluster mass fraction $f_{GC}^{13\,Gyr}$,
the mass fraction $F_M$ of surviving clusters at an age $t = 13$\,Gyr,
the mean star formation efficiency $\overline \epsilon$ and the 
mean bound star formation efficiency $\overline \epsilon _b$ are related through:
\begin{equation}
f_{GC}^{13\,Gyr} = \frac{F_M \times \overline {\epsilon _b}}{0.7 ~ \overline \epsilon}\;.
\label{eq:fGC2}
\end{equation}
$\overline {\epsilon _b}$ depends on the star formation efficiency distribution 
$\wp (\epsilon )$ and on the $F_{bound}$ vs. $\epsilon$ relation, while 
$\overline {\epsilon}$ depends on the former only 
(see equations \ref{eq_meansfe} and \ref{eq_meanbsfe}).
The top panel of Fig.~\ref{fig:isoQfGC_mlowmup} shows the locii of points 
with identical mean star formation efficiencies $\overline \epsilon$ (thick curves) 
and the locii of points with identical mean bound star formation efficiencies 
$\overline {\epsilon _b}$ (thin curves) 
in the $\delta$ vs. $r_c$ diagram.  This plot enables us to determine 
the function $\wp (\epsilon )$, via its parameters $\delta$ and $r_c$, for any given
couple ($\overline \epsilon$, $\overline \epsilon _b$).  Yet,  
equation \ref{eq:fGC2} shows that matching the distribution $\wp (\epsilon)$ and the present-day
cluster mass fraction $f_{GC}^{13\,Gyr}$ in the halo still requires an estimate of $F_M$. 
The mass fraction $F_M$ of surviving clusters depends on the initial mass function of the
clusters, on their initial distribution in space, on their age and on the assumed
cluster disruption time-scale $t_{dis}$.  The greater the contribution of the (more vulnerable)
low-mass clusters, the more concentrated around the Galactic centre the cluster spatial distribution, the older the cluster system and/or the shorter the cluster disruption time-scale, the smaller the ratio $F_M$.  We have detailed our assumptions regarding the last three quantities earlier in this section.  We now need to determine a cluster initial mass function compatible with equation \ref{eq:fGC1}.  
Since $F_M$ is required to determine, in turn, $\overline {\epsilon _b}$
(equation \ref{eq:fGC2}), $\wp (\epsilon )$ (via the top panel of 
Fig.~\ref{fig:isoQfGC_mlowmup}) and the cluster initial mass function itself (with
the model described in section \ref{subsec:PL_G}), we need to iterate.
The middle panel of Fig.~\ref{fig:isoQfGC_mlowmup} summarizes the outcomes of a first set
of simulations with ($\delta$, $r_c$)=(-2.4, 0.013).  This corresponds to
($\overline {\epsilon}$, $\overline {\epsilon _b}$) = (0.025, 0.002).
The $x$ and $y$-axes represent the lower and upper limits of the protoglobular
cloud mass spectrum, respectively.  The thick curves are the locii of points with
the same value for the incomplete gamma function $Q$.  This quantifies the goodness
of fit between the evolved cluster mass function, as predicted by our model, and
the Old Halo cluster mass function \footnote{We remind the reader
that: a $Q$ value of 0.1 or larger indicates a satisfactory agreement
between the model and the data; if $Q \geq 0.001$, the fit may be
acceptable if, e.g., the errors have been moderately underestimated;
if $Q < 0.001$, the model can be called into question (Press\ et
al.~1992).}.  The dashed-dotted, solid and dotted lines 
correspond to $Q$=0.001, 0.1 and 0.3 respectively.  The dotted region of the diagram,
at ${\rm log}\,m_{low} \gtrsim 5.9$, corresponds to simulations for which
the cluster mass at the turnover of the evolved mass function is larger than 
$3 \times 10^5\,{\rm M}_{\odot}$, that is, the turnover is located at too large 
a mass with respect to what is observed.  This result seems to contradict the
excellent goodness of fit ($0.1 < Q \lesssim 0.3$) in this region.  The top panel 
of Fig.~\ref{fig:OHMF} highlights the origin of that discrepancy.  The lower solid 
line is the modelled cluster mass function at an age of 13\,Gyr for our fiducial 
case (see below) and the plain squares depict the Old Halo cluster mass function.
Our model overestimates slightly the number of low-mass clusters.  As illustrated in
that panel, an increase in the lower limit $m_{low}$ of the cloud mass range 
shifts the cluster mass function toward larger mass.  This improves the goodness of 
fit of the low-mass wing of the evolved cluster mass function and increases
the incomplete gamma function $Q$ accordingly, even though the cluster mass at 
the turnover gets too large  (compare the lower solid line, for which 
$m_{low} = 6 \times 10^5\,{\rm M}_{\odot}$, to the lower dashed-dotted line,
corresponding to $m_{low} = 10^6\,{\rm M}_{\odot}$).  Coming back to the middle
panel of Fig.~\ref{fig:isoQfGC_mlowmup}, the two thin lines are the locii of 
points with given present-day cluster mass fractions in the Galactic halo 
$f_{GC}^{13\,Gyr}$.  The solid line (at the bottom left of the plot) and the dotted
line correspond to $f_{GC}^{13\,Gyr}=0.02$ and $f_{GC}^{13\,Gyr}=0.03$, respectively.
The region of the plane ($m_{low}$,$m_{up}$) where the evolved cluster
mass function fits well the observed one (i.e. $Q \gtrsim 0.1$, but outside the dotted
area) coincides with a slightly too large cluster mass fraction, that is, 
$f_{GC}^{13\,Gyr} \simeq 0.03$.  
As shown in Table \ref{tab:sfe_dist}, a steeper slope $\delta$ for the star formation
efficiency distribution $\wp (\epsilon)$ -while adjusting its core $r_c$
so as to retain the same mean star formation efficiency $\overline {\epsilon}$-
results into a smaller mean bound star formation efficiency $\overline {\epsilon _b}$ 
and, therefore, a smaller cluster mass fraction $f_{GC}^{13\,Gyr}$ at an age
of 13\,Gyr (see equation \ref{eq:fGC2}).  The bottom panel of Fig.~\ref{fig:isoQfGC_mlowmup} 
illustrates the 
effect of steepening $\wp (\epsilon)$, namely, $\delta =-2.9$ corresponding to 
$\overline {\epsilon _b}=0.001$ (top panel of Fig.~\ref{fig:isoQfGC_mlowmup}).
As explained in section \ref{subsubsec:delta} (see also Figs.~\ref{fig:alpha_delta} and 
\ref{fig:sfeth}), the cluster mass function does not strongly depend on $\delta$.  
Consequently, while the iso$f_{GC}$ curves are moved upwards when $\delta$ is decreased, 
the locus of points of any given value of $Q$ is little affected.  One can now see that
our new choice of $\delta$ enables us to fit the Old Halo cluster mass function 
(i.e. $Q \gtrsim 0.1$) while, at the same time, accounting for the present-day 
cluster mass fraction in the stellar halo $f_{GC}^{13\,Gyr} \simeq 0.02$. 
In what follows, we adopt $\delta =-2.9$, $r_c=0.025$, 
$m_{low}=6 \times 10^5 {\rm M}_{\odot}$ and $m_{up}=5 \times 10^6 {\rm M}_{\odot}$ 
as our fiducial case ($Q=0.1$, depicted as the plain four-branch star in the bottom panel of 
Fig.~\ref{fig:isoQfGC_mlowmup}) and we explore the sensitivity of our results to variations in 
the input parameters ($m_{low}$, $m_{up}$, $\delta$ and $r_c$).  For each simulation,
Table \ref{tab:OHMF_var_PL} lists the mean star formation efficiency $\overline {\epsilon}$ 
and the mean bound star formation efficiency $\overline {\epsilon _b}$, the mass fraction 
$f_{GC}^{13\,Gyr}$ of clustered stars in the halo at an age of 13\,Gyr, the ratio $F_M$ 
between the final total mass and the initial total mass in clusters, its number 
counterpart $F_N$ (i.e. the fraction of surviving clusters) and the goodness of fit 
$Q$ of the evolved cluster mass function to the observed one.  The corresponding modelled cluster mass
distributions, both initial and evolved to an age of 13\,Gyr, are shown in the different panels of 
Figs.~\ref{fig:OHMF} and \ref{fig:OH_mup}, together with the Old Halo cluster mass distribution. 
The solid lines correspond to the fiducial case.  In all these panels,
the initial cluster mass distributions have been corrected for the effect of
stellar evolutionary mass loss, that is, all cluster masses have been decreased by 30\% with respect to their actual initial value (see equation \ref{eq:m_GC(t)}).  
This correction, indicated by a leftward arrow of size 0.15 in ${\rm log}m$, enables us to estimate how the cluster mass function evolves with time due only to cluster dynamical evolution (i.e. evaporation and disruption). 

\begin{figure}
\begin{minipage}[t]{\linewidth}
\centering\epsfig{figure=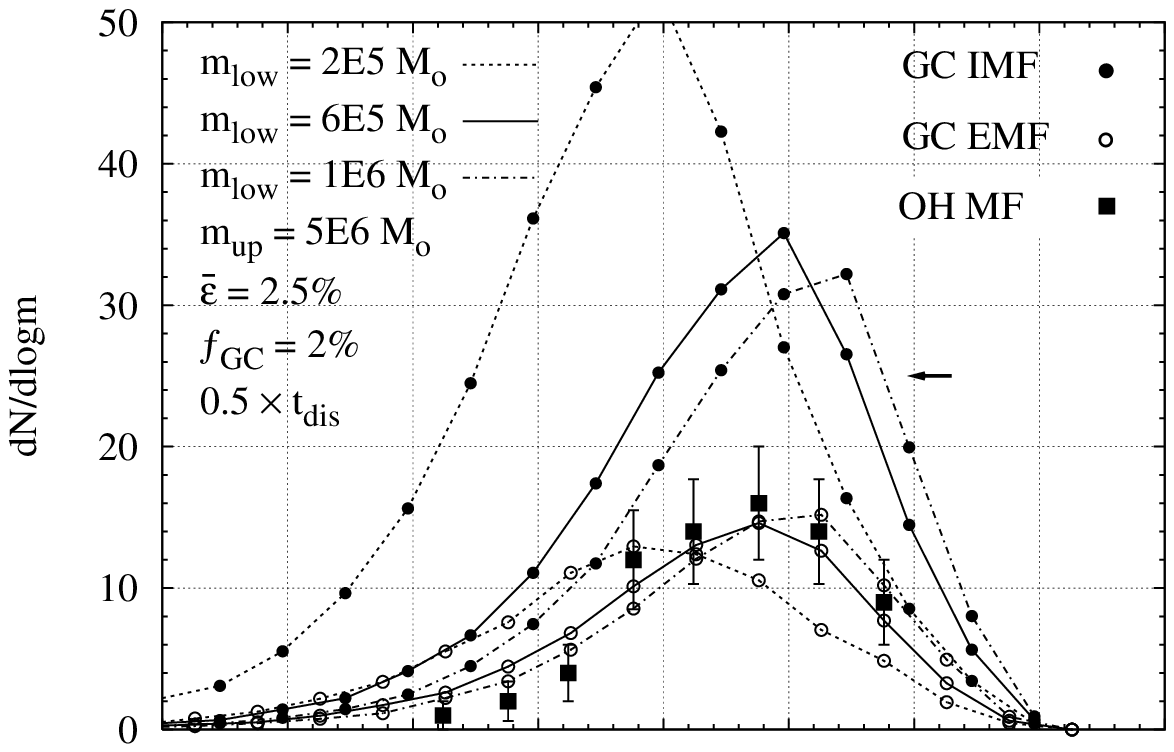, width=\linewidth} 
\end{minipage}
\vfill
\vspace{-10mm}
\begin{minipage}[t]{\linewidth}
\centering\epsfig{figure=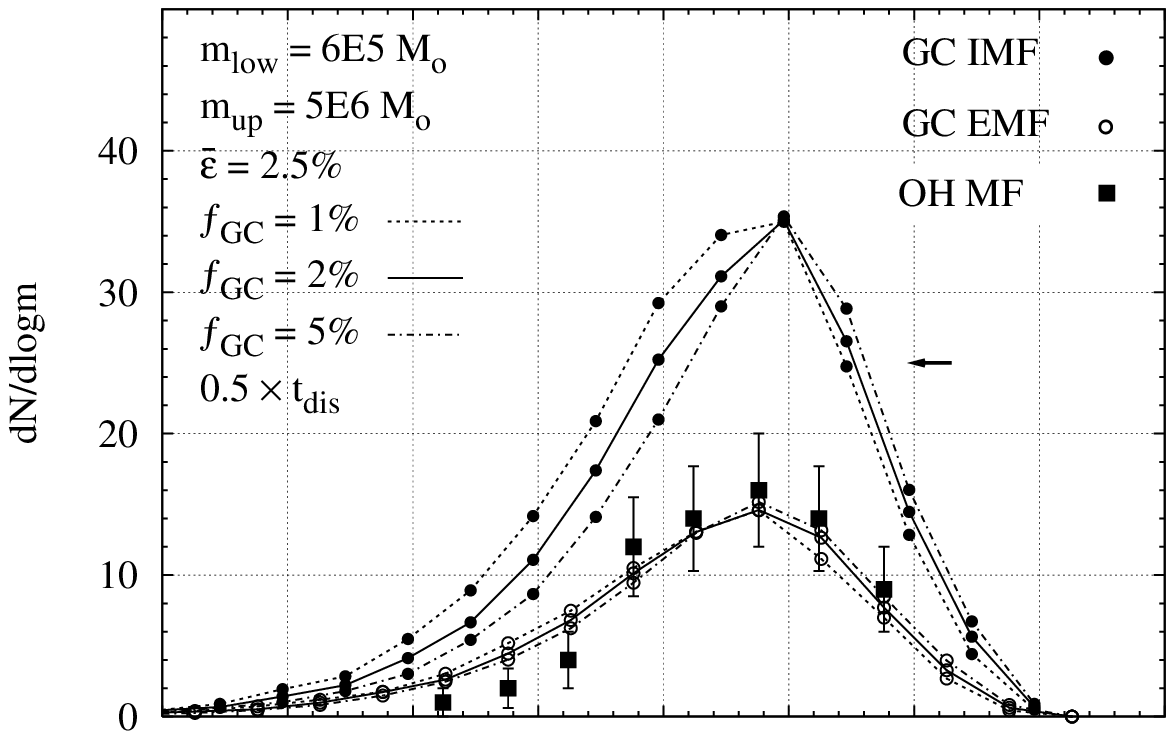, width=\linewidth} 
\end{minipage}
\vfill
\vspace{-10mm}
\begin{minipage}[t]{\linewidth}
\centering\epsfig{figure=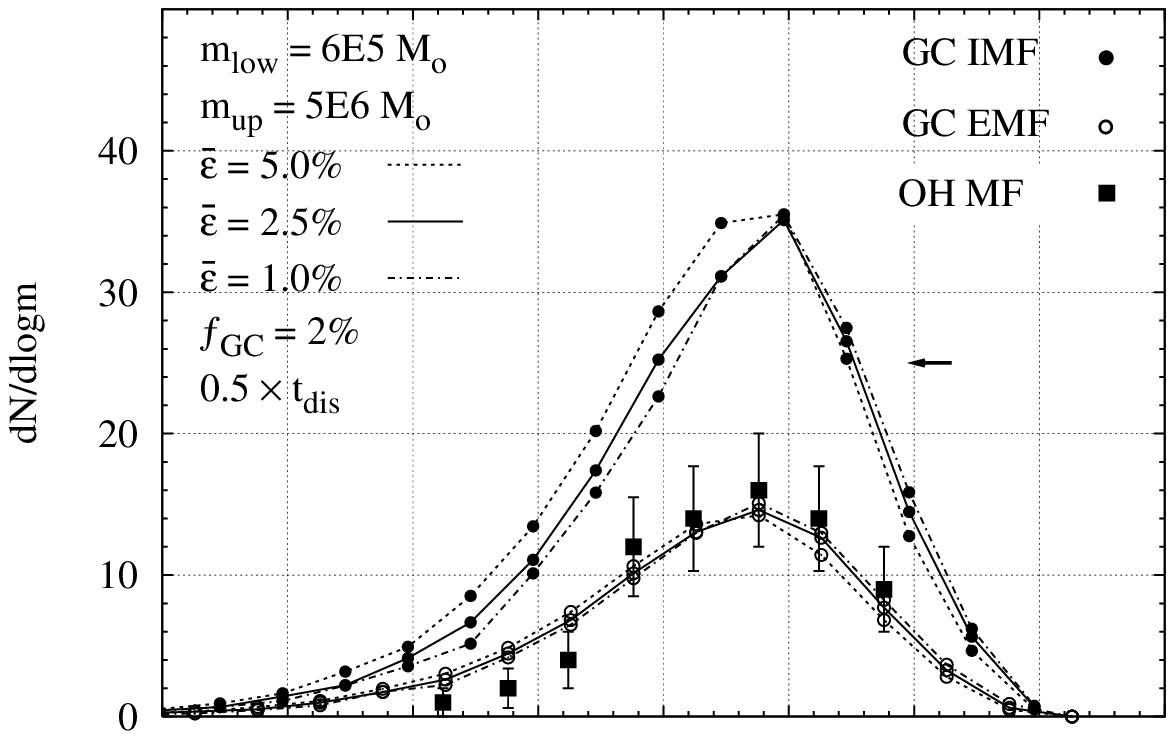, width=\linewidth}
\end{minipage}
\vfill
\vspace{-10mm}
\begin{minipage}[t]{\linewidth}
\centering\epsfig{figure=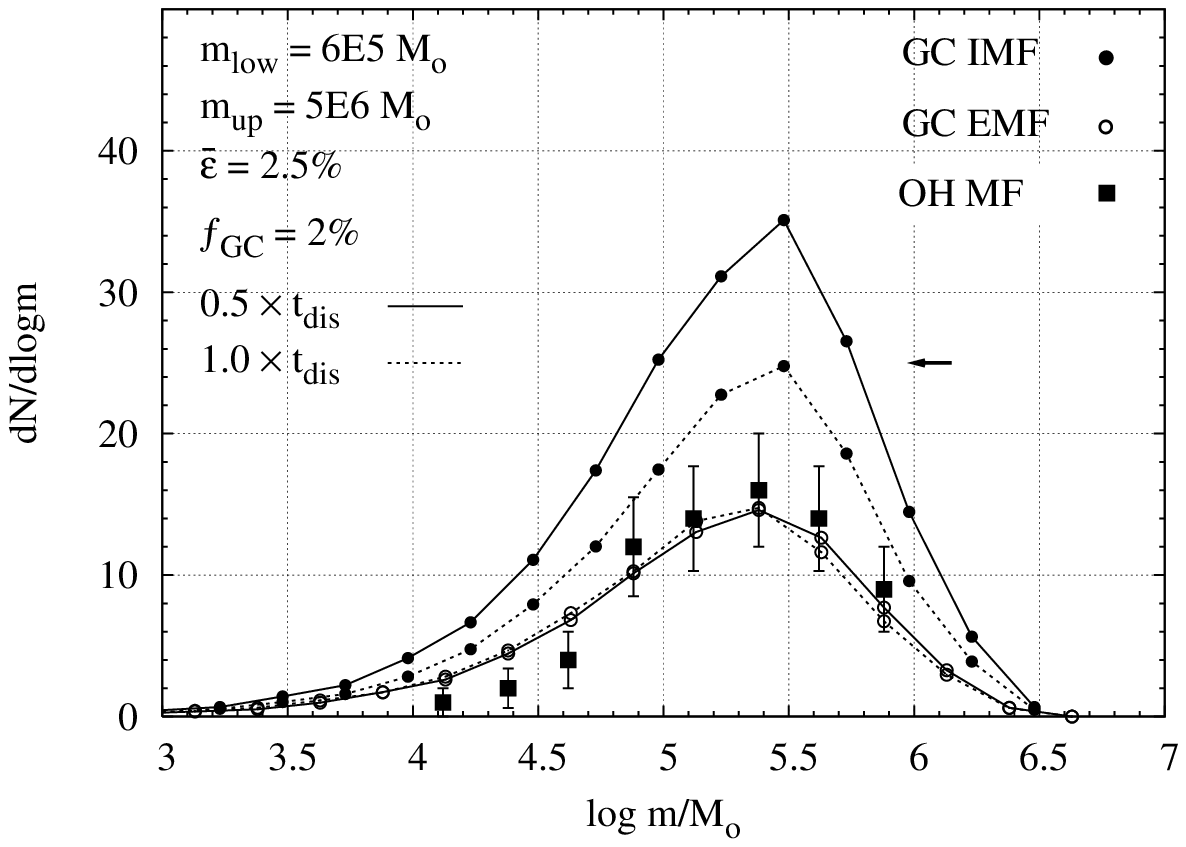, width=\linewidth}
\end{minipage}
\caption{Initial/evolved (plain/open circles) cluster mass functions corresponding to the cases listed in Table \ref{tab:OHMF_var_PL}.  In each panel, the solid lines are the fiducial case.  The arrow indicates that the initial mass functions have been shifted by -0.15, that is, corrected for stellar evolutionary mass loss.  The top and bottom panels show the effect of varying the lower cloud mass limit $m_{low}$ and the cluster disruption time-scale $t_{dis}$, respectively.  The middle panels display the results for different present-day cluster mass fractions $f_{GC}^{13Gyr}$ and different mean star formation efficiencies $\overline {\epsilon}$ (i.e. different $\wp (\epsilon)$).  The Old Halo mass function is depicted by the full squares. } 
\label{fig:OHMF} 
\end{figure}

\begin{table*}
\begin{center}
\caption[]{Impact of varying the protoglobular cloud mass range and the star formation efficiency distribution $\wp (\epsilon)$ on the model outcomes.   We successively explore how variations in the lower cloud mass limit $m_{low}$, the upper cloud mass limit $m_{up}$, the slope $\delta$ and core $r_c$ of $\wp ({\epsilon})$ affect the cluster mass fraction in the stellar halo $f_{GC}^{13Gyr}$ and the goodness of fit $Q$ of the cluster mass function at an age of 13\,Gyr, as well as the ratio $F_M$ between the final and initial masses in clusters and the fraction $F_N$ of surviving clusters, with respect to the fiducial case (first line of the Table).  We also consider a cluster disruption time-scale $t_{dis}$ twice as high as for the fiducial case, corresponding to clusters on circular orbits.  The mean (bound) star formation efficiency $\overline {\epsilon}$ ($\overline {\epsilon _b}$) is given for each couple ($\delta$, $r_c$).    } 
\label{tab:OHMF_var_PL}
\begin{tabular}{ l | c c c c c c c c c c } \hline 
 & $m_{low}/{\rm M}_{\odot}$ & $m_{up}/{\rm M}_{\odot}$ & $\delta$ & $r_c$ & $\overline {\epsilon}$ & $\overline {\epsilon}_b$ & $f_{GC}(13\,Gyr)$ & $F_M$  & $F_N$ &  Q \\ \hline
Fiducial case                   & $6 \times 10^5$ & $5 \times 10^6$ & $-2.9$ & $0.025$ & $0.025$ & $1.2 \times 10^{-3}$ & $0.019$ & $0.28$ & $0.44$ & $0.10$ \\ \hline
$\Delta (m_{low})$              & $2 \times 10^5$ & $5 \times 10^6$ & $-2.9$ & $0.025$ & $0.025$ & $1.2 \times 10^{-3}$ & $0.016$ & $0.24$ & $0.28$ & $10^{-10}$ \\
                                & $4 \times 10^5$ & $5 \times 10^6$ & $-2.9$ & $0.025$ & $0.025$ & $1.2 \times 10^{-3}$ & $0.018$ & $0.26$ & $0.37$ & $5 \times 10^{-4}$ \\
                                & $1 \times 10^6$ & $5 \times 10^6$ & $-2.9$ & $0.025$ & $0.025$ & $1.2 \times 10^{-3}$ & $0.020$ & $0.30$ & $0.49$ & $0.44$ \\ \hline
$\Delta (f_{GC})$               & $6 \times 10^5$ & $5 \times 10^6$ & $-1.9$ & $0.006$ & $0.025$ & $3.1 \times 10^{-3}$ & $0.052$ & $0.29$ & $0.46$ & $0.29$  \\ 
                                & $6 \times 10^5$ & $5 \times 10^6$ & $-3.9$ & $0.048$ & $0.025$ & $0.5 \times 10^{-3}$ & $0.008$ & $0.27$ & $0.40$ & $0.02$  \\ \hline
$\Delta (\overline {\epsilon})$ & $6 \times 10^5$ & $5 \times 10^6$ & $-2.4$ & $0.005$ & $0.010$ & $0.5 \times 10^{-3}$ & $0.021$ & $0.29$ & $0.45$ & $0.29$ \\ 
                                & $6 \times 10^5$ & $5 \times 10^6$ & $-4.4$ & $0.123$ & $0.050$ & $2.4 \times 10^{-3}$ & $0.019$ & $0.27$ & $0.41$ & $0.03$ \\ \hline
$\Delta (m_{up})$               & $6 \times 10^5$ & $3 \times 10^6$ & $-2.9$ & $0.025$ & $0.025$ & $1.2 \times 10^{-3}$ & $0.017$ & $0.25$ & $0.41$ & $0.02$ \\ 
                                & $6 \times 10^5$ & $1 \times 10^7$ & $-2.9$ & $0.025$ & $0.025$ & $1.2 \times 10^{-3}$ & $0.020$ & $0.36$ & $0.45$ & $0.23$ \\ 
                                & $6 \times 10^5$ & $2 \times 10^7$ & $-2.9$ & $0.025$ & $0.025$ & $1.2 \times 10^{-3}$ & $0.020$ & $0.29$ & $0.45$ & $0.25$ \\ \hline
$\Delta (t_{dis})$              & $6 \times 10^5$ & $5 \times 10^6$ & $-2.9$ & $0.025$ & $0.025$ & $1.2 \times 10^{-3}$ & $0.026$ & $0.38$ & $0.62$ & $0.05$ \\ \hline
\end{tabular}
\end{center}
\end{table*}   

\begin{enumerate}
\item Variations in $m_{low}$ \\ 
As noted in section \ref{subsubsec:mlow}, the turnover location
depends primarily on the lower limit of the cloud mass range.  A smaller
value of $m_{low}$ shifts the turnover downward, thereby raising the relative
fraction of low-mass clusters (see top panel of Fig.~\ref{fig:OHMF}, in which 
$m_{low} = 2 \times 10^5, 6 \times 10^5 ~{\rm and}~ 10^6 {\rm M}_{\odot}$).  The 
present-day mass fraction $f_{GC}^{13Gyr}$ of clusters in the halo and the 
survival ratios $F_M$ and $F_N$ are therefore decreased with respect to 
the fiducial case.

\item Variations in $f_{GC}^{13Gyr}$ \\ 
Varying the slope $\delta$ of the efficiency distribution $\wp (\epsilon )$,
while retaining the same mean star formation efficiency $\overline {\epsilon}$, 
alters the fraction of protoglobular clouds giving rise to bound star clusters 
following gas removal, that is, those for which $\epsilon > \epsilon _{th}$.  
As illustrated by the comparison between the middle and bottom panels of Fig.~\ref{fig:isoQfGC_mlowmup}, a smaller $\delta$ value lessens the present-day mass fraction of clusters $f_{GC}^{13Gyr}$
through a lower mean bound star formation efficiency $\overline {\epsilon _b}$
(see top panel and equation \ref{eq:fGC2}).  
Accordingly, the cluster initial mass function
reveals a larger fraction of low-mass clusters.  Yet, at an age of 13\,Gyr, 
the evolved cluster mass functions are undistinguishable (see the second panel 
of Fig.~\ref{fig:OHMF} for which $\delta = -3.9, -2.9, -1.9$,
corresponding to $f_{GC}^{13Gyr} \simeq 0.01, 0.02 0.05$, respectively).  

\item Variations in $\overline {\epsilon }$ \\ 
We have already emphasized how poorly constrained is our estimate $\overline {\epsilon }=2.5$\,\%
of the mean star formation efficiency.  However, the third panel in Fig.~\ref{fig:OHMF}
shows that a lower (1\,\%) or higher (5\,\%) value for $\overline {\epsilon }$,
while retaining the observational constraint $f_{GC}^{13Gyr} = 2\,\%$, 
hardly affects the shape of the evolved cluster mass function.

\item Variations in $t_{dis}$ \\
In the simulations described so far in the present section, we have assumed 
a mean eccentricity $e=0.5$ for the cluster orbits.  Equivalently, the disruption
time-scale of any cluster is half (i.e. $(1-e)$) that of the same cluster on a 
circular orbit.  The bottom panel of Fig.~\ref{fig:OHMF} compares our fiducial 
case to the case for which all clusters are on circular orbits.  It appears that 
the exact scaling of the cluster disruption time-scale (or, equivalently, the 
exact age of the cluster system) does not influence the shape of the cluster 
evolved mass function either.  This is so because the observed globular cluster mass 
function is a state of dynamical equilibrium, its shape being preserved over 
the cluster system evolution (see section \ref{subsec:eqMF}).  

Encounters between gas clouds and globular clusters during their early secular evolution, when they orbit through still gas-rich environments, constitute another source of uncertainty in the cluster disruption time-scale.  Gieles et al.~(2006) find that, when taking into account giant molecular cloud encounters, the disruption time-scale of clusters in the solar neighbourhood is shortened by a factor $\simeq 3.5$ compared to what is found in simulations taking into account the combined effect of Galactic tidal field and stellar evolution only.  Again, this uncertainty is of little relevance to the evolution with time of the globular cluster mass function, since this represents an equilibrium state.
This is exemplified by Fig.~3 in de Grijs, Parmentier \& Lamers (2005) which shows that the evolved counterpart of a Gaussian cluster initial mass function similar to that inferred in this section is practically the same, regardless of whether the cluster disruption time-scale is that predicted by Baumgardt \& Makino (2003) or 10 times shorter.  

The main effect of a shortened cluster disruption time-scale is a lowering of the cluster survival rates $F_M$ and $F_N$, equivalently, an increase of the initial number of clusters in order to explain any given present-day cluster population.

\item Variations in $m_{up}$ \\
How the upper limit $m_{up}$ of the cloud mass range affects
the results is discussed in the next section.
\end{enumerate}
Examination of the panels of Figs.~\ref{fig:OHMF} and \ref{fig:OH_mup} reveals that, over the mass range $m < 5 \times 10^4\,{\rm M}_{\odot}$, none of our models fit the observed mass distribution in the sense that they all slightly overpredict the number of clusters.  The discrepancy is far from severe, however, as it remains limited to less than a 2-$\sigma$ error.  Whether this effect truly resides in our model or reflects the incompleteness of the cluster sample is uncertain.  A few globular clusters located on the other side of the Galactic bulge
may have run undetected thus far (see, e.g., Fig.~1 of Larsen 2006 which shows the asymmetrical distribution of observed globular clusters in the Galactic plane).  Missing clusters tend to preferentially include low-mass ones, owing to the combination of their lower luminosity and lower concentration (see Fig.~4 of Larsen 2006 for the mass-concentration correlation).  They may thus account for the discrepancy noticed in Figs.~\ref{fig:OHMF} and \ref{fig:OH_mup}.  This possibility gets strengthened by a comparison between the mass spectrum of the Old Halo cluster system (plain squares in the top panel of Fig.~\ref{fig:OH_mup}) and that of the more populous M31 globular cluster system (see Fig.~1 of McLaughlin \& Pudritz 1996).
While the former tends to steadily decrease with decreasing cluster mass over the mass range $m < 5 \times 10^4\,{\rm M}_{\odot}$, the latter remains flat, possibly even slightly increasing with decreasing cluster mass, thereby agreing with what is predicted by our model.    

\begin{figure}
\begin{minipage}[t]{\linewidth}
\centering\epsfig{figure=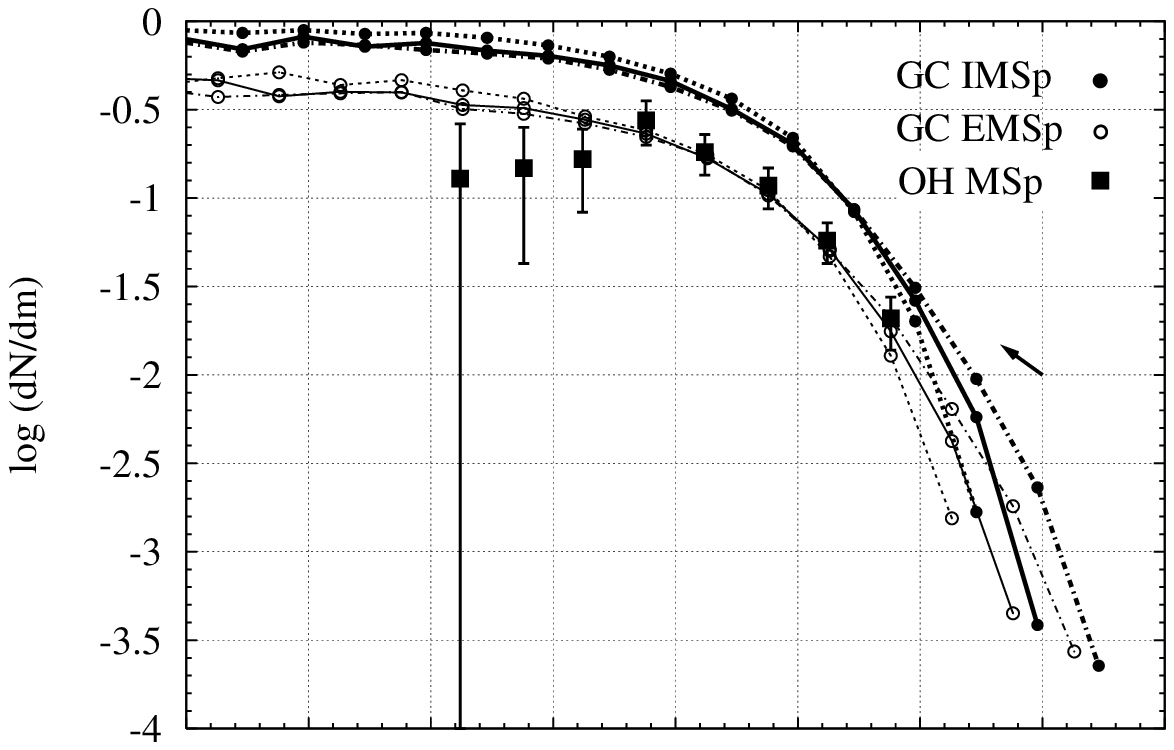, width=1.02\linewidth} 
\end{minipage}
\vfill
\vspace{-10mm}
\begin{minipage}[t]{\linewidth}
\hspace*{1.8mm}\epsfig{figure=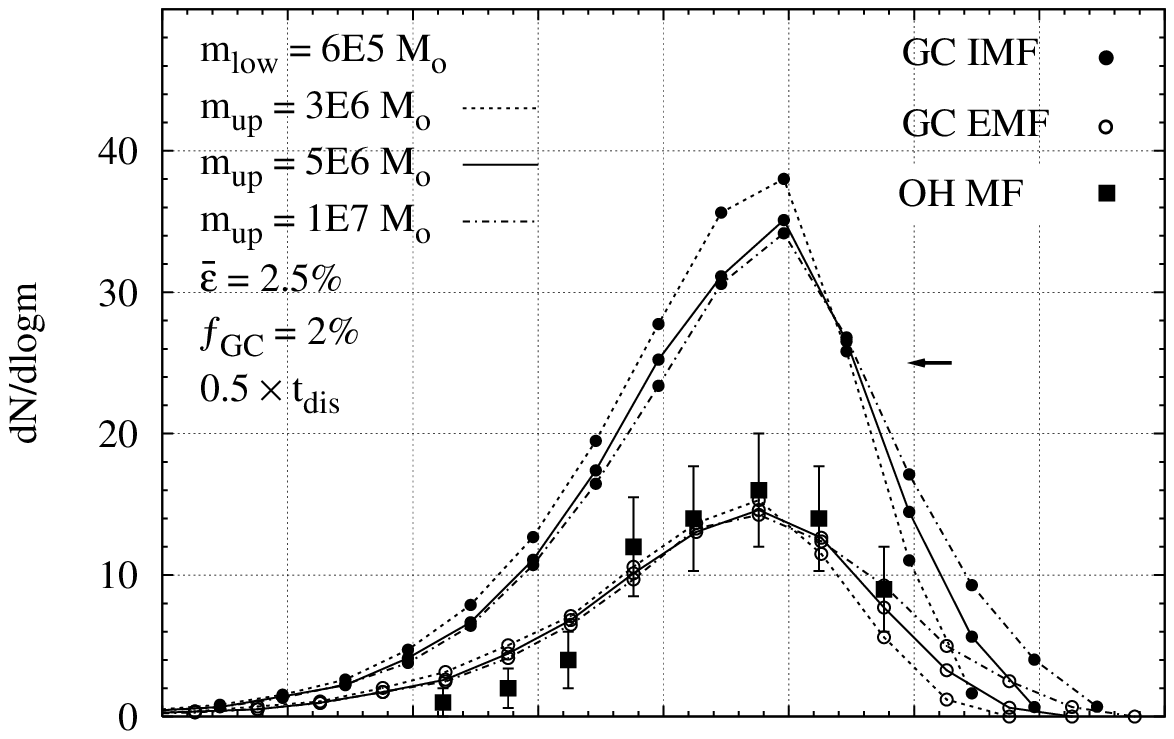, width=\linewidth}
\end{minipage}
\vfill
\vspace{-10mm}
\begin{minipage}[t]{\linewidth}
\hspace*{1.8mm}\epsfig{figure=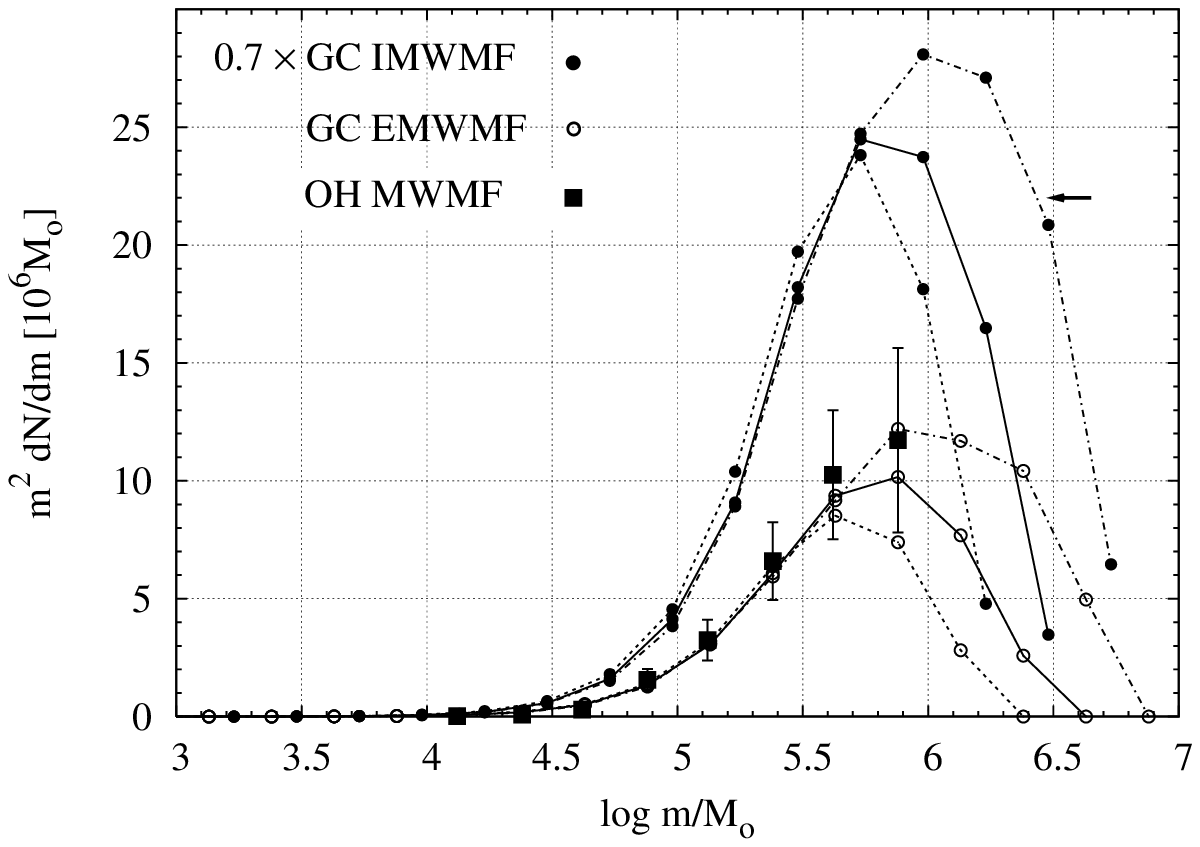, width=\linewidth}
\end{minipage}
\caption{Initial/evolved (solid/open circles) cluster mass distributions for three distinct upper cloud mass limits $m_{up}$ (see Table \ref{tab:OHMF_var_PL}).  In each panel, the solid lines are the fiducial case.  The arrow indicates that the initial mass functions have been shifted by -0.15, that is, corrected for stellar evolutionary mass loss.  The Old Halo mass distribution is depicted as the plain squares.  The top, middle and bottom panels show the mass spectra, the mass functions and the mass-weighted mass functions, respectively.   } 
\label{fig:OH_mup} 
\end{figure}

\begin{figure}
\begin{minipage}[t]{\linewidth}
\centering\epsfig{figure=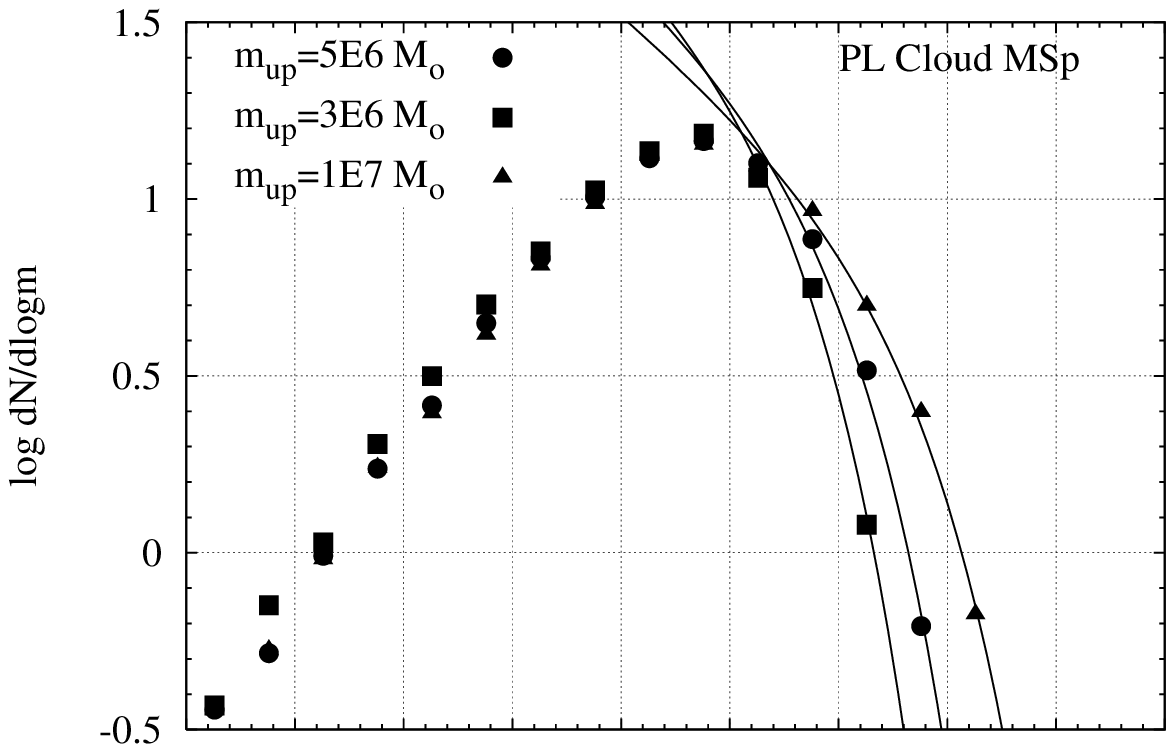, width=\linewidth} 
\end{minipage}
\vfill
\vspace{-10mm}
\begin{minipage}[t]{\linewidth}
\centering\epsfig{figure=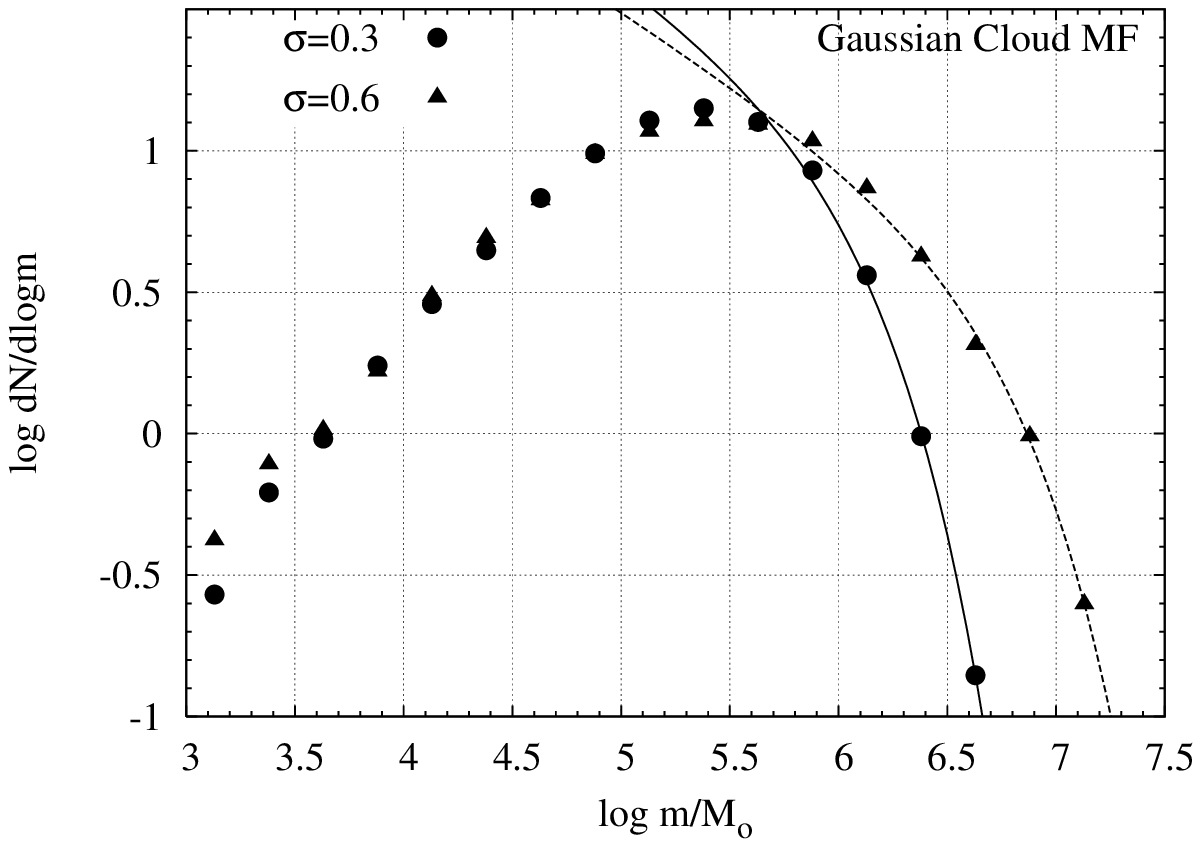, width=\linewidth}
\end{minipage}
\caption{Evolved cluster lognormal mass functions generated by our model (solid symbols) for various upper limits $m_{up}$ of the power-law cloud mass spectrum ({\it Top panel}, see Table \ref{tab:OHMF_var_PL}) and various standard deviations $\sigma$ of the Gaussian cloud mass function ({\it Bottom panel}, see Table \ref{tab:OHMF_var_G}).  The solid lines correspond to the functional form given by equation \ref{eq:msp_cutoff} and introduced by Burkert \& Smith (2000) to describe globular cluster mass distributions above the turnover of the mass function.  {\it Top panel}: $m_c = 0.5\times 10^6 {\rm M}_{\odot}, 0.9\times 10^6 {\rm M}_{\odot}, 2 \times 10^6 {\rm M}_{\odot}$.  {\it Bottom panel}: $m_c = 1.1\times 10^6 {\rm M}_{\odot}, 6 \times 10^6 {\rm M}_{\odot}$.   } 
\label{fig:lognorm} 
\end{figure}

\subsection{Substructures in the high-mass regime of the cluster mass distribution}
\label{subsec:mup}
Although the globular cluster mass at the turnover of the present-day Gaussian mass function is practically constant from one massive galaxy to another, the overall shape of the globular cluster mass function is {\it not} strictly universal.  For instance, McLaughlin \& Pudritz (1996) note that the high-mass regime of the mass function of the globular cluster systems of the Milky Way and M31 combined differs from that of the giant elliptical M87.  These differences are best revealed by the mass-weighted mass function 
$m^2 {\rm d}N/{\rm d}m \propto m {\rm d}N/{\rm dlog}m$,  
which is derived by summing individual cluster masses in evenly spaced logarithmic mass bins.  
The mass-weighted mass function therefore describes how the total mass of the cluster system is distributed between the high- and low-mass members of the cluster population.  
If the spectral index of the cluster mass spectrum ${\rm d}N/{\rm d}m$ is $\alpha = -2$, 
the mass-weighted mass function is flat and any two identical-size logarithmic mass bins 
contribute equally to the total mass of the cluster system (e.g. the contributions to the system total mass of the clusters in the mass range $10^3 < m < 10^4 {\rm M}_{\odot}$ and of those
in the mass range $10^5 < m < 10^6 {\rm M}_{\odot}$ are the same).  A shallower (steeper) mass spectrum implies that the total mass is mostly contained in the high-mass (low-mass) objects and that the mass-weighted mass function is increasing (decreasing) with cluster mass.    

McLaughlin \& Pudritz (1996) note that there is a cluster mass $m_{peak}$, larger than that at the mass function turnover, at which the mass-weighted mass function peaks.  That is, the spectral index of the mass spectrum steepens from $\alpha > -2$ to $\alpha < -2$ and, above the turnover, the globular cluster mass spectrum cannot be accurately fitted with a single power-law.  

The mass-weighted mass function of the Old Halo, as well as those predicted by our model for various upper protoglobular cloud mass limits $m_{up}$, are shown in the bottom panel of Fig.~\ref{fig:OH_mup}.  The observed distribution reveals that the contribution of low-mass ($\lesssim 10^5 {\rm M}_{\odot}$) globular 
clusters to the total mass of the Old Halo cluster system is negligible.  The absence of any peak in the mass-weighted mass function of the Old Halo likely arises from the limited size of the cluster sample (72 clusters).  In contrast, the mass-weighted mass function of the globular cluster system of the giant elliptical M87 shows a prominent peak (McLaughlin \& Pudritz 1996, their Fig.~2).  Owing to a cluster number about two orders of magnitude larger than that of the Old Halo, the M87 mass distribution is probed up to higher cluster mass and the substructures of the high-mass regime are therefore better highlighted.

Interestingly enough, our model predicts a peak in the mass-weighted mass function.  The bottom panel of Fig.~\ref{fig:OH_mup} shows that its location depends on the upper limit $m_{up}$ of the protoglobular cloud mass range, the higher $m_{up}$, the higher $m_{peak}$.  The dotted, solid and dashed-dotted lines correspond to $m_{up} = 3 \times 10^6, 5 \times 10^6, 10^7 {\rm M}_{\odot}$, respectively.  The corresponding mass spectra and mass functions are shown in the top and middle panels of Fig.~\ref{fig:OH_mup}, respectively.  We note that the cluster mass at the mass function turnover is well-preserved, regardless of $m_{up}$.  The upper limit of the cloud mass range controls mostly the high-mass regime of the cluster mass distribution, that is, the variations with the cluster mass of the slope of the mass spectrum, the extent of the tail of the mass function above the turnover and the cluster mass at the peak of the mass-weighted mass function.

Burkert \& Smith (2000) revisit the issue of how different among galaxies is 
the high-mass regime of the cluster mass distribution.  They show that the bright
end of the luminosity distribution of globular cluster systems in five elliptical 
and spiral galaxies (M87, NGC5846, NGC4472, the Milky Way and M31) is well-fitted with an underlying mass spectrum of the form
\begin{equation}
\frac{{\rm d}N}{{\rm d}m} \propto m^{-3/2} e^{-m/m_c}\,,
\label{eq:msp_cutoff}
\end{equation}  
under the assumption that the cluster mass-to-light ratio is constant within each system.  In equation \ref{eq:msp_cutoff}, the exponential cutoff causes the mass spectrum to drop off toward high masses more steeply than a pure power-law of the form $\frac{{\rm d}N}{{\rm d}m} \propto m^{-3/2}$.  This approach thus supersedes that of McLaughlin \& Pudritz (1996), replacing their double index power-law description of the mass spectrum high-mass regime 
(i.e. $\alpha > -2$ and $\alpha < -2$ below and above $m_{peak}$, respectively) by a continuously varying slope.  Figure 1 of Burkert \& Smith (2000) shows that equation \ref{eq:msp_cutoff} fits well the clear curvature of the considered globular cluster mass distributions above the turnover, the cutoff mass $m_c$ being in the range $10^6 {\rm M}_{\odot} < m_c < 5 \times 10^6 {\rm M}_{\odot}$.

\begin{figure}
\begin{minipage}[t]{\linewidth}
\centering\epsfig{figure=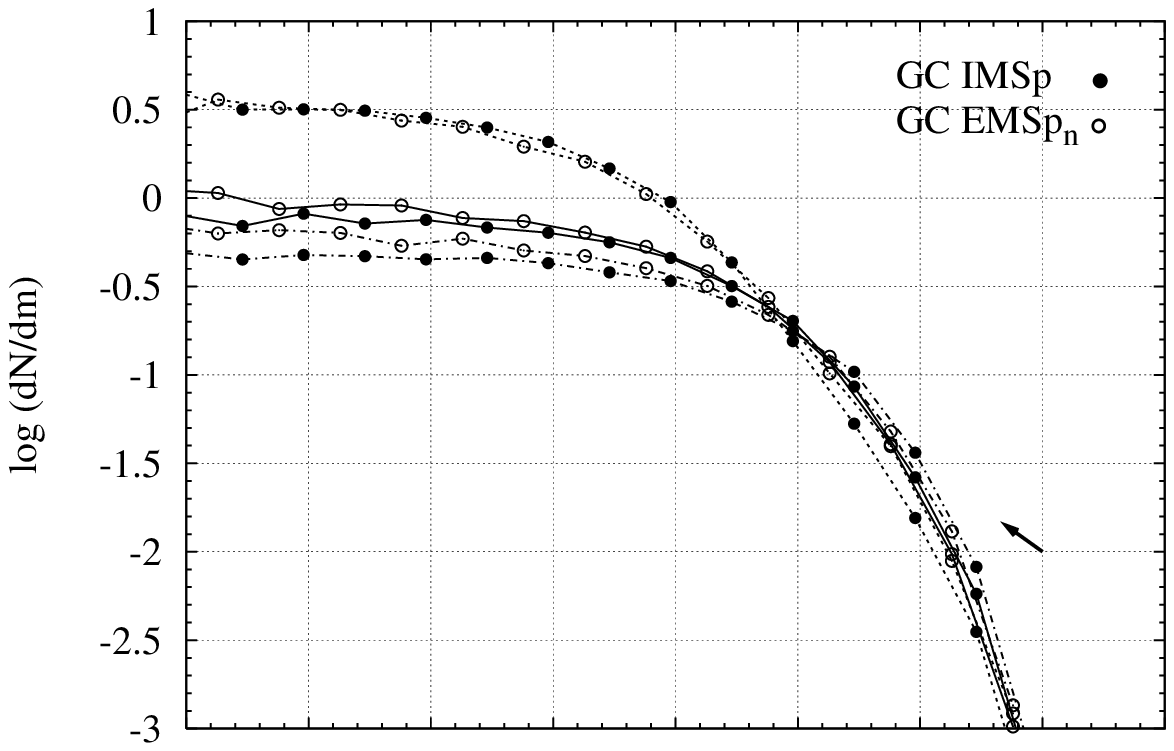, width=1.02\linewidth} 
\end{minipage}
\vfill
\vspace{-10mm}
\begin{minipage}[t]{\linewidth}
\hspace*{1.8mm}\epsfig{figure=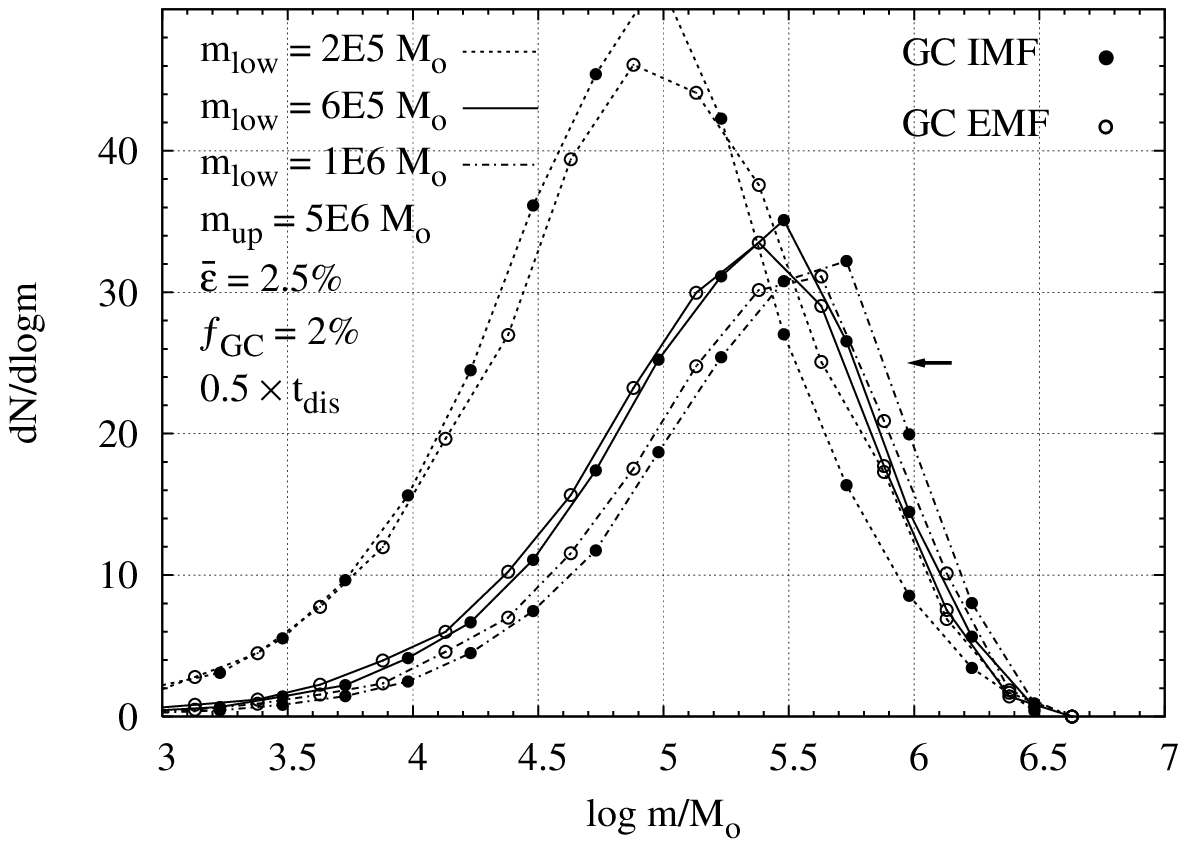, width=\linewidth}
\end{minipage}
\caption{Initial/evolved (solid/open circles) cluster mass distributions for three distinct lower cloud mass limits $m_{low}$.  The top and bottom panels correspond to the mass spectra and the mass functions,  respectively.  The solid lines are the fiducial case.  The arrow indicates that the initial mass distributions have been shifted by -0.15 in ${\rm log}m$, that is, corrected for stellar evolutionary mass loss.  The evolved mass distributions have been normalized so as to contain the same number of clusters as the inital ones.  The initial and evolved distributions are alike, that is, the cluster mass distribution is preserved over the course of the cluster system dynamical evolution.     } 
\label{fig:eqMF} 
\end{figure}

Figure \ref{fig:lognorm} compares the mass distributions of evolved cluster systems as obtained by our model (plain symbols) to those derived from equation \ref{eq:msp_cutoff} (solid lines, with $m_c = 0.5 \times 10^6, 0.9 \times 10^6, 2 \times 10^6\,{\rm M}_{\odot}$ in the top panel and $m_c = 1.1 \times 10^6, 6 \times 10^6\,{\rm M}_{\odot}$ in the bottom panel).  For the sake of comparison with Burkert \& Smith 's (2000) figure 1, we use the same cluster mass distribution, namely, the log-normal mass function, which is the logarithm of the bell-shaped mass function, i.e. ${\rm log}({\rm d}N/{\rm dlog}m)$.  The bottom and top panels consider the cases of a power-law mass spectrum and of a Gaussian mass function for the protoglobular clouds, respectively (see section \ref{subsec:cloud_G} for the latter).  The evolved cluster mass distributions match well the curvature characteristic of equation 10, which is the curvature observed for the mass distribution of globular cluster systems above the turnover.  We therefore conclude that our model is able {\it to account for the detailed shape of the globular cluster mass distribution beyond the turnover}.  Besides, varying the upper limit $m_{up}$ of the cloud mass range enables us to reproduce different exponential cutoff $m_c$ at high-masses, {\it while preserving the location of the turnover of the mass function, as is observed}.  These variations may originate from a correlation between $m_{up}$ and the mass of the parent galaxy and/or its environment.  Observations of present-day systems of molecular clouds actually reveal
a dependence of the upper cloud mass limit on the host galaxy (e.g. Wilson et al.~2003, Williams \& McKee 1997).

\subsection{An equilibrium cluster mass function}
\label{subsec:eqMF}
A striking feature of the (practically) universal globular cluster mass function is its ability to preserve its shape over a Hubble-time of dynamical evolution in the tidal field of its host galaxy, an effect firstly highlighted by Vesperini (1998) (see section \ref{sec:intro}).  The cluster mass distributions predicted by our model obey this property, as illustrated in Fig.~\ref{fig:eqMF}
(top panel: mass spectra, bottom panel: mass functions).  The plain symbols depict the cluster initial mass distributions after correction for stellar evolutionary mass loss.  The open symbols show the corresponding cluster mass distributions evolved to an age of 13\,Gyr and normalized so as to contain the same number of clusters as the initial ones.  We can thus straightforwardly estimate whether the shape of the mass distributions evolve significantly over a Hubble-time owing to the evaporation and dissolution of globular clusters.  The initial and evolved cluster mass distributions for the fiducial case ($m_{low} = 6 \times 10^5 {\rm M}_{\odot}$, solid lines) appear indistinguishable in shape.  Slightly different lower cloud mass limits ($m_{low} = 2 \times 10^5 {\rm M}_{\odot}$ and $m_{low} = 10^6 {\rm M}_{\odot}$, dotted and dashed-dotted lines, respectively) also result in fairly well-preserved cluster mass functions.

\subsection{Another cloud mass distribution}
\label{subsec:cloud_G}

\begin{figure}
\centering\epsfig{figure=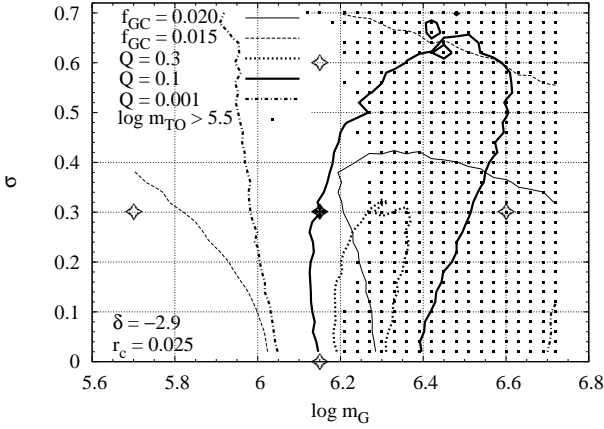, width=\linewidth}
\caption{Incomplete gamma functions $Q$ of the fit of the evolved cluster mass functions to the observed one (thick curves) and cluster mass fractions $f_{GC}^{13Gyr}$ at an age of 13\,Gyr as predicted by the model (thin curves) versus the mean ${\rm log}m_G$ and standard deviation $\sigma$ of the Gaussian cloud mass function.  The dotted area corresponds to evolved cluster mass functions for which the cluster mass at the turnover is larger than $3 \times 10^5\,{\rm M}_{\odot}$.  The four-branch stars represent different cases presented in Fig.~\ref{fig:OH_G} and Table \ref{tab:OHMF_var_G}, the solid symbol being our fiducial case. }
\label{fig:isoQfGC_mGsig} 
\end{figure}

\begin{figure}
\begin{minipage}[t]{\linewidth}
\centering\epsfig{figure=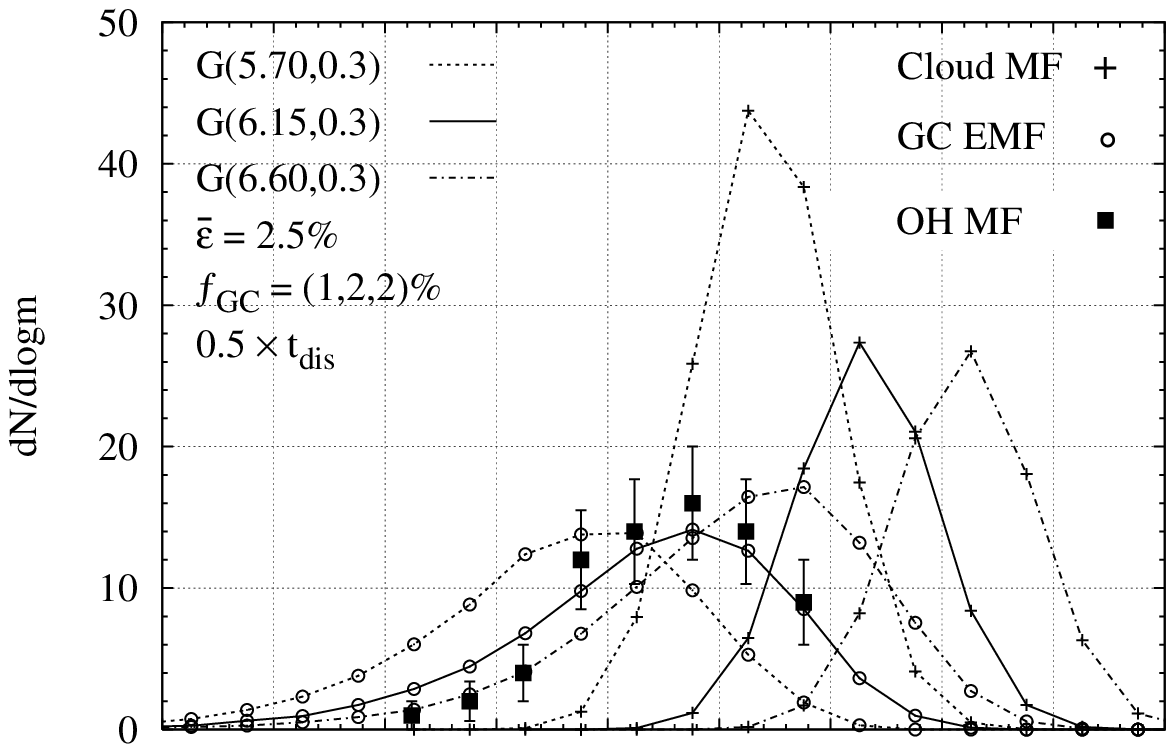, width=\linewidth} 
\end{minipage}
\vfill
\vspace{-10mm}
\begin{minipage}[t]{\linewidth}
\centering\epsfig{figure=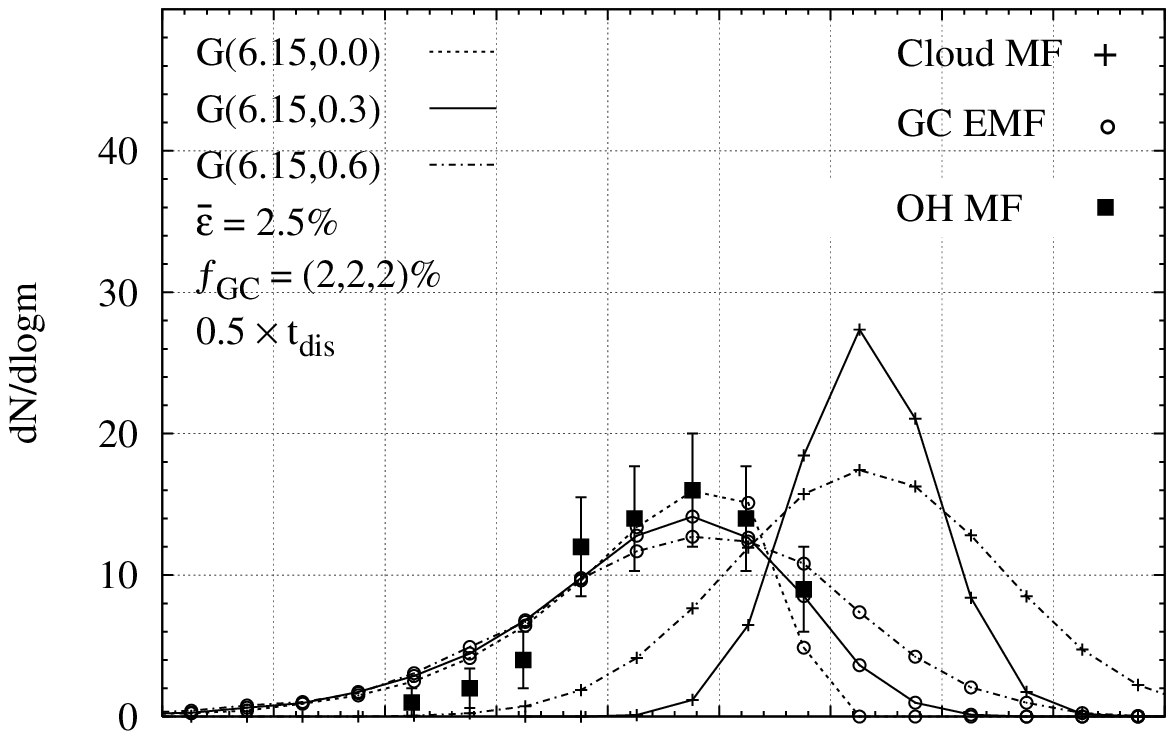, width=\linewidth}
\end{minipage}
\vfill
\vspace{-10mm}
\begin{minipage}[t]{\linewidth}
\centering\epsfig{figure=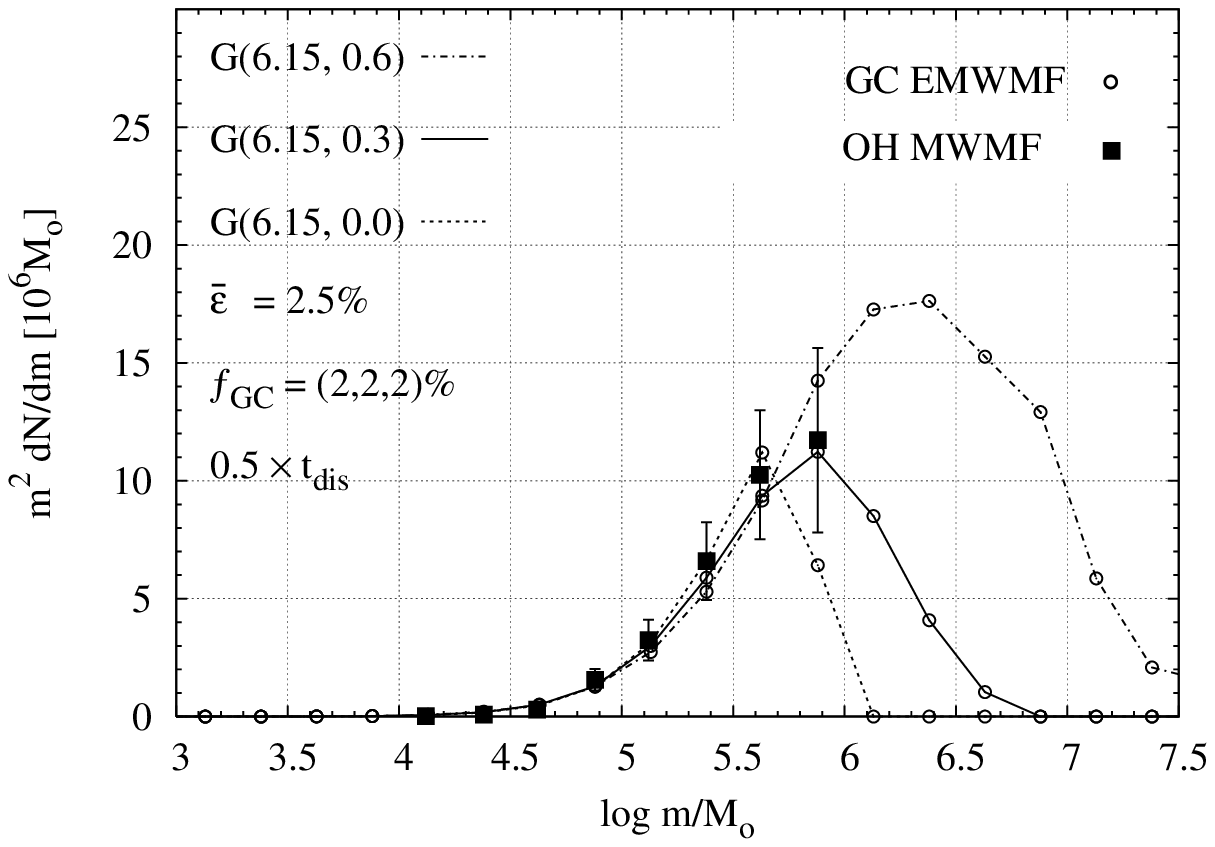, width=\linewidth}
\end{minipage}
\caption{Assumed Gaussian cloud mass functions: Cloud mass distributions (crosses) and evolved (open circles) cluster mass distributions corresponding to some of the cases listed in Table \ref{tab:OHMF_var_G}.  In each panel, the solid lines are the fiducial case.   The top and middle panels show how the evolved cluster mass functions respond to variations in the mean ${\rm log }m_G$ and the standard deviation $\sigma$ of the Gaussian cloud mass function, respectively.  The bottom panel displays the evolved mass-weighted mass functions corresponding to the cluster mass functions in the middle panel.  The Old Halo mass distribution is shown as the full squares. } 
\label{fig:OH_G} 
\end{figure}

\begin{table*}
\begin{center}
\caption[]{Same as Table \ref{tab:OHMF_var_PL}, but for a Gaussian cloud mass function with a mean ${\rm log }m_G$ and a standard deviation $\sigma$. } 
\label{tab:OHMF_var_G}
\begin{tabular}{ l | c c c c c c c c c c } \hline 
 & $log(m_G/{\rm M}_{\odot})$ & $\sigma$ & $\delta$ & $r_c$ & $\overline {\epsilon}$ & $\overline {\epsilon}_b$ & $f_{GC}(13\,Gyr)$ & $F_M$  & $F_N$ &  Q \\ \hline
Fiducial case                   & $6.15$ & $0.3$ & $-2.9$ & $0.025$ & $0.025$  & $1.2 \times 10^{-3}$ & $0.019$ & $0.28$ & $0.43$ & $0.10$      \\ \hline
$\Delta (log~m_G)$              & $5.70$ & $0.3$ & $-2.9$ & $0.025$ & $0.025$  & $1.2 \times 10^{-3}$ & $0.013$ & $0.19$ & $0.26$ & $10^{-15}$  \\
                                & $6.60$ & $0.3$ & $-2.9$ & $0.025$ & $0.025$  & $1.2 \times 10^{-3}$ & $0.021$ & $0.31$ & $0.53$ & $0.04$ \\ \hline
$\Delta (f_{GC})$               & $6.15$ & $0.3$ & $-1.9$ & $0.006$ & $0.025$  & $3.1 \times 10^{-3}$ & $0.053$ & $0.29$ & $0.46$ & $0.30$  \\ 
                                & $6.15$ & $0.3$ & $-3.9$ & $0.048$ & $0.025$  & $0.5 \times 10^{-3}$ & $0.008$ & $0.28$ & $0.40$ & $0.02$  \\ \hline
$\Delta (\overline {\epsilon})$ & $6.15$ & $0.3$ & $-2.4$ & $0.005$ & $0.010$  & $0.5 \times 10^{-3}$ & $0.021$ & $0.29$ & $0.44$ & $0.34$ \\ 
                                & $6.15$ & $0.3$ & $-4.4$ & $0.123$ & $0.050$  & $2.4 \times 10^{-3}$ & $0.020$ & $0.28$ & $0.41$ & $0.03$ \\ \hline
$\Delta (\sigma )$              & $6.15$ & $0.0$ & $-2.9$ & $0.025$ & $0.025$  & $1.2 \times 10^{-3}$ & $0.018$ & $0.26$ & $0.45$ & $0.13$ \\ 
                                & $6.15$ & $0.6$ & $-2.9$ & $0.025$ & $0.025$  & $1.2 \times 10^{-3}$ & $0.018$ & $0.27$ & $0.39$ & $0.03$ \\ \hline
$\Delta (t_{dis})$              & $6.15$ & $0.3$ & $-2.9$ & $0.025$ & $0.025$  & $1.2 \times 10^{-3}$ & $0.026$ & $0.38$ & $0.61$ & $0.05$ \\ \hline
\end{tabular}
\end{center}
\end{table*}   

The bottom panel of Fig.~\ref{fig:isoQfGC_mlowmup} demonstrates that a power-law protoglobular cloud mass spectrum with a narrow mass range, e.g. $10^6 {\rm M}_{\odot} \lesssim m_{cloud} \lesssim 2 \times 10^6 {\rm M}_{\odot}$, leads to a good fit ($Q \simeq 0.1$) of the modelled cluster mass function onto the observed one.    
This suggests that the present-day halo cluster mass distribution, which covers two decades in mass, may equally-well arise from a characteristic mass for the protoglobular clouds.
In order to investigate this point more closely, we now assume that the protoglobular cloud mass function obeys a Gaussian of mean ${\rm log}m_G$ and of standard deviation $\sigma$.  As for the power-law hypothesis presented in section \ref{subsec:isoQfGC_mlowmup}, the outcomes of the simulations are summarized as the iso-$Q$ and iso-$f_{GC}^{13Gyr}$ curves in the (${\rm log}m_G$, $\sigma$) plane
(see Fig.~\ref{fig:isoQfGC_mGsig}).  We have considered a probability distribution $\wp (\epsilon)$ for the star formation efficiency identical to that derived in the frame of the power-law hypothesis, namely,
$\delta =-2.9$ and $r_c=0.025$.  A Gaussian protoglobular cloud mass function centered at ${\rm log}m_G = 6.15$ with a standard deviation $\sigma$ less than 0.4 fits well ($Q \simeq 0.1$) the Old Halo cluster mass function, while predicting the correct halo cluster mass fraction, namely, $f_{GC}^{13Gyr} \simeq 2$\,\%.  Although a higher ${\rm log}m_G$ improves the goodness of fit, it also shifts the turnover of the evolved mass function toward unacceptably high cluster mass, an effect illustrated by the dotted
area at the right of the (${\rm log}m_G$, $\sigma$) plane, for which ${\rm log}m_{TO} \gtrsim 5.5$ as in Fig.~\ref{fig:isoQfGC_mlowmup}.  As for the power-law cloud mass spectrum, this artifact comes from the slight overestimate of the number of low-mass clusters by our model (see section \ref{subsec:isoQfGC_mlowmup}).

In Table \ref{tab:OHMF_var_G}, we investigate how model parameter variations affect our results with respect to the adopted fiducial case ${\rm log}m_G = 6.15$ and $\sigma =0.3$ (solid four-branch star in Fig.~\ref{fig:isoQfGC_mGsig}).  
The top and middle panels of Fig.~\ref{fig:OH_G} illustrate the effect of varying ${\rm log}m_G$ and $\sigma$, respectively.  The protoglobular cloud mass functions and the evolved cluster mass functions are depicted as the crosses and the open circles.  The shape of the cluster mass function being little affected by dynamical evolution (see section \ref{subsec:eqMF}), the cluster initial mass functions are not displayed for the sake of clarity.  The main controlling parameter of the turnover location is the mean protoglobular cloud mass, i.e., the higher ${\rm log}m_G$, the higher ${\rm log}m_{TO}$.  In case of a lower ${\rm log}m_G$ (e.g. ${\rm log}m_G \simeq 5.7$), the cluster initial mass function favours lower-mass clusters and the present-day cluster mass fraction $f_{GC}^{13Gyr}$ and the survival rates $F_M$ and $F_N$ get smaller (see Table \ref{tab:OHMF_var_G}).
As for the standard deviation $\sigma$ of the Gaussian cloud mass function, it controls the high-mass regime of the cluster mass distribution, without affecting the turnover location of the cluster mass function significantly.  A larger standard deviation $\sigma$ results in a more extended high-mass tail for the cluster mass function and a larger cluster mass $m_{peak}$ at the peak of the mass-weighted mass function (middle and bottom panels of Fig.~\ref{fig:OH_G}, respectively).              
In that sense, $\sigma$ plays a role similar to the upper limit $m_{up}$ of the power-law cloud mass spectrum (see section \ref{subsec:mup}).  We have also investigated the effect of variations in the star formation efficiency distribution $\wp (\epsilon)$ and in the cluster disruption time-scale $t_{dis}$, while keeping the fiducial cloud mass function.  The evolved cluster mass functions, the survival rates $F_M$ and $F_N$ and the present-day cluster mass fractions $f_{GC}^{13Gyr}$ are very similar to those derived in the frame of the power-law hypothesis (compare Table \ref{tab:OHMF_var_PL} and Table \ref{tab:OHMF_var_G}).  

\subsection{The universal globular cluster mass at the turnover and the need for a high protoglobular cloud mass-scale}
\label{subsec:proto_gx}

In order for our model to explain the universal Gaussian globular cluster mass function, the protoglobular clouds 
must be characterized by an almost invariant high-mass scale, either in the form of a lower truncation of the power-law cloud mass 
spectrum ($m_{low} \simeq 6 \times 10^5 {\rm M}_{\odot}$) or in the form of a high mean mass ${\rm log}m_G \simeq 6.2$ 
if the clouds are distributed following a Gaussian in log-mass.

The old age of GCs has prompted the development of models for the formation of protoglobular clouds before or during the formation of galaxies, in which the clouds actually show a high mass-scale of order $10^5-10^6\,{\rm M}_{\odot}$.
Peebles \& Dicke (1968) assign a pregalactic origin to globular clusters in that their gaseous precursors were the first gravitationally bound systems to form out of the expanding Universe.  They infer a typical protoglobular cloud mass of 
order $10^5\,{\rm M}_{\odot}$.  On the other hand, Fall \& Rees (1985, see also Kang, Lake \& Ryu 2000) suggest that globular clusters formed during the collapse of the protogalaxy out of gas that cools slowly below $10^4$\,K, thereby imprinting a preferred mass-scale of $\simeq 10^6\,{\rm M}_{\odot}$ for the protoglobular clouds.  Although these characteristic cloud masses match well those we have derived independently in sections \ref{subsec:isoQfGC_mlowmup} and \ref{subsec:cloud_G}, the pregalactic and protogalactic scenarios of Peebles \& Dicke (1968) and Fall \& Rees (1985), respectively, are hampered with several problems.  For instance, a pregalactic origin for globular clusters is hardly reconciled with the paucity of intergalactic globular clusters, while a metallicity [Fe/H] larger than $-2$ (i.e. the metallicity range of most globular clusters in the Galactic halo) prevents the gas temperature from hanging at $10^4$\,K.  As a result, considerable efforts have gone into analysing the protoglobular cloud cooling histories.  External sources of UV or X-ray flux (a background of hot halo gas, an active galactic nucleus or population III stars) have been invoked for the gas to retain those characteristic temperature and mass (Kang et al.~1990, Murray \& Lin 1992).  
More recently, Bromm \& Clarke (2002) explored a mechanism for the formation of the first globular clusters, operating during the assembly of dwarf galaxies at high redshift, $z \gtrsim 10$.  Adopting the $\Lambda$CDM model for structure formation, they showed that the collapse and virialization of a dwarf galaxy result in the formation of a handful of high-density gas clumps, these being identified as globular cluster progenitors.  
Because, in these models, dwarf galaxies formed before the epoch of reionization ($z \gtrsim 10$), no UV background prevents the gas temperature from dropping below $10^4$\,K and the gas thus falls into the dark matter subhalos.  The resulting clump masses are determined by the mass spectrum of the dark matter substructure.  Bromm \& Clarke 's (2002) three-dimensional numerical simulations of the collapsing dark matter and gaseous components lead to a clump mass range of $10^5 \lesssim m_{cloud} \lesssim 10^7\,M_{\odot}$.  Therefore, although the hypothesis that the gaseous progenitors of old globular clusters are characterized by a mass-scale of order $\simeq 10^6\,{\rm M}_{\odot}$, as found in our simulations, still remains debated, this appears to be a real possibility.

In contrast, it is clear that present-day mass distributions of giant molecular clouds and of their cluster forming cores show no evidence for a low-mass cut-off.  Giant molecular cloud masses are well described by a featureless power-law with $\alpha \simeq -1.7$ down to the completeness limit, this being of order $10^4\,{\rm M}_{\odot}$ in the Large Magellanic Cloud (Blitz et al.~2006).  If the mass distribution of the cluster gaseous progenitors were genuinely truncated at $\simeq 10^4\,{\rm M}_{\odot}$, this would lead to a bell-shaped cluster initial mass function with a cluster mass at the turnover of a few $10^3\,{\rm M}_{\odot}$ only.  Considering a system of young massive star clusters located beyond the Local Group of galaxies whose detection limit is at a similar cluster mass, only the high-mass regime of the bell-shaped mass function is probed and the inferred mass function is a power-law.  Therefore, our model does not contradict the observations of power-law cluster mass spectra, with spectral indices $-2 \lesssim \alpha \lesssim -1.8$ down to $10^3\,{\rm M}_{\odot}-10^4\,{\rm M}_{\odot}$, for systems of young star clusters in present-day starbursts and mergers.  \footnote{That the slope of power-law cluster mass spectra in starbursts ($-2 \lesssim \alpha \lesssim -1.8$) is often slightly steeper than that of cluster forming cores ($\alpha \simeq -1.7$) -assuming that the value deduced for the Galactic disc is valid in other large galaxies- may tell us something about an $\epsilon - m_{cloud}$ relation.  To investigate that particular point is beyond the scope of the present study, however.}

The exact shape of their mass distribution remains disputed, however.  This is exemplified by
the heavily studied system of star clusters hosted by the Antennae galaxies (NGC 4038/4039).
While Zhang \& Fall (1999) report a featureless power-law mass spectrum with a slope of $-2$ down to  $10^4\,{\rm M}_{\odot}$, Fritze v. Alvensleben (1998, 1999) finds a Gaussian mass function similar to that of old globular clusters.  Because the formation duration of young star cluster sytems formed in ongoing or recent starbursts may be a significant fraction of the system's median age, deriving their mass function is hampered by the strong variations in time of the integrated cluster mass-to-light ratio.  
That is, the observed cluster luminosity distribution may not be a faithfull mirror of the underlying cluster mass distribution and reliable cluster age estimates are required to convert the former into the latter (Meurer 1995).  Moreover, avoiding bright star contamination and accounting for completeness effects properly prove equally challenging
and may lead to discrepant cluster luminosity functions from one study to another.  For instance, Anders et al.~(2007)
report the first detection of a turnover in the Antennae cluster luminosity function at $M_V \simeq -8.5$. 
The young age of their sample ($<$100\,Myr), however, implies a cluster mass at the turnover roughly an order of magnitude lower than what is observed for old globular cluster systems.  

As for the galaxies NGC~3310 and NGC~6745, de Grijs et al.~(2003) report power-law cluster mass spectra with $\alpha \simeq -2$.  The cluster mass ranges probed corresponding to the high mass regime of the globular cluster Gaussian mass function ($>10^5\,{\rm M}_{\odot}$ and $> 4 \times 10^5\,{\rm M}_{\odot}$, respectively), one cannot distinguish the observed cluster mass functions from that of old globular clusters.  

Region "{\sl B}" of the starburst galaxy M82 hosts an almost  coeval cluster population.  In spite of an intermediate age ($\simeq$1\,Gyr), its bell-shaped mass function shows a turnover already located at a cluster mass similar to that of old globular cluster systems, i.e. $1.5 \times 10^5\,{\rm M}_{\odot}$.  de Grijs, Parmentier \& Lamers (2005) find that that cluster mass function is inconsistent with an initial power-law mass spectrum with $\alpha \simeq -2$, since the initial number of clusters would be unphysically high in that case \footnote{This result is driven by the combination of the high cluster mass at the turnover and the 1-Gyr age of the clusters.  Ongoing {\sl HST} {\sl U}-band observations will enable to refine these cluster mass and age distributions as well as the initial number of clusters required if starting from a power-law.}.  In contrast, a Gaussian initial cluster mass function similar to that observed today (hence, similar to the mass function characteristic of old globular clusters) constitutes an acceptable solution.  

An additional concern worth being borne in mind when dealing with the mass distribution of young clusters is whether they are in virial equilibrium or not.  Actually, in case of instantaneous gas removal, a protocluster settles back into virial equilibrium at an age of 40-50\,Myr only (see Figs.~2 and 5 of Goodwin \& Bastian 2006).  That is, any population of clusters younger than that age is necessarily contaminated by unbound clusters whose stellar content is being scattered into the field (i.e. those characterized by $\epsilon < \epsilon _{th}$) as well as by clusters which are still experiencing star escapes due to gas expulsion (i.e., when $\epsilon > \epsilon _{th}$).  The masses of these out-of-equilibrium/expanding clusters are in the ranges  0 - $\epsilon \times m_{cloud}$ and $F_{bound} \times \epsilon \times m_{cloud}$ - $\epsilon \times m_{cloud}$, respectively.  Yet, the mass of relevance here is 
that of gas-free bound star clusters, i.e. $F_{bound} \times \epsilon \times m_{cloud}$.  Consequently, the mass function of clusters younger than 40-50\,Myr, even if corrected for all the above mentioned caveats (temporal cluster mass-to-light ratio variations, bright star contamination and completeness effects), is {\it not necessarily} the cluster initial mass function inferred by studies of cluster sytem secular dynamical evolution (e.g., Baumgardt 1998, Vesperini 1998, Fall \& Zhang 2001 and Parmentier \& Gilmore 2005).  For instance, Bik et al.~(2003) report a power-law cluster mass spectrum with $\alpha \simeq -2$ and going down to $10^3\,{\rm M}_{\odot}$ for the cluster population in the inner spiral arms of the interacting galaxy M51.  That mass spectrum is "pre-initial", however, as it covers clusters younger than 10\,Myr, i.e. clusters still experiencing infant mortality and infant weight-loss. 
Despite being less than 10\,Myr old, Bik et al.~'s (2003) observed cluster mass spectrum will stand for the cluster initial mass spectrum which secular dynamical evolution models build on, provided that both the star formation efficiency $\epsilon$ and the star bound fraction $F_{bound}$ are independent of the cluster gaseous progenitor mass, as is explicitely assumed in our model.  Indications that infant mortality actually constitutes a mass-independent process, at least for observed cluster masses less than $10^4\,{\rm M}_{\odot}$, are given in Bastian et al.~(2005) for M51 and Fall et al.~(2005) for NGC4038/39 (see also Anders et al.~2007).  In that case, a featureless power-law mass distribution for the gas-embedded clusters is turned into the same featureless power-law for the cluster initial mass distribution (except for a vertical shift driven by infant mortality and infant weight-loss).  Bik et al.~'s (2003) result may thus actually points to a featureless power-law {\it initial} cluster mass spectrum with the "canonical" slope $\alpha \simeq -2$ in M51.

The above discussion shows that it is not straightforward to derive the initial mass distributions of young star cluster systems located beyond the Local Group.  In contrast, the Large Magellanic Cloud and the Small Magellanic Cloud are close enough so that a detailed survey of even faint clusters can be carried out.  Building on the framework of Boutloukos \& Lamers (2003), de Grijs \& Anders (2006) derive a disruption time-scale ${\rm log}(t_4^{dis}/{\rm yr})=9.9\pm0.1$ for a $10^4\,{\rm M}_{\odot}$ cluster (the disruption time-scale for any other cluster mass being obtained from $t^{dis} = t_4^{dis} (m/10^4{\rm M}_{\odot})^{0.62}$, see Boutloukos \& Lamers (2003) for details).  So long a cluster disruption time-scale stems from the low-density environment of the Clouds and guarantees that the cluster mass distributions derived by de Grijs \& Anders (2006) have not yet been significantly altered by dynamical secular evolution.  Considering clusters older than 60\,Myr and more massive than $10^3\,{\rm M}_{\odot}$, the cluster mass spectrum slope is $\alpha \simeq -2$, i.e. fully consistent with what is generally inferred for young star cluster systems.  Similar results are obtained by Hunter et al. (2003).  Yet, the mass spectrum slope of clusters younger than 60\,Myr appears significantly shallower, with $\alpha \simeq -1.7$ for the same mass range and $\alpha \simeq -1$ for clusters less massive than $10^3\,{\rm M}_{\odot}$ (see Fig.~8 and Table 3 of de Grijs \& Anders 2006).  They ascribe these substructures in the young age/low-mass regime to ongoing infant mortality.  As already quoted, at so young an age, the surveyed population of clusters necessarily consists of a mix of bound and unbound clusters.

The observation of power-law mass spectra for young massive star clusters has led to the idea that the initial mass spectrum of old globular clusters was also a power-law with a similar slope.  Yet, when evolving an initial power-law, most current models fail to reproduce the present-day globular cluster Gaussian mass function, except under very specific conditions, generally not consistent with observations (e.g. a narrow range of perigalactic distances, see Fall \& Zhang 2001 vs. Vesperini et al.~2003).  In an attempt to reconcile the Gaussian mass function of old globular clusters with the power-law cluster mass spectrum of young massive star clusters, Vesperini \& Zepf (2003) build on the observed trend between the mass of Galactic globular clusters and their concentration (i.e. the more massive the cluster, the higher its concentration; van den Bergh 1994, see also Fig.~4 of Larsen 2006).  This rough correlation being likely of primordial origin (Bellazini et al.~1996), they investigate how the dissolution of low-mass low-concentration clusters affects the temporal evolution of the cluster mass function.  Their results suggest that it may be possible to reproduce the 13\,Gyr-old bell-shaped globular cluster mass function starting either from an initial power-law mass spectrum or from a Gaussian mass function similar to that today.  A detailed study of this effect must still be carried out, however.   \\

If the protogalactic era actually sets a characteristic mass ${\rm log}m_G$ for the protoglobular clouds (or a lower mass limit $m_{low}$ in case of a power-law protoglobular cloud mass spectrum) independent of the host galaxy,
then our gas removal model naturally explains why the initial mass function of old globular clusters is a Gaussian similar to that observed today, while the mass function of present-day massive star clusters is a power-law.  
That the initial mass functions of old globular clusters and of present-day massive star clusters are different is also suggested by the work carried out by Hunter et al.~(2003). 
In their study of about 1,000 star clusters of the Large and Small Magellanic Clouds, they determine the cluster initial mass function by fitting cluster population models to the mass distribution integrated over age and to the age distribution integrated over mass.  They derive a power-law cluster mass spectrum with a slope $\gtrsim -2$.  They note however that the distributions of the oldest clusters, i.e. those corresponding to Galactic halo globular clusters with respect to their age and mass, do not appear to be extrapolations of the distributions for the less massive and younger clusters.  To explain why the old and massive clusters form a distinct population, Hunter et al.~(2003) propose that they formed under a different cluster initial mass function, presumably a Gaussian similar to that of the Old Halo cluster system.  

In our model, old globular clusters formed out of exclusively massive ($\simeq 10^6\,{\rm M}_{\odot}$) gaseous precursors, spanning a one decade or so mass range (sections \ref{subsec:isoQfGC_mlowmup} and \ref{subsec:cloud_G}).  This is significantly narrower than the 4-decade mass range of present-day giant molecular clouds and giant molecular cloud cores.  That narrow mass range is then broadened and turned into a bell-shaped cluster initial mass function by gas removal, while the $\simeq 10^6\,{\rm M}_{\odot}$ protoglobular cloud mass-scale guarantees that the turnover settles at the observed value.
This difference in the mass distribution of cluster gaseous progenitors does not necessarily imply markedly distinct locations for old globular clusters and young massive star clusters in the fundamental plane of star clusters.  The fundamental plane of star clusters describes how they are distributed in terms of mass, compactness and velocity dispersion.  Since gas removal leads to the expansion of the protocluster and to the loss of a fraction of its stars, specifically those with the hotest kinematics, it certainly plays a key role in assigning star clusters to their initial location in the fundamental plane.  The high mass-scale of protoglobular clouds may result from a different distribution of their sources of support against gravitational collapse.  That is, the contributions of thermal motions, turbulence and magnetic fields to the protoglobular cloud overall velocity dispersion may be different from what is observed today.  For instance, Fall \& Rees (1985) build on the hypothesis that protoglobular clouds are thermally supported only, while Harris \& Pudritz (1994) propose that the high mass of the protoglobular clouds (the star forming cores of their "supergiant molecular clouds") stems from strong non-thermal motions driven by magnetic fields and turbulence. While gas removal depends on gas velocity dispersion, the lower the velocity dispersion, the quicker the gas removal (e.g. Elmegreen \& Efremov 1997, Kroupa \& Boily 2002), the source of the velocity dispersion itself is of little relevance.  As a result, cluster forming clouds/cores differing with respect to how they are gravitationally supported should not lead to markedly different initial locations in the star cluster fundamental plane.  Young massive star clusters do
occupy a wider region of the so-called $\kappa$-space (i.e. the fundamental plane described in terms of cluster mass and cluster compactness) than the old globular clusters do, even when the former are aged to 10\,Gyr using Simple Stellar Population models in order to account for their stellar evolution-driven fading (see Fig.~8 of Bastian et al.~2006).
This difference is most probably the imprint of the cluster dynamical evolution (rather than differences in the formation process), in the sense that only a narrow region of the mass-radius plane of globular clusters is stable against a Hubble-time of dynamical evolution (Gnedin \& Ostriker 1997).

While our simulations suggest the existence of a high-mass scale for the protoglobular clouds, which the protogalactic era may actually be responsible for, we emphasise that our model still lacks a 
crucial ingredient which prevents us from drawing any definitive conclusion in this respect.  
The $F_{bound}$ vs. $\epsilon$ relation used in our Monte-Carlo simulations has been 
derived for the specific case of isolated clusters, that is, it ignores the effect of 
an external tidal field upon the protocluster evolution.  Since residual gas expulsion 
leads to protocluster expansion (see e.g. Fig.~7 of Goodwin 1997, Fig.~1 of Geyer \& Burkert 2001), a 
fraction of otherwise supposedly bound stars may be driven beyond the protocluster tidal 
radius.  That is, an external tidal field unbinds a larger fraction of 
protocluster stars than in the case of an isolated group of stars (i.e. the
$F_{bound}$ vs. $\epsilon$ relations of Fig.~\ref{fig:Fb_SFE} define upper limits rather 
than one-to-one relations).  At a given galactocentric distance, a low-mass cluster 
has a smaller tidal radius and, therefore, a greater probability of additional star loss 
(with respect to Fig.~\ref{fig:Fb_SFE}) than a high-mass one.  The galactic tidal field 
may thus well lead to additional destruction of low-mass clusters. 
This effect may also be enhanced
by the fact that low-mass clusters preferentially arise from low star formation 
efficiencies, inducing in turn a greater spatial expansion of their stellar component
(see e.g. Fig.~3 of Geyer \& Burkert 2001).  Detailed $N$-body simulations are therefore 
required to investigate whether an external tidal field may influence the turnover
location.  For instance, let us consider the case of a small lower limit for the
cloud mass range, say, ${\rm log}m_{low} \simeq 4$ (or a characteristic protogobular
cloud mass ${\rm log}m_G << 10^6\,{\rm M}_{}\odot$).  In the present stage of our model,
in which clusters are regarded as isolated star systems, the resulting turnover 
location is predicted at ${\rm log}m_{TO} \lesssim {\rm log}m_{low} \simeq 4$, 
more than an order of magnitude below what is observed.  The question is whether 
a model accounting for the galactic tidal field properly may drive that low-mass 
turnover towards higher cluster masses, possibly up to the equilibrium location 
at ${\rm log}m_{TO} \simeq 5.3$, as a result of protocluster tidal truncation.
Should this be the case, a lower-mass limit $m_{low}$ for the cloud mass range 
significantly less than $6 \times 10^5\,{\rm M}_{\odot}$ (or a characteristic 
protoglobular cloud mass $<< 10^6\,{\rm M}_{\odot}$), not necessarily 
universal among galaxies, may actually be acceptable.  We note however that if an external tidal field
contributes significantly to the shape of the cluster initial mass function by removing 
low-mass protoclusters and creating a turnover 
even before secular evolution sets in, then the effect should also be observed for the
mass function of young massive star clusters formed in the dense environment of starbursts galaxies.
Clearly, we need more detailed protocluster gas removal modelling (taking into account external
effects) as well as more surveys of young star cluster systems located beyond the Local Group, 
going deep and correcting for bright star contamination and incompleteness effects.

\subsection{The issue of discreteness}
Up to now, each of our Monte-Carlo simulations has encompassed $10^7$ clouds.  With the star formation efficiency distribution $\wp (\epsilon)$ inferred in section \ref{subsec:isoQfGC_mlowmup}  ($\delta = -2.9$, $r_c = 0.025$), this corresponds to an initial number of clusters of about 44,000 (i.e. $\simeq$ 44,000 clouds achieve $\epsilon > \epsilon _{th}$).  In Figs.~\ref{fig:isoQfGC_mlowmup}-\ref{fig:OH_G}, the evolved cluster mass functions have been normalised so as to contain 72 clusters with $10^4 < m < 10^6\,{\rm M}_{\odot}$, namely the number of Old Halo clusters in that mass range.  The initial cluster mass functions have been normalised accordingly. 
With so high an initial number of clusters in each simulation, discreteness issues inherent to any size-limited sample are avoided and the results (cluster mass functions, survival rates $F_M$ and $F_N$ and present-day cluster mass fractions $f_{GC}^{13Gyr}$) are independent of the random seeds.  Nevertheless, Table \ref{tab:OHMF_var_PL} and Table \ref{tab:OHMF_var_G} show that the fraction $F_N$ of surviving clusters  is $\lesssim 0.5$, implying an initial number of Old Halo clusters $\gtrsim$ 150 only.  Therefore, it is interesting to investigate how the fiducial cases behave if the number of protoglobular clouds is limited so as to lead to a cluster system containing as many members as the Old Halo at an age of 13\,Gyr.  

\begin{figure}
\begin{minipage}[t]{\linewidth}
\centering\epsfig{figure=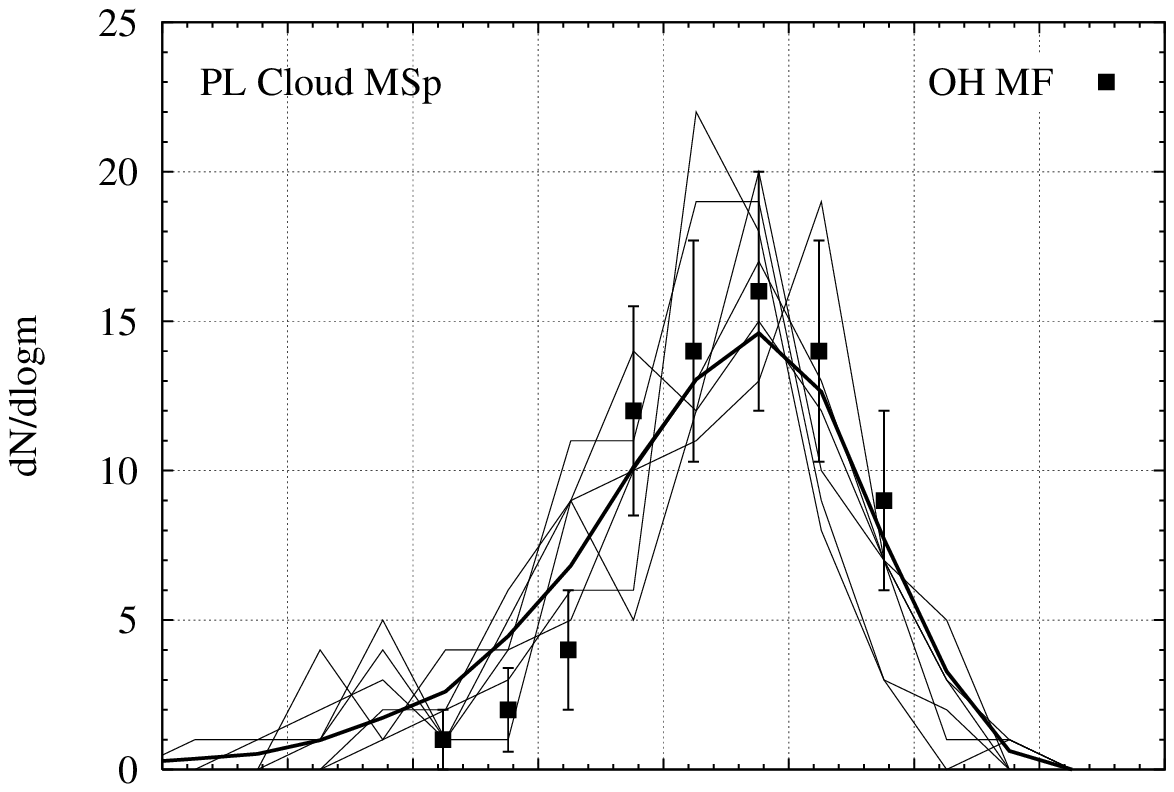, width=\linewidth} 
\end{minipage}
\vfill
\vspace{-8mm}
\begin{minipage}[t]{\linewidth}
\centering\epsfig{figure=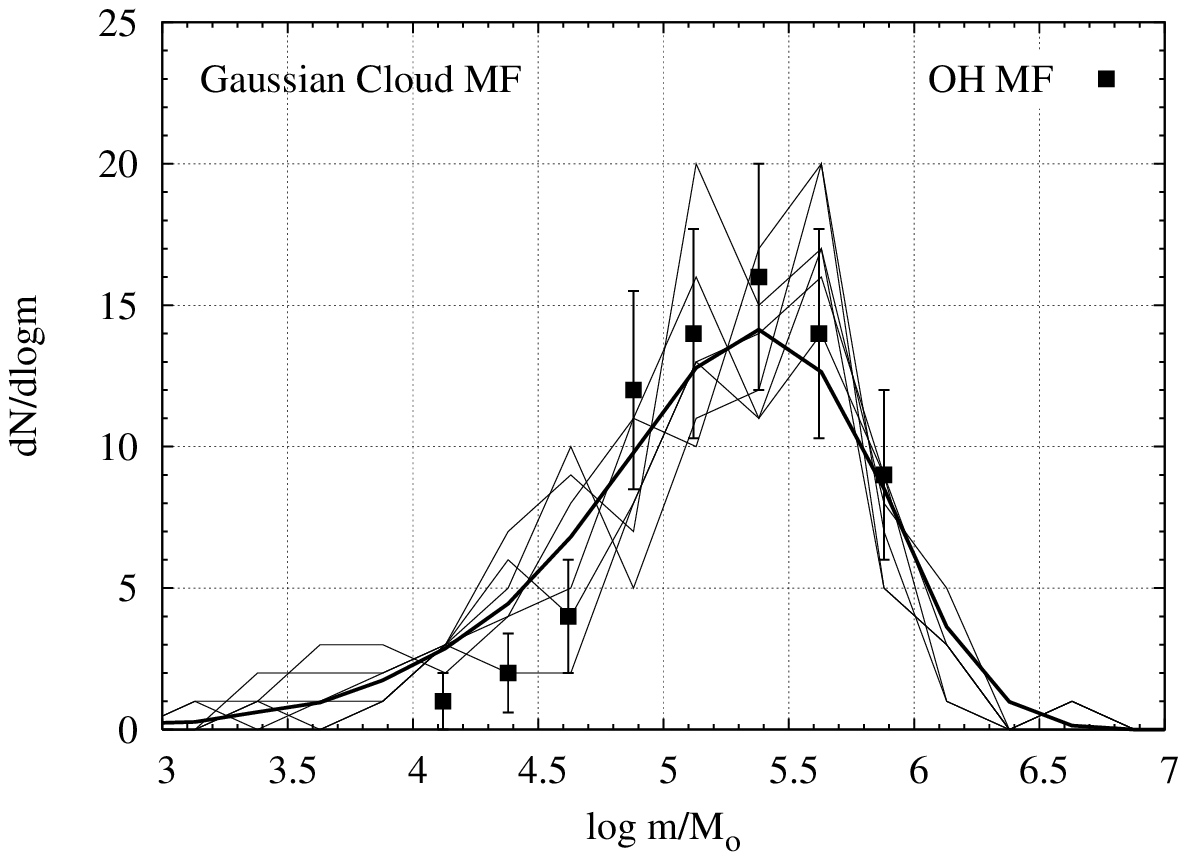, width=\linewidth}
\end{minipage}
\caption{{\it Top panel:} Evolved cluster mass functions at an age of 13\,Gyr for the fiducial case of a power-law cloud mass spectrum.  The thick and thin curves correspond to protoglobular cloud numbers of $N_{cloud}=10^7$ and $N_{cloud}=43,000$, respectively.  In the second case, the limited size of the sample results in a different mass function for each simulation.  The Old Halo mass distribution is shown as the full squares.  {\it Bottom panel:} Same in case of an underlying Gaussian cloud mass function.    } 
\label{fig:OH_disc} 
\end{figure}

This section presents the results of simulations for which the number of protoglobular clouds is $N_{cloud} \simeq 43,000$. We have reconsidered both fiducial cases, namely, the power-law cloud mass spectrum and the Gaussian cloud mass function.  For each cloud mass distribution, Tables \ref{tab:discretness_PL} and \ref{tab:discretness_G} present the goodness of fit $Q$, the present-day cluster mass fraction $f_{GC}^{13Gyr}$, the survival rates $F_M$ and $F_N$ and the number of clusters $N_{GC}$ at an age of 13\,Gyr for a sequence of six simulations.  The mass range of the Old Halo clusters being $10^4 < m < 10^6\,{\rm M}_{\odot}$, the column $N_{GC}$ quotes both the total number of clusters and the number of clusters in the range $10^4 < m < 10^6\,{\rm M}_{\odot}$.  This second number is used to compare the final size of the modelled cluster system to that of the Old Halo (i.e. 72 clusters) consistently.  The corresponding cluster evolved mass functions are shown and compared to the observed one in Fig.~\ref{fig:OH_disc}.  In contrast to the previous simulations, no vertical shift has been applied to the modelled distributions.  Also shown as the thick solid lines are the mass functions derived if $N_{cloud}=10^7$ (i.e., no discreteness issue, see sections \ref{subsec:isoQfGC_mlowmup} and \ref{subsec:cloud_G}).   

The successive simulations lead to widely different values for the incomplete gamma function, ranging from unacceptable fits ($Q=10^{-5}$) up to excellent ones ($Q \gtrsim 0.1$).  Globally however, most fits are acceptable with $Q>0.001$.  

\begin{table}
\begin{center}
\caption[]{Effects of sampling noise: Results of different simulations run with $N_{cloud} \simeq 43,000$ in the fiducial case of the power-law protoglobular cloud mass spectrum.  The limited initial number of clusters ($\simeq 190$) implies that each run provides a different cluster mass function, as shown by the thin solid lines in Fig.~\ref{fig:OH_disc}.  In addition to the values of $Q$, $f_{GC}^{13Gyr}$, $F_M$ and $F_N$, the table also lists the number $N_{GC}$ of clusters which survive over a Hubble-time, both total and in the mass range $10^4 < m < 10^6\,{\rm M}_{\odot}$, i.e. the mass range of the Old Halo clusters.} 
\label{tab:discretness_PL}
\begin{tabular}{ l | c c c c c  } \hline 
 & $Q$         & $f_{GC}$ & $F_M$  & $F_N$  & $N_{GC}$   \\ \hline
 & $0.068$     & $0.019$  & $0.30$ & $0.46$ & ($87,75$)                \\ 
 & $0.001$     & $0.016$  & $0.25$ & $0.39$ & ($74,71$)                \\ 
 & $0.420$     & $0.018$  & $0.29$ & $0.42$ & ($79,70$)                \\ 
 & $0.010$     & $0.013$  & $0.21$ & $0.38$ & ($72,68$)                \\ 
 & $0.226$     & $0.019$  & $0.29$ & $0.43$ & ($81,71$)                \\ 
 & $10^{-5}$   & $0.014$  & $0.25$ & $0.45$ & ($86,80$)                \\ \hline
\end{tabular}
\end{center}
\end{table}   

\begin{table}
\begin{center}
\caption[]{Same as Table \ref{tab:discretness_PL}, but for the fiducial case of the Gaussian cloud mass function. } 
\label{tab:discretness_G}
\begin{tabular}{ l | c c c c c  } \hline 
 & $Q$                 & $f_{GC}$ & $F_M$  & $F_N$  & $N_{GC}$   \\ \hline
 & $0.065$             & $0.019$  & $0.30$ & $0.44$ & ($85,75$)                \\ 
 & $2 \times 10^{-4}$  & $0.018$  & $0.28$ & $0.40$ & ($76,73$)                \\ 
 & $0.326$             & $0.017$  & $0.27$ & $0.38$ & ($72,67$)                \\  
 & $4 \times 10^{-5}$  & $0.016$  & $0.25$ & $0.47$ & ($90,83$)                \\ 
 & $0.139$             & $0.019$  & $0.30$ & $0.47$ & ($90,77$)                \\ 
 & $0.021$             & $0.016$  & $0.25$ & $0.38$ & ($73,68$)                \\ \hline
\end{tabular}
\end{center}
\end{table}   

\subsection{The origin of halo field stars}
 
In the present-day Galaxy, the formation of unbound stellar groups is the rule and not the exception
(see section \ref{subsubsec:delta}).  Most field stars in the Galactic disc likely originate from embedded clusters which either lost a fraction of their original members or were disrupted while emerging out of their natal clouds.  Embedded clusters may therefore well be the fundamental units of star formation in the sense that they may account for the formation of the vast majority of all stars in the Galaxy (Lada \& Lada 2003).      

It is likely that that paradigm also characterizes the formation of the stellar halo.  Actually, 98\,\% of its mass consists of field stars and so large a mass fraction cannot be accounted for by the secular evaporation and disruption of globular clusters over a Hubble-time, regardless of the shape of the cluster initial mass function (e.g. Vesperini 1998, Baumgardt 1998, Parmentier \& Gilmore 2005).  For instance, Tables \ref{tab:discretness_PL} and \ref{tab:discretness_G} demonstrate that the initial total mass in halo clusters is at most $10^8\,{\rm M}_{\odot}$ (see the survival rate $F_M$), which is an order of magnitude lower than the mass of the stellar halo.  The present-day overwhelming fraction of field stars thus constitutes an "{\it ab initio}" feature of the Galactic halo.  It is worth noting that this may help understand why field stars and globular cluster stars sometimes show different patterns in their light element abundances.  For instance, the group of O-poor and Na-rich stars observed in some globular clusters is hardly detected in the field (Carretta, Gratton \& Sneden 2000).  Chemical abundance anomalies observed in globular cluster stars are usually ascribed to the accretion onto their surface of stellar winds of intermediate mass stars ascending the asymptotic giant branch, a process made feasible by the dense stellar environment characteristic of globular clusters.  With the possible exception of binary systems, {\it ab initio} field stars remain unaffected by this external pollution, their external layers having never been exposed to any accretion process.  If most present-day field stars have actually never belonged to bound globular clusters, then the absence of abundance anomalies characteristic of the asymptotic giant branch phase for these stars is not surprising.   

The violent relaxation phase affecting protoclusters following the expulsion of their residual star forming gas constitutes a prime candidate to explain the origin of field stars in the Galactic halo, without conflicting with the well-accepted paradigm following which most stars form in clusters.  In our fiducial cases, at an age of 13\,Gyr, the total mass fraction of field stars $f_{FS}=98\,\%$ in the halo arises from the three following contributions (results for the power-law cloud mass spectrum and the Gaussian cloud mass function are similar): 
\begin{enumerate}
\item the disruption of protoclusters in the $\epsilon < \epsilon _{th}$ regime (infant mortality): $f_{FS}^a=91\,\%$,
\item the loss of stars by protoclusters for which $\epsilon > \epsilon _{th}$ (infant weight-loss): $f_{FS}^b=4\,\%$, 
\item the evaporation and the disruption of globular clusters over a Hubble-time (adult mortality):  $f_{FS}^c=3\,\%$.
\end{enumerate}

Most halo field stars would thus be given off by star forming regions whose efficiency $\epsilon$ is less than the star formation efficiency threshold $\epsilon _{th}$.  In this respect, the large value for $f_{FS}^a$ directly results from the steep slope of $\wp (\epsilon)$, namely, $\delta =-2.9$ (combined with the assumption of instantaneous gas removal, that is, $\epsilon _{th}$ takes its highest possible value and, thus, strongly limits the number of clouds with $\epsilon > \epsilon _{th}$).  We caution however that this result is utterly model-dependent, having been derived under the assumption of a single functional form for the star formation efficiency distribution $\wp (\epsilon)$ valid over the entire range $0 < \epsilon <1$, which may not be true.  

\section{Discussion and Conclusions}
\label{sec:conclu}

In this paper, we have presented simulations highlighting how the mass function of 
protoglobular clouds evolves into that of gas-free bound star clusters as a result of
the expulsion of the residual star forming gas due to supernova activity.  The initial 
mass of a star cluster depends on the mass fraction of its gaseous precursor which is 
turned into {\it bound} stars.  To retain a bound core of stars, a star forming region 
must achieve a star formation efficiency threshold $\epsilon _{th}$ of about 33\% 
(although this threshold gets lower if the gas is removed non-explosively, 
see section \ref{subsubsec:sfeth}, and/or if the number of stars in the embedded cluster 
is a few thousands only).  Above this critical 
value, the fraction $F_{bound}$ of bound stars steadily increases from 0 
(when $\epsilon \simeq \epsilon _{th}$) up to 1 (when $\epsilon \lesssim 1$) 
(see Fig.~\ref{fig:Fb_SFE}).  Therefore, the initial mass of a star cluster does not 
depend on the star formation efficiency only.  It is the bound star formation efficiency 
$F_{bound} \times \epsilon$ which is important (equation \ref{eq:minit}).  By virtue of 
the large variations in $F_{bound} \times \epsilon$ (i.e. between 0 and 1), the cluster 
initial mass function does {\it not} necessarily mirror the cluster forming cloud mass function.    
In order to investigate this effect, we have convolved cloud mass spectra ${\rm d}N/{\rm d}m$ 
with star formation efficiency distributions $\wp (\epsilon)$, each value of the star formation 
efficiency $\epsilon$ being coupled to the corresponding fraction $F_{bound}$ of stars remaining 
bound following gas removal.  We have explicitely assumed that $\epsilon$ is independent of the 
cloud mass (section \ref{subsec:PL_G}).

Starting the simulations with power-law cloud mass spectra truncated at a lower-mass limit $m_{low}$, 
we have shown that the resulting cluster initial mass functions are bell-shaped.  
Since the gaseous precursors of old globular clusters (i.e. those formed during the protogalactic era) 
may actually be characterized by a lower mass limit (section \ref{subsec:proto_gx}), 
this result provides support to the early suggestion made by Vesperini (1998) that the 
globular cluster initial mass function is a Gaussian similar to that observed today
(i.e. an equilibrium cluster mass function, see section \ref{subsec:eqMF}).  
That is, the present-day Gaussian mass function is the preserved imprint of the 
cluster formation process rather than the imprint of cluster system dynamical evolution.  Consequently, our model also provides a new approach to investigate 
the origin of the universal location of the globular cluster mass function turnover.

In this respect, we have discussed how the cluster initial mass function responds 
to variations in the input parameters of our model (see section \ref{subsec:TO}).  
We have successively varied the slope $\alpha$ of the cloud mass spectrum,
its lower and upper limits $m_{low}$ and $m_{up}$, the slope $\delta$ and the scale-length $r_c$
of the star formation efficiency distribution $\wp (\epsilon)$, and the efficiency threshold 
$\epsilon _{th}$ required to retain a bound core of stars (equivalently the gas removal time-scale 
$\tau _{gr}$ measured in units of the protocluster crossing-time $\tau _{cross}$).  We
find that the turnover location is mostly sensitive to the lower limit $m_{low}$ of the 
protoglobular cloud mass spectrum.  The observed universality of the turnover of the old globular 
cluster mass function would therefore originate from a common value among galaxies for 
the lower mass limit of protoglobular clouds, possibly with second-order variations 
driven by differences in the slopes of the cloud mass spectrum, that of the efficiency 
distribution $\wp (\epsilon)$, as well as by differences in the gas removal time-scale. 
 
Combining our model generating cluster initial mass functions with the cluster 
evolutionary model of Baumgardt \& Makino (2003), we have subsequently 
investigated which input parameters reproduce both the present-day
mass function of the Old Halo clusters and the present-day mass fraction
in the stellar halo (section \ref{subsec:isoQfGC_mlowmup}).  
Considering the case of a power-law mass spectrum
for the protoglobular clouds with a spectral index $\alpha \simeq -1.7$ (as is often
reported for the present-day giant molecular clouds and their cores) and explosive gas removal
(i.e. the $F_{bound}$ vs. $\epsilon$ relation considered is the solid curve in 
Fig.~\ref{fig:Fb_SFE}), we find a good fit for $\delta \simeq -2.9$, $r_c \simeq 0.025$, 
$m_{low} \simeq 6 \times 10^5\,{\rm M}_{\odot}$ and $m_{up} \ge 5 \times 10^6\,{\rm M}_{\odot}$.
With respect to the cluster mass function, the most critical parameter is the cloud
mass lower limit $m_{low}$, since it sets the turnover location.  The 
cloud mass upper limit $m_{up}$ affects the detailed shape of the mass function beyond 
the turnover without altering the location of the latter.  Variations in $m_{up}$
may thus explain why the high-mass regime of the cluster mass function 
differs among galaxies, even though the turnover remains unaffected 
(see section \ref{subsec:mup}).  Variations in the slope $\delta$ and core $r_c$ of the
probability distribution $\wp (\epsilon)$ of the star formation efficiency weakly
influence the shape of the cluster mass function at an age of 13\,Gyr 
(see e.g. Table \ref{tab:OHMF_var_PL} and Fig.~\ref{fig:OHMF}).
These parameters are mostly used to fix the global star formation efficiency
$\overline {\epsilon}$ (i.e. the mass fraction of gas turned into stars at the scale
of the protoglobular cloud system) and the mean bound star formation efficiency
$\overline {\epsilon _b}$ (i.e., the mass fraction of gas ending up in clustered stars 
following gas removal) (see the top panel of Fig.~\ref{fig:isoQfGC_mlowmup}).
Specifically, for a given value of $\overline {\epsilon}$,
the slope $\delta$ sets, in turn, the fraction of protoglobular clouds achieving
$\epsilon > \epsilon _{th}$, $\overline {\epsilon _b}$ and the present-day mass 
fraction of clusters in the
halo (see equation \ref{eq:fGC2}).  The total mass of the stellar halo is dominated
by field stars and this overwhelming mass fraction (98\,\%) cannot be accounted for 
by the secular dynamical evolution of globular clusters.  The Galactic stellar halo 
has thus been dominated by field stars since its earliest stages.
In our model, this feature is accounted for by a steep value for the slope of $\wp (\epsilon)$,
i.e. $\delta =-2.9$.  This results in the vast majority of the protoglobular clusters
going into the infant mortality regime, that is, they are dislocated following gas removal
and their stars are scattered into the field.  

Finally, we have also investigated whether a bell-shaped cluster initial mass function
similar to that observed in old clusters today may arise from a characteristic cloud mass, 
the cluster mass spread resulting from the range in the bound star formation efficiency 
$F_{bound} \times \epsilon$.  Actually, an ensemble of protoglobular clouds with a mass function
centered around ${\rm log}m_G \simeq 6.1-6.2$ and a standard deviation $\sigma \lesssim 0.4$
reproduces the Old Halo cluster mass function, the turnover location being mostly driven 
by ${\rm log}m_G$ (see section \ref{subsec:cloud_G}).

In order to generate a Gaussian globular cluster initial mass function similar to that today,
our model thus requires a protoglobular cloud mass-scale of order $10^6\,{\rm M}_{\odot}$.  The shape 
of the $\simeq$ 1-decade cloud mass distribution (truncated power-law or Gaussian) is of little relevance, 
as gas removal broadens it and turns it into a bell-shaped cluster initial mass function
(although the shape of the cloud mass distribution does influence the detailed shape of the cluster initial mass function).  The observed location
of the turnover at a cluster mass of $\simeq 2 \times 10^5\,{\rm M}_{\odot}$ stems from the 
$10^6\,{\rm M}_{\odot}$ mass-scale.  Therefore, the universality of the turnover of the globular 
cluster mass function
would originate mostly from a common protoglobular cloud mass-scale  among galaxies.  That such a 
high mass-scale did actually exist remains debated (section \ref{subsec:proto_gx}).
In contrast, present-day giant molecular clouds and their cores show no such cut-off at high mass.
If their power-law mass spectrum is truncated at, say, $100\,{\rm M}_{\odot}$, the turnover 
generated by our model gets located at similar a cluster mass and the inferred cluster initial mass 
spectrum is a power-law down to $\lesssim 100\,{\rm M}_{\odot}$.  

Therefore, with the mass-scale of the cluster gaseous progenitors as its key-parameter, our model 
can account for both the Gaussian globular cluster initial mass function and the observed
power-law mass spectrum of young massive star clusters formed in starbursts and mergers.  

We emphasize however that the gas-removal phase of our model does not include the impact of external 
effects, e.g., the tidal field of the host galaxy.

While the mass function of globular clusters is a bell-shape, that of  open clusters is a power-law 
(van den Bergh \& LaFontaine 1984, Battinelli, Brandimarti, Capuzzo-Dolcetta 1994).  As for massive young 
clusters formed beyond the Local Group, the difference may be driven by the presence or not of a lower mass
limit for the cluster forming clouds/cores.  Additionally, 
we caution that the application of our model is necessarily 
restricted to the initial mass function of {\it massive} star clusters.  Actually, 
as for low-mass clusters (e.g. open clusters), their relaxation time-scale is significantly 
shorter by virtue of their smaller number of stars.  As a result, the gravitational 
interactions between stars contribute to cluster dynamical evolution at an earlier time,
while the exposed cluster is attempting to settle to a new state of equilibrium.  
This eases the formation of a bound core of stars, thereby altering the 
$F_{bound}$ vs. $\epsilon$ relation with respect to what is shown in Fig.~\ref{fig:Fb_SFE}.  
For instance, despite applying a tidal field and despite considering instantaneous gas
removal, Kroupa, Aarseth \& Hurley (2001) find that a substantial bound core of stars 
$F_{bound} \simeq 0.33$ remains even if $\epsilon$ is as low as $\simeq 0.33$.  
Accounting for the variations in the bound fraction of stars as a function of the 
protocluster mass will eventually lead to a picture of the overall cluster mass function,
from the low-mass regime ($100$ - $1,000\,{\rm M}_{\odot}$) up to the high-mass one 
($10^6$-$10^7\,{\rm M}_{\odot}$) (Kroupa \& Boily 2002).

\bsp

\section*{Acknowledgments}
This research was supported by a Marie Curie Intra-European
Fellowships within the $6^{th}$ European Community Framework 
Programme.  GP also acknowledges support from the Belgian 
Science Policy Office in the form of a Return Grant.

\label{lastpage}

\end{document}